\documentclass[twoside,journey]{IEEEtran}
\usepackage{makecell}
\usepackage{array}
\usepackage{graphicx,amssymb,amsmath}
\usepackage{multicol}
\usepackage[noadjust]{cite}
\usepackage{setspace}
\usepackage{subfigure}
\usepackage{graphicx}
\usepackage{float}
\usepackage {url}
\usepackage{stfloats}
\usepackage{amsthm,pifont}
\usepackage{flushend}
\usepackage{cases,subeqnarray}
\usepackage{bm,multirow,bigstrut}
\usepackage{amsmath, amsthm, amssymb}
\usepackage{textcomp}
\usepackage{latexsym,bm}
\usepackage{booktabs}
\usepackage{xcolor}
\usepackage{mathtools}
\usepackage{dsfont}
\usepackage{extarrows}
\usepackage{epsfig}
\usepackage{epsfig}
\usepackage{epstopdf}
\usepackage[noend]{algpseudocode}
\usepackage{algorithmicx,algorithm}

\usepackage{cite}
\usepackage{bm}
\usepackage{multicol}       
\usepackage{multirow}       
\usepackage{array}          
\usepackage{colortbl}
\usepackage{makecell}
\definecolor{crimson}{RGB}{192,0,0}         
\definecolor{navy}{RGB}{47,85,151}         
\usepackage{bbding}
\usepackage{graphicx}
\usepackage{booktabs}
\usepackage{algorithm}
\usepackage{algpseudocode}
\usepackage{diagbox}

\usepackage{graphicx}
\usepackage{setspace}
\usepackage{cleveref}

\usepackage{url}
\usepackage{mathtools}
\def\BibTeX{{\rm B\kern-.05em{\sc i\kern-.025em b}\kern-.08em
 T\kern-.1667em\lower.7ex\hbox{E}\kern-.125emX}}

\usepackage{stackengine}
\usepackage{enumerate}

\newcommand{\tabitem}{~~\llap{\textbullet}~~}

\usepackage{multicol}
\usepackage[noend]{algpseudocode}
\usepackage{pgfplots}
\makeatletter
\def\BState{\State\hskip-\ALG@thistlm}
\makeatother
\usepackage{multirow}
\definecolor{Blues}{RGB}{0,0,0}
\definecolor{mygray}{gray}{.8}
\definecolor{mygray2}{gray}{.7}
\definecolor{mygray3}{gray}{.6}

\algdef{S}[FOR]{ForEach}[1]{\algorithmicforeach\ #1\ \algorithmicdo}

\theoremstyle{plain}

\theoremstyle{plain}

\usepackage{amsmath}

\IEEEoverridecommandlockouts

\begin{document}
\title{A Tutorial on Extremely Large-Scale MIMO for 6G: Fundamentals, Signal Processing, and Applications

\thanks{Z. Wang, J. Zhang, and B. Ai are with the School of Electronic and Information Engineering and Frontiers Science Center for Smart High-speed Railway System, Beijing Jiaotong University, Beijing 100044, China, and Z. Wang is also with the School of Computer Science and Engineering, Nanyang Technological University, Singapore 639798 (e-mail: \{zhewang\_77, jiayizhang, boai\}@bjtu.edu.cn).}

\thanks{H. Du and D. Niyato are with the School of Computer Science and Engineering, Nanyang Technological University, Singapore 639798 (e-mail: hongyang001@e.ntu.edu.sg; dniyato@ntu.edu.sg).}

\thanks{S. Cui is with the School of Science and Engineering (SSE) and the Future Network of Intelligent Institute (FNii), the Chinese University of Hong Kong (Shenzhen), Shenzhen 518172, China. He is also affiliated with the Department of Broadband Communication, Peng Cheng Laboratory, Shenzhen 518000, China (email: shuguangcui@cuhk.edu.cn).}

\thanks{M. Debbah is with the Center for 6G Technology, the Khalifa University of Science and Technology, Abu Dhabi, United Arab Emirates (e-mail: merouane.debbah@ku.ac.ae).}

\thanks{K. B. Letaief is with the Department of Electrical and Computer Engineering, Hong Kong University of Science and Technology (HKUST), Hong Kong (e-mail: eekhaled@ust.hk).}

\thanks{H. V. Poor is with the Department of Electrical and Computer Engineering, Princeton University, Princeton, NJ 08544 USA (e-mail: poor@princeton.edu).}}

\author{Zhe Wang, Jiayi Zhang,~\IEEEmembership{Senior Member, IEEE}, Hongyang Du, Dusit Niyato,~\IEEEmembership{Fellow,~IEEE}, Shuguang~Cui,~\IEEEmembership{Fellow,~IEEE}, Bo Ai,~\IEEEmembership{Fellow,~IEEE},  M{\'e}rouane~Debbah,~\IEEEmembership{Fellow,~IEEE},  Khaled~B.~Letaief,~\IEEEmembership{Fellow,~IEEE}, and H.~Vincent~Poor~\IEEEmembership{Fellow,~IEEE}}
\maketitle
\vspace{-1cm}
\begin{abstract}
Extremely large-scale multiple-input-multiple-output (XL-MIMO), which offers vast spatial degrees of freedom, has emerged as a potentially pivotal enabling technology for the sixth generation (6G) of wireless mobile networks. With its growing significance, both opportunities and challenges are concurrently manifesting. This paper presents a comprehensive survey of research on XL-MIMO wireless systems. In particular, we introduce four XL-MIMO hardware architectures: uniform linear array (ULA)-based XL-MIMO, uniform planar array (UPA)-based XL-MIMO utilizing either patch antennas or point antennas, and continuous aperture (CAP)-based XL-MIMO. We comprehensively analyze and discuss their characteristics and interrelationships. Following this, we introduce several electromagnetic characteristics and general distance boundaries in XL-MIMO. Given the distinct electromagnetic properties of near-field communications, we present a range of channel models to demonstrate the benefits of XL-MIMO. We further discuss and summarize signal processing schemes for XL-MIMO. It is worth noting that the low-complexity signal processing schemes and deep learning empowered signal processing schemes are reviewed and highlighted to promote the practical implementation of XL-MIMO. Furthermore, we explore the interplay between XL-MIMO and other emergent 6G technologies. Finally, we outline several compelling research directions for future XL-MIMO wireless communication systems.
\end{abstract}
\begin{IEEEkeywords}
XL-MIMO, channel modeling, near-field communications, deep learning, signal processing.
\end{IEEEkeywords}
\IEEEpeerreviewmaketitle
\section{Introduction}
\subsection{Motivation}
Since 2020, the fifth generation (5G) of wireless communication networks has witnessed widespread deployment and development globally. The corresponding application scenarios include enhanced mobile broadband (eMBB), massive machine type communications (mMTC), and ultra-reliable and low latency communications (URLLC). 5G has been designed to meet ambitious performance benchmarks, including 10 Gb/s uplink peak data rate, 20 Gb/s downlink peak data rate, 5 ms end-to-end latency, and 99.999\% end-to-end reliability, among others \cite{7894280,6824752,7414384}. Several promising key technologies, such as massive multiple-input-multiple-output (mMIMO), millimeter wave (mmWave)~\cite{du2021millimeter}, and ultra-dense networking (UDN), have been implemented to meet these 5G requirements~\cite{zheng2021eicic}.
The anticipated sixth generation (6G) of wireless networks, expected to serve communication needs beyond 2030, have sparked significant research interest due to the advent of novel requirements following the rapid evolution of wireless applications \cite{you2021towards,9390169,9170653,8869705,9040264,rajatheva2020white,8808168}. Compared with 5G networks, 6G networks are anticipated to achieve a 100-fold increase in peak data rate (reaching the Tb/s level), a tenfold reduction in latency, and an end-to-end reliability requirement of 99.99999\%. These new requirements have inspired novel application scenarios by extending conventional 5G eMBB, mMTC, and URLLC into their enhanced 6G counterparts: further eMBB, ultra-mMTC, and extensive variations of URLLC \cite{8869705,yang2023bu}.
Addressing these extremely high expected demands on 6G, several promising technologies are garnering substantial attention, such as extremely large-scale multiple-input-multiple-output (XL-MIMO) \cite{ZheMag,9903389,2022arXiv221201257G}, reconfigurable intelligent surfaces (RIS) \cite{8811733,8910627,enyusurvey,9743355,du2022semantic}, and Terahertz-band communications \cite{6882305,8732419,kurner2014towards}. All these technologies are promising and can be integrated with each other. In this survey, we focus on XL-MIMO technology, which can significantly enhance the spectral efficiency (SE) and spatial degrees of freedom (DoF) of wireless networks \cite{ZheMag,9903389,2022arXiv221201257G}.

As an emerging paradigm evolution of MIMO, XL-MIMO is the next step in developing MIMO technology. MIMO technology has been a key enabler of high-rate wireless communication since the fourth generation (4G) of wireless networks \cite{biglieri2007mimo,10144733,1266912,foschini1996layered,paulraj1994increasing,1197843}. With the rapid development of wireless communications and more critical communication demands of users~\cite{du2022attention}, the concept of mMIMO was proposed in \cite{5595728}. Compared with basic MIMO technology, mMIMO involves much larger numbers of antennas (up to hundreds of antennas) at a wireless base station (BS). With the aid of many antennas, the so-called channel hardening phenomenon and the favorable propagation conditions can be exploited in mMIMO to reduce inter-cell interference and achieve better SE performance. MIMO systems can be divided into two implementation paradigms: centralized and distributed. As for the distributed implementation, the multi-cell scenario is a general case where distributed antennas are deployed among multiple serving cells. This concept includes coordinated multipoint (CoMP) \cite{5706317,boldi2011coordinated} and distributed antenna systems (DAS) \cite{5490977,4114255}. However, the concepts of cell and cell boundary still exist, and users at cell boundaries may suffer from strong inter-cell interference. To cope with this problem, a distributed mMIMO paradigm called cell-free massive MIMO (CF mMIMO) has been proposed in \cite{7827017}. Compared with cellular mMIMO, the concept of cell boundary has been removed in CF mMIMO systems, where a large number of distributed access points (APs) are deployed in a wide area to serve the user equipment (UE) by joint transmission and reception \cite{9113273,[162],04962,9586055,9416909,10013728,9174860}. It is worth noting that, in this setting, all APs are connected to central processing units (CPUs), enabling various processing schemes with different levels of cooperation between the APs and CPUs to improve macro-diversity gain and achieve high SE performance uniformly within the coverage area. CF mMIMO is considered to be a key technology for beyond 5G and future wireless communications. 

\begin{table*}[t!]
  \centering
  \fontsize{8}{12}\selectfont
  \caption{Differences between XL-MIMO and conventional mMIMO and their challenges/potentials.}
  \label{Comparisons}
   \begin{tabular}{ !{\vrule width1.2pt}  m{3.2 cm}<{\centering} !{\vrule width1.2pt}  m{5.6 cm}<{\centering} !{\vrule width1.2pt}  m{4.8 cm}<{\centering} !{\vrule width1.2pt}}

    \Xhline{1.2pt}
        \rowcolor{gray!50} \bf New characteristics and features & \bf Challenges  & \bf Potentials \cr
    \Xhline{1.2pt}
    More flexible hardware designs & New hardware designs based on emerging design ideas and materials: ULA-based, UPA-based, and CAP-based XL-MIMO & The specific XL-MIMO design can be chosen based on their specific needs, network constraints, and desired performance metrics.\cr\hline
    A much larger number of antennas & Additional design requirements, higher power consumption, much higher signal processing complexity, and consequent EM characteristics, such as the spatial non-stationarity and severe mutual coupling   & Excellent SE and DoF performance \cite{ZheMag} \cr\hline
    The Much smaller antenna spacing & Increased hardware complexity and severe mutual coupling property  & Ability to deploy an extremely large number of antennas\cr\hline
    New EM characteristics, which are omitted or inapplicable to be considered in mMIMO &The consideration of new EM characteristics such as the spherical wave characteristic \cite{[32]} and spatial non-stationarity  \cite{9170651}, and brand-new near-field based channel models such as the Fourier plane-wave representation based channel models \cite{[41]} & The capture of actual EM characteristics and the potential to implement near-field based signal processing  \cr\hline
    Near-field based signal processing schemes & The urgency to design near-field based signal processing schemes for XL-MIMO due to the mismatch of the conventional far-field based signal processing schemes \cite{ZheMag,9903389} & New promising near-field based applications: near-field beam focusing \cite{9723331} and near-field wireless energy transfer \cite{[102]}, etc.  \cr\hline

    \Xhline{1.2pt}
    \end{tabular}
  \vspace{0cm}
\end{table*}

With the further development of wireless communications and much higher communications demands~\cite{du2023enabling}, the number of antennas and the array aperture are expected to be much larger than those of existing mMIMO technologies. Thus, XL-MIMO stands as a critical facilitator for next-generation wireless networks, offering substantial improvements in spectral efficiency, spatial resolution, and degrees of freedom. These advancements, driven by its deployment of a large array of antennas and substantial array aperture, are essential to satisfying the increasing demands for high-capacity communications \cite{ZheMag,9903389,2022arXiv221201257G}. The basic idea of XL-MIMO is to deploy an extremely large number of antennas in a compact space. Two mainstream implementation approaches to this idea have been investigated in existing works. The first implementation approach is to deploy a huge number of antennas in a compact manner, where an antenna spacing much smaller than the widely used half-wavelength is applied. In this approach, thousands of antennas can be implemented to achieve excellent system performance. The array aperture is discrete due to the discrete deployment of antennas, like that of mMIMO. The second implementation approach is to embed an approximately infinite number of antennas in a compact space in an extremely dense style so that the array aperture can be viewed as being approximately continuous with the aid of meta-materials. This approach is called continuous aperture (CAP) MIMO. An ideal CAP MIMO adopts the form of a spatially-continuous electromagnetic (EM) volume, where an infinite number of infinitesimal antennas are embedded. From a mathematical perspective, this implementation method can also be viewed as the case of the first approach with an infinitesimal antenna spacing. Note that the analytical modeling and practical implementation for the first realization approach are relatively mature. Thus, the first implementation approach has been widely considered in existing works. However, due to the compact deployment, the mutual coupling effect between the discrete antenna elements is severe. As for the second implementation approach, the emerging meta-materials can help to construct the approximately continuous aperture. However, only a few works have investigated the modeling and implementation of emerging meta-materials. Thus, the second approach needs to be further explored. Based on these two implementation approaches, many XL-MIMO designs with different architectures and terminologies have been studied, e.g., \emph{holographic MIMO} \cite{[50],[41],[44],[54]}, \emph{large intelligent surfaces (LISs)} \cite{[23],[24],[56],[11]}, \emph{extremely large-scale antenna arrays (ELAAs)} \cite{9903389,[7]}, and \emph{CAP MIMO} \cite{[70]}. All these XL-MIMO designs are promising enablers for 6G and future wireless communications. However, the characteristics and mutual relationships among these designs have yet to be extensively discussed. Therefore, in this survey, we introduce all these promising XL-MIMO designs and present their features and their similarity and difference.

Compared with conventional mMIMO, the introduction of XL-MIMO involves not only a sharp increase in the number of antennas but also fundamental changes in EM characteristics, which can be relied on to significantly improve the SE performance and the spatial DoF \cite{ZheMag,9903389,2023arXiv230304003L}. We summarize the differences between XL-MIMO and conventional mMIMO and their challenges/potentials in Table~\ref{Comparisons} as:

$\bullet$ \emph{More flexible hardware designs}:
As discussed above, two mainstream realization approaches for XL-MIMO can be implemented. Moreover, many XL-MIMO designs over different architectures and terminologies can be regarded as promising implementation schemes, e.g., \emph{holographic MIMO}, \emph{LISs}, \emph{ELAAs}, and \emph{CAP MIMO}. All these XL-MIMO designs provide a much more flexible implementation than conventional mMIMO. This flexibility arises from the availability of multiple XL-MIMO designs~\cite{[26],[30]}. The specific XL-MIMO design can be chosen based on specific needs, network constraints, and desired performance metrics~\cite{[27]}.
 
$\bullet$ \emph{A Much larger number of antennas}:
Compared with conventional mMIMO, the number of antennas for XL-MIMO is expected to have a higher order of magnitude~\cite{[2],[3]}. Many existing works focused on XL-MIMO consider thousands or even ten thousand antennas for XL-MIMO with discrete aperture, while for the ideal CAP-based XL-MIMO, an infinite number of infinitesimal antennas can theoretically be considered. Accompanied by the extremely large number of antennas, hardware design in XL-MIMO is also different from that of conventional mMIMO~\cite{[102]}. Besides, the extremely large number of antennas would lead to extremely high signal processing complexity and some other EM characteristics, such as the spatial non-stationarity and severe mutual coupling effect, which all need to be paid attention to when studying XL-MIMO systems.  

$\bullet$ \emph{The Much smaller antenna spacing}:
Note that half-wavelength antenna spacing is widely considered in conventional mMIMO~\cite{[31]}. However, in XL-MIMO, the antenna spacing is expected to be much smaller than the conventional half-wavelength. Much smaller antenna spacing can allow for the deployment of an extremely large number of antennas to achieve better performance than conventional mMIMO. However, the small antenna spacing can increase the hardware complexity and design precision when producing XL-MIMO~\cite{ZheMag}.

$\bullet$ \emph{New EM characteristics}:
Several new EM characteristics, which are unnecessary to consider in conventional mMIMO, must be taken into account in the design and deployment of XL-MIMO. For instance, the spherical wave characteristics instead of the planar wave characteristics should be considered when modeling the channel in XL-MIMO due to the near-field property \cite{ZheMag,9903389}. Moreover, other important EM characteristics, such as channel spatial non-stationarity \cite{9170651}, mutual coupling \cite{940505}, and polarization \cite{[65]}, should also be considered in XL-MIMO. Based on these EM characteristics, the channel models for XL-MIMO are much different from those of conventional mMIMO \cite{ZheMag,9903389,2023arXiv230304003L}.

$\bullet$ \emph{Near-field based signal processing}:
Existing signal processing schemes for conventional mMIMO may not apply to XL-MIMO due to the different channel features of XL-MIMO. Thus, the near-field based signal processing schemes should be designed to exploit the full capability of XL-MIMO. Some new and near-field based processing schemes, such as the polar domain channel estimation \cite{[6]} and near-field beam focusing \cite{9723331}, are of interest.

As discussed above, challenges and opportunities of XL-MIMO mainly arise from two aspects: the extremely large number of antennas and near-field characteristics. 
The extremely large number of antennas results in significant signal processing complexity and unique EM characteristics not present in conventional mMIMO. 
With this increase in antennas, near-field communication traits become significant, making far-field-based channel models and signal processing schemes used in conventional mMIMO ineffective. 
Therefore, it is important to develop near-field channel models and signal processing schemes for XL-MIMO. These aspects and their associated characteristics present the main challenges for XL-MIMO systems that require further study.

\begin{table*}
\small
 \centering
 \caption{{Comparisons between relevant survey papers on XL-MIMO and our paper.}}
 \begin{tabular}{|c|m{0.08cm}|m{7.5cm}|m{7.5cm}|}
  \hline
  \multicolumn{2}{|c|}{} & \bfseries Key contributions  & \bfseries Main limitations \\
  \hline
\cite{2022arXiv221201257G} & \raisebox{-7.4\normalbaselineskip}[0pt][0pt]{\rotatebox{90}{{{Comprehensive survey}}}}  & {{\quad}} \newline \vspace{ -0.6cm} \newline \tabitem Review basic physics aspects of HMIMOS and provide a panoramic view \newline \tabitem Introduce the theoretical foundations and enabling technologies for HMIMO communications \newline
 \tabitem Summarize signal processing schemes for HMIMO communications \newline \tabitem Demonstrate various extensions of HMIMOs and future directions & { {\quad}} \newline \vspace{ -0.6cm} \newline \tabitem Focus mainly on HMIMOS, which is one of the XL-MIMO technologies \newline \tabitem Review the theoretical foundations of the HMIMOS physical layer simply, without further excavating the unique characteristics for XL-MIMO\\ \cline{1-1} \cline{3-4}
 \cite{10220205} & & {{\quad}} \newline \vspace{ -0.6cm} \newline \tabitem Introduce near-field channel models for both the SPD and CAP antennas \newline \tabitem Highlight the near-field beamfocusing and antenna architectures  \newline \tabitem Discuss performance analysis framework for near-field LoS and statistical channels & {{\quad}} \newline \vspace{ -0.6cm} \newline \tabitem   Omit the review and comparison for XL-MIMO hardware designs (so-called SPD and CAP antennas in \cite{10220205}) \newline \tabitem More low-complexity near-field aided signal processing schemes should be investigated\\ \hline \hline
 
 \multicolumn{2}{|c|}{} &  \multicolumn{2}{m{15cm}|}{\bfseries Key contributions} \\
  \hline
 \cite{9903389}  & \raisebox{-5.3\normalbaselineskip}[0pt][0pt]{\rotatebox{90}{{{Short brief}}}}  & \multicolumn{2}{m{15cm}|}{{\quad} \newline \vspace{ -0.6cm} \newline \tabitem Review the near-field features of the ELAA-based system \newline \tabitem Introduce some key challenges of near-field communications, including near-field channel estimation and beam split} \\ \cline{1-1} \cline{3-4}
\cite{[101]} &  &  \multicolumn{2}{m{15cm}|} {{\quad} \newline \vspace{ -0.6cm} \newline \tabitem Discuss the opportunities and challenges in radiating near-field communications \newline \tabitem Highlight the spherical waves based beamfocusing schemes}\\ \cline{1-1} \cline{3-4}
 \cite{2023arXiv230304003L}&  & \multicolumn{2}{m{15cm}|} {{\quad} \newline \vspace{ -0.6cm} \newline \tabitem Address the technical problem ``What will be different between near-field communications and far-field communications?" from four perspectives: channel modeling, performance analysis, beamforming, and applications} \\ \cline{1-1} \cline{3-4}
\cite{ZheMag} &  &  \multicolumn{2}{m{15cm}|} {{\quad} \newline \vspace{ -0.6cm} \newline \tabitem Summarize several XL-MIMO schemes and introduce their characteristics and relationships \newline \tabitem Review the fundamentals of XL-MIMO from channel modeling, performance analysis, and signal processing}\\ \hline \hline

\raisebox{-8\normalbaselineskip}[0pt][0pt]{\rotatebox{90}{{{Our paper}}}} & \raisebox{-10.1\normalbaselineskip}[0pt][0pt]{\rotatebox{90}{{{Comprehensive survey}}}}& \bfseries Overlapping Contributions & \bfseries Distinct Contributions\\ \cline{3-4}
& &{\quad} \newline \vspace{ -0.6cm} \newline \tabitem Summarize four XL-MIMO hardware designs and introduce their characteristics and relationships 
\newline \tabitem Review fundamentals of the near-field channel modeling from the perspective of LoS propagation, NLoS propagation, and hybrid propagation
\newline \tabitem Discuss the near-field signal processing schemes for XL-MIMO systems
\newline \tabitem Outline a series of applications and future directions for XL-MIMO 
& {\quad} \newline \vspace{ -0.6cm}  \newline \tabitem Comprehensively summarize the system implementation features and antenna characteristics for four XL-MIMO hardware designs \newline \tabitem Motivate the performance comparison framework for four XL-MIMO hardware designs 
\newline \tabitem Summarize distinct EM characteristics and distance boundaries for XL-MIMO 
\newline \tabitem Provide many tutorials for the near-field channel modeling 
\newline \tabitem Review many low-complexity signal processing schemes to promote the practical implementation of XL-MIMO \newline \tabitem Summarize and motivate the deep learning empowered signal processing schemes for XL-MIMO  \\
 \hline 
 \end{tabular}
 \label{Comparison}
\end{table*}

\subsection{Comparisons and Key Contributions}
XL-MIMO technology has attracted considerable research, and some review papers have emerged to focus on this topic. The authors in \cite{2022arXiv221201257G} focused on the holographic MIMO surface (HMIMOS). The physics aspects of HMIMOS were reviewed from the perspectives of hardware structures, holographic design methodologies, tuning mechanisms, and aperture shapes. Then, the theoretical foundations for HMIMO communications were comprehensively introduced from channel modeling, performance analysis, EM field sampling, and EM information theory inspired by a first-principles perspective. To promote the practical implementation of XL-MIMO, some signal processing schemes for HMIMO communications were reviewed regarding holographic channel estimation and holographic beamforming/beam focusing. However, the authors focus mainly on HMIMOS, which is one of the XL-MIMO technologies. Thus, a survey focusing on a more general XL-MIMO technology is needed. Also, theoretical foundations of the HMIMOS physical layer were reviewed in \cite{2022arXiv221201257G} briefly. Some critical issues for the practical implementation of XL-MIMO also need elaboration, such as the low-complexity signal processing schemes and practical implementation paradigm.

In particular, the authors in \cite{10220205} studied near-field communication for XL-MIMO, presenting basic near-field channel models for both spatially-discrete (SPD) antennas and CAP antennas, including uniform spherical wave (USW) and non-uniform spherical wave (NUSW) models for the former, and Green’s function models for the latter.
It emphasizes near-field beamfocusing and advanced antenna designs, detailing the hybrid beamforming architectures, metasurface-based antennas for approximating CAP antennas, DoF analysis framework for near-field communications, and near-field beam training.
Additionally, the authors explored the performance analysis framework for line-of-sight (LoS) and statistical channels, respectively. For LoS channels, the signal-to-noise ratio (SNR) and power scaling laws were studied. For the statistical channels, the authors analyzed the outage probability (OP), ergodic channel capacity (ECC), and ergodic mutual information (EMI).
However, this study lacks a comparative analysis of XL-MIMO hardware design (so-called SPD antennas and CAP antennas in \cite{10220205}), which calls for further research on low-complexity, near-field signal processing schemes.

\begin{table*}[t!]
  \centering
  \fontsize{8}{13.75}\selectfont
  \caption{Important Abbreviations.}
  \label{Abbreviations}
    \begin{tabular}{ !{\vrule width1.2pt}  m{1.7cm}<{\centering} !{\vrule width1.2pt}  m{5.3 cm}<{\centering} !{\vrule width1.2pt} m{1.7cm}<{\centering} !{\vrule width1.2pt}  m{5.3 cm}<{\centering} !{\vrule width1.2pt}}

    \Xhline{1.2pt}
        \rowcolor{gray!50} \bf Abbreviation & \bf Definition & \bf Abbreviation & \bf Definition  \cr
    \Xhline{1.2pt}
    3GPP  & The third-Generation Partnership Project & LTE   & Long-term evolution \\\hline
    ADC   & Analog-to-digital converter   & MIMO  & Multiple-Input-Multiple-Output \\\hline
    ANN   & Artificial neural network & mMIMO   & Massive MIMO \\\hline
    AoA   & Angle-of-arrival & MMSE  & Minimum mean-squared error \\\hline
    AoD   & Angle-of-departure & mMTC   &Massive machine type communications \\\hline
    AP    & Access point & MR   & Maximum ratio \\\hline
    BS    & Base station & mmWave   &Millimeter wave \\\hline
    CAP   & Continuous aperture & NLoS & Non-line-of-sight \\\hline
    CE    & Channel estimation & NMSE & Normalized mean-squared error \\\hline
    CF    & Cell-free & OMP   & Orthogonal matching pursuit \\\hline
    CNN   & Convolutional Neural Network & PLS & Physical layer security \\\hline
    CS    & Compressed sensing & QoS & Quality of Service \\\hline
    CSI   & Channel state information & RIS & Reconfigurable intelligent surface\\\hline
    DAS   & Distributed antenna system & SCA  & Successive convex approximation\\\hline
    DoF   & Degrees of Freedom & SE   & Spectral efficiency \\\hline
    ELAA  & Extremely large-scale antenna array & THz & Terahertz \\\hline
    EM    & Electromagnetic &UAV & Unmanned aerial vehicles \\\hline
    eMBB  & Enhanced mobile broadband & UE  &User equipment\\\hline
    EP    & Expectation propagation & ULA & Uniform linear array\\\hline
    HMIMOS& Holographic MIMO surface & UPA & Uniform planar array \\\hline
    IoT   & Internet of Things & URLLC  &Ultra-reliable and low latency communications \\\hline
    LISs   & Large intelligent surfaces  & VMP  & Variational message passing\\\hline
    LoS   & Line-of-sight  & VR   &Visibility region\\\hline
    LPU   & Local processing unit & XL-MIMO & Extremely large-scale MIMO \\\hline
    LS    & Least square & ZF  & Zero-forcing \\\hline
    \Xhline{1.2pt}
    \end{tabular}
  \vspace{0cm}
\end{table*}

In addition to \cite{2022arXiv221201257G} and \cite{10220205}, there are several short overview papers \cite{9903389,ZheMag,2023arXiv230304003L,[101]} giving details of XL-MIMO from different perspectives. The authors in \cite{9903389} reviewed the near-field features of the ELAA-based system. The fundamental differences between near-field and far-field communications were clarified. Then, some key challenges of near-field communication were introduced, including near-field channel estimation and near-field beam split. The authors in \cite{[101]} highlighted the near-field beam focusing aspect, where the spherical wave characteristics, near-field channel models, and promising near-field beam focusing features were presented. Alternatively, the technical problem ``What will be different between near-field communications and far-field communications?" was addressed in \cite{2023arXiv230304003L} from four perspectives: channel modeling, performance analysis, beamforming, and applications.
Moreover, the authors in \cite{ZheMag} summarized several XL-MIMO schemes and introduced their characteristics and relationships. Furthermore, the fundamentals of XL-MIMO were reviewed from channel modeling, performance analysis, and signal processing. All these studies provide important insights for XL-MIMO and near-field communications. However, they mainly focus on a single XL-MIMO technology and design or provide reviews from a particular perspective. They lack a holistic presentation and comparison of the technical aspects and technical tutorials of XL-MIMO systems.

\begin{figure*}[t]
\setlength{\abovecaptionskip}{-0.1cm}
\centering
\includegraphics[width = 1\textwidth]{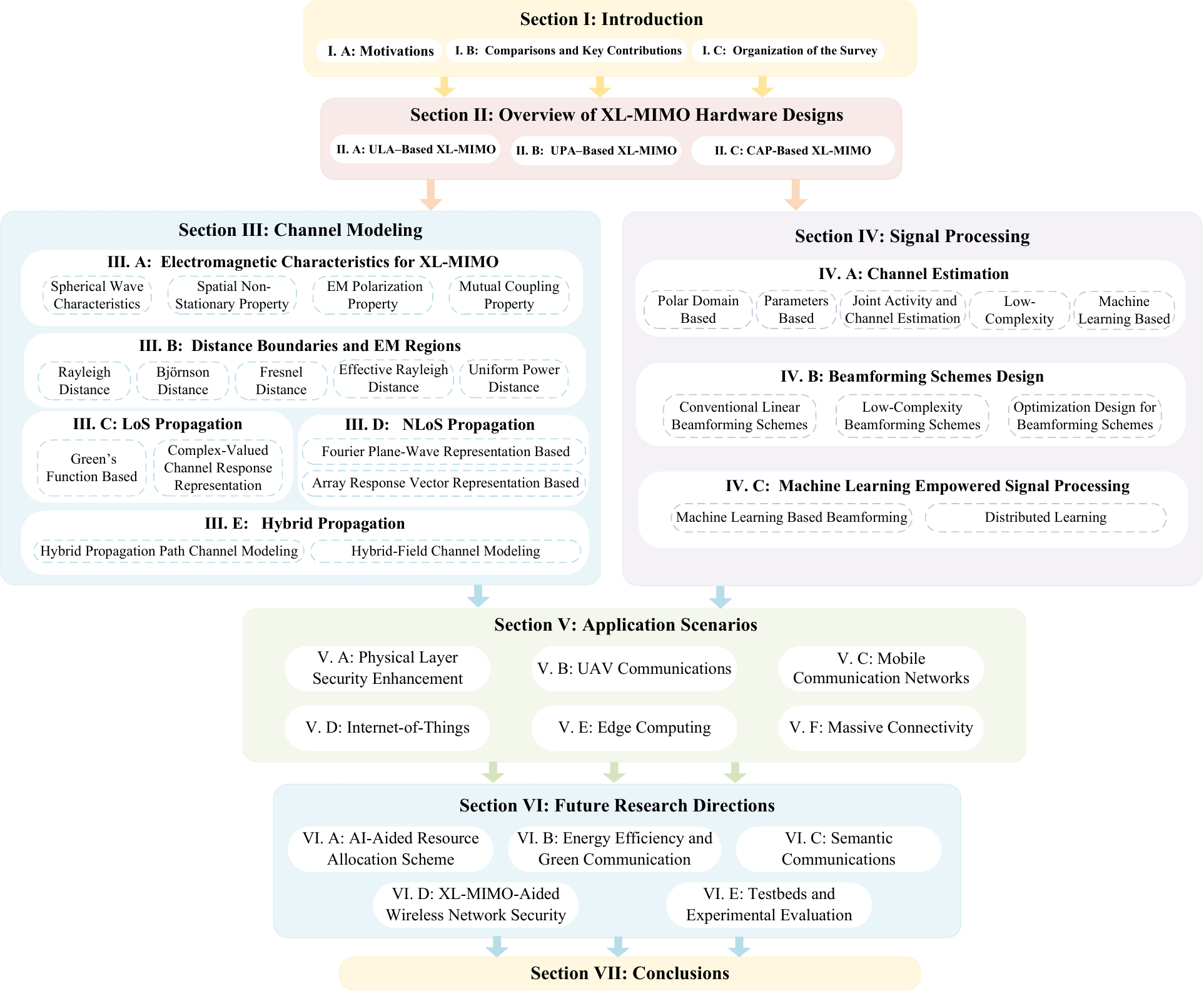}
\caption{The organization structure of the survey.}
\label{Structure}
\end{figure*}

To this end, in this paper, we review XL-MIMO technology extensively. The major contributions of our survey are summarized as follows.
\begin{itemize}
\item We introduce four XL-MIMO designs: uniform linear array (ULA)-based XL-MIMO, uniform planar array (UPA)-based XL-MIMO with patch antennas or point antennas, and CAP-based XL-MIMO. More importantly, we present their characteristics from the perspective of antenna spacing, antenna characteristics, communication scenarios, and implementation carrier frequencies. Moreover, we demonstrate the mutual relationships of these four XL-MIMO designs to provide important insights for the analysis and implementation of XL-MIMO.
\item The fundamentals of the channel modeling for XL-MIMO are reviewed thoroughly including line-of-sight (LoS) propagation channel modeling, non-line-of-sight (NLoS) propagation channel modeling, and hybrid propagation channel modeling. We also provide guidelines for the channel modeling of XL-MIMO. Notably, we also discuss the electromagnetic characteristics of XL-MIMO and distance boundaries and EM regions, which systematically summarize the EM features of XL-MIMO.
\item We review signal processing schemes for XL-MIMO from the perspective of channel estimation, beamforming schemes design, and machine learning empowered processing. Notably, the low-complexity channel estimation and beamforming designs are presented and motivated to gain insights into the practical implementation of XL-MIMO.
\item Finally, we discuss XL-MIMO application scenarios, e.g., physical layer security enhancement, integrated sensing and communications, and internet-of-things, and provide many directions for future research on XL-MIMO, such as AI-aided resource allocation scheme, energy efficiency, and green communication, and semantic communications.
\end{itemize}

We provide a more detailed comparison between our paper and the existing survey papers and short overview papers in Table~\ref{Comparison}.

\subsection{Organization of the Survey}
The organization of this survey is illustrated in Fig. \ref{Structure}. In Section~\ref{systemModel}, we first present hardware designs for XL-MIMO. ULA-based XL-MIMO, UPA-based XL-MIMO, and CAP-based XL-MIMO are introduced. Then, channel modeling for XL-MIMO is reviewed in Section~\ref{channel}. The EM characteristics for XL-MIMO are first introduced in Section~\ref{EMCharacteristics}, and distance boundaries are presented in Section~\ref{Distances}. More significantly, channel modeling for XL-MIMO is introduced from the perspective of LoS propagation channel modeling, NLoS channel propagation channel modeling, and hybrid propagation channel modeling in Section~\ref{LOSchannel}, Section~\ref{NLOSchannel}, and Section~\ref{HybridChannel}, respectively. Section~\ref{SignalProcessing} focuses on low-complexity signal processing schemes designed for XL-MIMO. In Section~\ref{ChannelEstimation}, we review channel estimation for XL-MIMO and introduce many low-complexity channel estimation schemes for XL-MIMO. Section~\ref{Processing} highlights the low-complexity beamforming schemes designed for XL-MIMO. Moreover, the deep learning-empowered signal processing design is presented in Section~\ref{LearningEmpoweredProcessing}, while Section~\ref{application} discusses major XL-MIMO application scenarios. Finally, a series of directions for future research on XL-MIMO are discussed in Section~\ref{future}. Some important abbreviations are summarized in Table~\ref{Abbreviations}.

\section{Overview of XL-MIMO Hardware Designs}\label{systemModel}
This section introduces major XL-MIMO hardware designs and reviews their characteristics. More specifically, four general XL-MIMO hardware designs are introduced: \emph{Uniform Linear Array (ULA)–based XL-MIMO}, \emph{Uniform Planar Array (UPA)–based XL-MIMO} with patch antenna elements or point antenna elements, and \emph{Continuous Aperture (CAP)-based XL-MIMO} \cite{ZheMag}. Note that ULA-based XL-MIMO, UPA-based MIMO with patch antenna elements, and UPA-based XL-MIMO with point antenna elements follow the first implementation approach for XL-MIMO, utilizing a discrete array aperture. In contrast, CAP-based XL-MIMO adheres to the second implementation approach, employing a continuous array aperture for construction.

\begin{figure}[t]
\setlength{\abovecaptionskip}{-0.1cm}
\centering
\includegraphics[width = 0.51\textwidth]{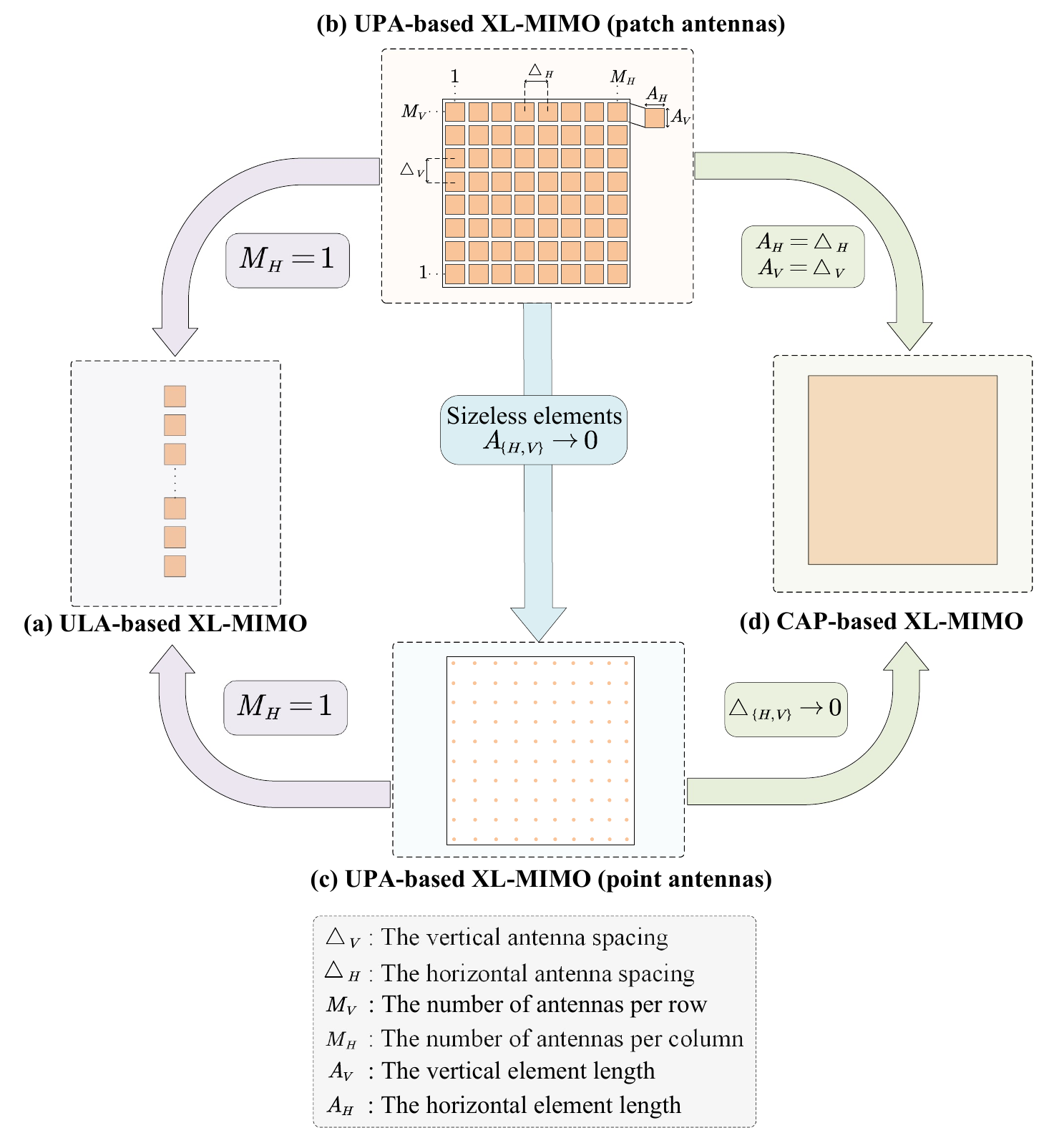}
\caption{Hardware design characteristics for the general XL-MIMO schemes and their relationships among each other.}
\label{fig1:four_models}
\end{figure}

Fig. \ref{fig1:four_models} illustrates their hardware design diagram and relationships. More importantly, we provide the characteristics and comparisons among these hardware designs in Table~\ref{System_Comparison}. 

\begin{table*}[t!]
  \centering
  \fontsize{8}{12}\selectfont
  \caption{The characteristics and comparisons among different hardware designs. ``(a)$\rightarrow$(b)" denotes the interconversion from the XL-MIMO design (a) to the XL-MIMO design (b).}
  \label{System_Comparison}
   \begin{tabular}{ !{\vrule width0.7pt}  m{2 cm}<{\centering} !{\vrule width0.7pt}  m{2.8 cm}<{\centering} !{\vrule width0.7pt}  m{3.6 cm}<{\centering} !{\vrule width0.7pt}  m{3.1 cm}<{\centering} !{\vrule width0.7pt} m{3.2 cm}<{\centering} !{\vrule width0.7pt}}

    \Xhline{0.7pt}
        \bf Schemes & \bf (a) ULA-based XL-MIMO  & \bf (b) UPA-based XL-MIMO (patch antennas)  & \bf (c) UPA-based XL-MIMO (point antennas)  & \bf (d) CAP-based XL-MIMO \cr\Xhline{0.7pt}

        \bf References & \cite{[6],[2],[17],[53],[55],[36],[21],[1],[35],[52],[74],[78],[31],[48],[33],[59],[81],[79],[75],[5],[7],[8],[16],[20],[32],[72],[73]}  & \cite{[3],[4],[15],[18],[19],[32],[34],[49],[50],[54],[56],[93],[97]} & \cite{[14],[20],[27],[28],[47],[56],[58],[60],[63],[65],[66],[67],[68],[69],[71],[76],[83],[84],[85],[87],[88],[37]} & \cite{[9],[10],[11],[12],[13],[16],[22],[23],[25],[29],[32],[38],[39],[41],[44],[51],[61],[64],[70],[89]} \cr\Xhline{0.7pt}
        \bf Characteristics & Discrete aperture  & Discrete aperture &Discrete aperture &Continuous aperture \cr\Xhline{0.7pt}
        \bf Antenna features & Patch/Point antennas & Patch antenna element with the a certain size &Sizeless point antenna element & An infinite number of infinitely sizeless antenna elements\cr\Xhline{0.7pt}

        \bf Design interconversion  &(b) or (c)$\rightarrow$(a): Only one column or one row of antenna elements. & Other schemes can be viewed as special cases of Scheme (b). &(b)$\rightarrow$(c): The antenna element size is infinitesimal. & (b)$\rightarrow$(d): The antenna element size equals the antenna spacing. (c)$\rightarrow$(d): The antenna spacing is infinitesimal.\cr\Xhline{0.7pt}

        \bf Benefits &1) Simple hardware design; 2) Easy for analysis  & General hardware design  & Easy for the analysis for the plane-based XL-MIMO  & Every point can manipulate EM waves in real-time\cr\Xhline{0.7pt}
        \bf Limitations &Low spatial degree of freedom & Difficult for performance analysis due to the consideration of element size &Idealistic assumption with sizeless antenna elements & The research for the meta-material, which is necessary for the design of CAP-based XL-MIMO, is still in its early stage.\cr
    \Xhline{0.7pt}
    \end{tabular}
  \vspace{0cm}
\end{table*}

\subsection{Uniform Linear Array (ULA)--Based XL-MIMO}

\begin{table*}[tp]
  \centering
  \fontsize{8}{12}\selectfont
  \caption{System implementation features and antenna characteristics for ULA-based XL-MIMO. SB-MU and SB-SU denote the scenario with single BS and multiple users and the scenario with single BS and single user, respectively.}
  \label{System_ULA}
    \begin{tabular}{ !{\vrule width1.2pt}  m{1 cm}<{\centering} !{\vrule width1.2pt}  m{2.1 cm}<{\centering} !{\vrule width1.2pt}  m{1 cm}<{\centering} !{\vrule width1.2pt} m{1.7 cm}<{\centering} !{\vrule width1.2pt} m{1.8 cm}<{\centering} !{\vrule width1.2pt} m{3.3 cm}<{\centering} !{\vrule width1.2pt} m{1.4cm}<{\centering} !{\vrule width1.2pt} m{1.4cm}<{\centering} !{\vrule width1.2pt} }

    \Xhline{1.2pt}
        \rowcolor{gray!30} \bf Ref.  &  \bf Communication scenario &  \bf UL/DL &  \bf Sub-array architecture  &  \bf Number of BS antennas &  \bf Antenna spacing &  \bf Aperture size & \bf Carrier frequency \cr
    \Xhline{1.2pt}

        \cite{[6]} & \makecell[c]{SB-MU} & \makecell[c]{UL} & \makecell[c]{\XSolidBrush} & \makecell[c]{$256$}  & \makecell[c]{Half-wavelength} & \makecell[c]{$0.38\ \mathrm{m}$} & \makecell[c]{$100\ \mathrm{GHz}$} \cr\hline
        \cite{[2]} & \makecell[c]{SB-MU} & \makecell[c]{UL} & \makecell[c]{\Checkmark} & \makecell[c]{$256$}  & \makecell[c]{$0.0578\ \mathrm{m}$} & \makecell[c]{$14.8\ \mathrm{m}$} & \makecell[c]{--} \cr\hline
        \cite{[5]} & \makecell[c]{SB-SU} & \makecell[c]{UL} & \makecell[c]{\Checkmark} & \makecell[c]{$256$}  & \makecell[c]{Half-wavelength} & \makecell[c]{$0.38\ \mathrm{m}$} & \makecell[c]{$100\ \mathrm{GHz}$} \cr\hline
        \cite{[17]} & \makecell[c]{SB-SU} & \makecell[c]{UL} & \makecell[c]{\XSolidBrush} & \makecell[c]{$512$}  & \makecell[c]{Half-wavelength} & \makecell[c]{$21.94\ \mathrm{m}$} & \makecell[c]{$3.5\ \mathrm{GHz}$}\cr\hline
        \cite{[53]} & \makecell[c]{SB-MU} & \makecell[c]{DL} & \makecell[c]{\XSolidBrush} & \makecell[c]{$1000$} & \makecell[c]{Half-wavelength} & \makecell[c]{$50\ \mathrm{m}$} & \makecell[c]{$3\ \mathrm{GHz}$}\cr\hline
        \cite{[55]} & \makecell[c]{SB-SU} & \makecell[c]{UL} & \makecell[c]{\XSolidBrush} & \makecell[c]{$512$}  & \makecell[c]{Half-wavelength} & \makecell[c]{$2.56\ \mathrm{m}$} & \makecell[c]{$30\ \mathrm{GHz}$}\cr\hline
        \cite{[36]} & \makecell[c]{SB-MU} & \makecell[c]{UL} & \makecell[c]{\XSolidBrush} & \makecell[c]{$512$} & \makecell[c]{Uniformly distributed in \\ a particular length} & \makecell[c]{$30\ \mathrm{m}$} & \makecell[c]{$2.6\ \mathrm{GHz}$}\cr\hline
        \cite{[21]} & \makecell[c]{SB-SU} & \makecell[c]{UL} & \makecell[c]{\Checkmark} & \makecell[c]{$1024$} & \makecell[c]{${1}/{2}$} & \makecell[c]{--} & \makecell[c]{--}\cr\hline
        \cite{[7]} & \makecell[c]{SB-SU} & \makecell[c]{UL} & \makecell[c]{\XSolidBrush} & \makecell[c]{$256$}  & \makecell[c]{Half-wavelength} &  \makecell[c]{$0.38\ \mathrm{m}$} & \makecell[c]{$100\ \mathrm{GHz}$}\cr\hline
        \cite{[8]} & \makecell[c]{SB-SU} & \makecell[c]{UL} & \makecell[c]{\XSolidBrush} & \makecell[c]{$512$} & \makecell[c]{Half-wavelength} & \makecell[c]{$1.28\ \mathrm{m}$} & \makecell[c]{$60\ \mathrm{GHz}$}\cr\hline
        \cite{[35]} & \makecell[c]{SB-SU} & \makecell[c]{--} & \makecell[c]{\XSolidBrush} & \makecell[c]{$128$}  & \makecell[c]{Half-wavelength} & \makecell[c]{$1.92\ \mathrm{m}$} & \makecell[c]{$10\ \mathrm{GHz}$}\cr\hline
        \cite{[74]} & \makecell[c]{SB-MU} & \makecell[c]{UL} & \makecell[c]{\Checkmark} & \makecell[c]{$500$} & \makecell[c]{Uniformly distributed in \\ a particular length} & \makecell[c]{$100\ \mathrm{m}$} & \makecell[c]{--}\cr\hline
        \cite{[31]} & \makecell[c]{SB-SU} & \makecell[c]{UL} & \makecell[c]{\XSolidBrush} & \makecell[c]{$2000$}  & \makecell[c]{Half-wavelength} & \makecell[c]{$85.71\ \mathrm{m}$} & \makecell[c]{$3.5\ \mathrm{GHz}$}\cr\hline
        \cite{[48]} & \makecell[c]{SB-MU} & \makecell[c]{DL} & \makecell[c]{\Checkmark} & \makecell[c]{$2048$} & \makecell[c]{--} & \makecell[c]{--} & \makecell[c]{--}\cr\hline
    \Xhline{1.2pt}
    \end{tabular}
  \vspace{0cm}
\end{table*}

Due to its simple hardware design and analysis, the ULA structure is well-investigated in conventional mMIMO. The BS in conventional mMIMO is typically equipped with 64 or 128 antennas. However, the number of antennas for ULA-based XL-MIMO is expected to be increased to another order of magnitude compared with that of conventional mMIMO, e.g., 512 antennas or thousands of antennas. Many existing works considered ULA-based XL-MIMO \cite{[6],[2],[17],[53],[55],[36],[21],[1],[35],[52],[74],[78],[31],[48],[33],[59],[81],[79],[75],[5],[7],[8],[16],[20],[32],[72],[73]}. We summarize the system implementation features and antenna characteristics for ULA-based XL-MIMO in Table~\ref{System_ULA}. The significant aspects are discussed as follows:

$\bullet$ \emph{Number of BS antennas}:
Many works studied ULA-based XL-MIMO with thousands of antennas \cite{[48],[53],[21],[33],[81],[78],[31]}. For instance, the authors in \cite{[21]} and \cite{[48]} investigated a ULA-based XL-MIMO system where the BS is equipped with $1024$ and $2048$ antennas, respectively. Besides, other works mainly considered ULA-based XL-MIMO systems with $256$ \cite{[6],[2],[7],[59]} or $512$ \cite{[17],[55],[36],[8],[1]} antennas. The number of BS antennas in XL-MIMO is much larger than that of mMIMO, which means that a much higher hardware design complexity for XL-MIMO should be considered compared with mMIMO.

$\bullet$ \emph{Antenna spacing}:
Although the number of antennas for ULA-based XL-MIMO is much larger than that of mMIMO, the antenna spacing is widely assumed to be half-wavelength \cite{[8],[35],[7]}. Moreover, another choice of antenna spacing is determined by the total length of the array and the number of antennas, which can be adjusted accordingly \cite{[36],[74],[52]}. For this choice, the antennas are uniformly distributed in a particular length, and the antenna spacing is equal to the total length of the array divided by the number of antennas. In this setting, the antenna spacing can be smaller than the conventional half-wavelength.

$\bullet$ \emph{Communication scenarios}:
As for the communication scenarios, all existing works studied a single BS scenario. Also, a single user \cite{[6],[36],[21],[35]} or multiple users \cite{[55],[52],[78],[81]} were served. The majority of existing works assumed that each user was equipped with only a single antenna \cite{[2],[5],[53],[1]}. However, some works assumed that the user was also equipped with a ULA \cite{[35],[59]}. More specifically, the authors in \cite{[35]} and \cite{[59]} considered a single ULA-based user with $64$ and $256$ antennas, respectively. Compared with the single-user scenario, the scenario with multiple users is more practical. However, the interference introduced by multiple users should be accurately modeled in XL-MIMO. Besides, compared with the single-antenna user setting, the scenario with the user equipped with ULA-based XL-MIMO presents a more complicated system model and signal processing scheme but can achieve more capability for XL-MIMO in excellent spatial resolution.

$\bullet$ \emph{Carrier frequencies}:
ULA-based XL-MIMO can be implemented at various carrier frequencies from $2.4 \ \mathrm{GHz}$ to $100 \ \mathrm{GHz}$ \cite{[33],[17],[81],[35],[55],[8],[6]}. More specifically, in \cite{[33],[36],[31],[17],[78],[81]}, ULA-based XL-MIMO systems were implemented on the Sub-6GHz. For instance, ULA-based XL-MIMO systems in \cite{[33]} and \cite{[31]} were operated on the carrier frequency of $2.4 \ \mathrm{GHz}$ and $3.5 \ \mathrm{GHz}$, respectively. Besides, in \cite{[55]}, the authors considered the operation frequency at a mmWave band with the carrier frequency of $30 \ \mathrm{GHz}$. ULA-based XL-MIMO system can also operate at the THz band as \cite{du2022performance,[5],[7],[59]} with the $100 \ \mathrm{GHz}$. These observations provide intuitive insights for further research of XL-MIMO that XL-MIMO can operate over wide ranges of carrier frequency.

$\bullet$ \emph{Sub-array architecture}:
When the BS is equipped with ULA-based XL-MIMO, the array aperture is much larger compared with that of mMIMO, so it is helpful to divide the array into a few sub-arrays, which have the signal processing capability, to reduce the signal processing complexity \cite{[2],[5],[21],[1],[74],[48]}. For instance, the authors in \cite{[5]} considered that the BS was equipped with $M$-antenna ULA, and $M_{\mathrm{RF}}$ RF chains were employed. The antenna array was divided into $M_{\mathrm{RF}}$ sub-arrays, and each sub-array with $N={M}/{M_{\mathrm{RF}}}$ antennas was connected to an RF chain. In the simulation, the authors considered ULA-based XL-MIMO system with $M=256$ antennas, and the antenna array was composed of $M_{\mathrm{RF}}=4$ sub-arrays. As discussed later, this sub-array architecture can be relied on to implement the distributed processing and describe the spatial non-stationarity.

$\bullet$ \emph{Aperture size}: The aperture size of ULA-based XL-MIMO can be depicted by the array length $L=M\bigtriangleup$ with $M$ and $\bigtriangleup$ being the number of antennas and the antenna spacing, respectively. As shown in Table~\ref{System_ULA}, the antenna spacing $\bigtriangleup$ is assumed to be half-wavelength in most works. A large aperture size can be implemented when the BS has thousands of antennas. For instance, the ULA with the length of $85.71\ \mathrm{m}$ was investigated in \cite{[31]} with $M=2000$ and the carrier frequency of $3.5\ \mathrm{GHz}$. 
Note that the aperture size is an important factor for implementing XL-MIMO practically. For ULA-based XL-MIMO, the aperture size depends on the array length, which is determined by the number of antennas and the antenna spacing. A high carrier frequency and small antenna spacing can support the implementation of ULA-based XL-MIMO with moderate aperture size.

As for another promising XL-MIMO scheme, UPA-based XL-MIMO is also regarded as an emerging scheme \cite{[19],[3],[15],[27],[34]}. Most of these works considered that the BS was equipped with a rectangular or square plane. However, the plane with the other shapes, such as the circular plane \cite{[37]}, was also feasible. Without loss of generality, for the rectangular plane-based UPA, the antenna elements are densely packed along the horizontal and vertical directions of the plane. The numbers of antennas per column and row are denoted by $M_V$ and $M_H$, respectively. Moreover, the UPA has $M=M_VM_H$ antennas. Besides, the vertical and horizontal antenna spacing is $\bigtriangleup _V$ and $\bigtriangleup _H$, respectively. Thus, the side lengths of the planar array are denoted as $L_V=M_V\bigtriangleup _V$ and $L_H=M_H\bigtriangleup _H$, respectively. More importantly, distinguished from the modeling and analysis, two types of antenna elements can be considered:

$\bullet$ \textbf{Patch antenna with a particular size}: One well-investigated antenna type is the patch antenna with a certain size \cite{[18],[19],[34],[32],[49],[56]}. For the generality, as shown in Fig. \ref{fig1:four_models}, we consider the rectangle patch antennas with $A_H\times A_V$ element size, where $A_H$ and $A_V$ are the horizontal and vertical antenna spacing, respectively. Note that most of the related works considered square patch antennas, where thousands of square patch antennas are incorporated to construct a plane \cite{[18],[19],[34],[32],[49]}. As discussed in \cite{[32]}, it is necessary and practical to consider the antenna element size to exactly model the EM waves impinged from various directions, especially with large array sizes.

$\bullet$ \textbf{Sizeless point antenna}: For its convenience of modeling and analysis, another antenna scheme for UPA-based XL-MIMO is to consider the sizeless point antenna \cite{[14],[28],[47],[60],[65],[66]}. Note that the sizeless point antenna is mainly motivated from the perspective of channel modeling and performance analysis. Moreover, from the mathematical perspective, the sizeless point antenna is a special case of the patch antenna by letting $A\rightarrow 0$ or assuming that only a particular point across the patch antenna receives the impinging EM wave.

Compared with the sizeless antenna, the patch antenna is more practical and can be implemented in practice. However, it is also more difficult for analysis since many integrals across each patch antenna element need to be calculated due to the consideration of element size. Although the sizeless antenna choice is easy to analyze, this choice is an idealistic assumption without considering element size, which may not be practical. The element architecture for UPA-based XL-MIMO can be selected based on different requirements. The sizeless antenna modeling and analysis ideas can also be implemented for the patch antenna-based scenario by assuming that only one receiving point across each antenna element is impinged. The system implementation features and antenna characteristics for UPA-based XL-MIMO are summarized in Table~\ref{System_UPA}. Some main issues are clarified as follows:

$\bullet$ \emph{Number of BS antennas}:
Compared to ULA-based XL-MIMO with thousands or hundreds of antennas, the BS in UPA-based XL-MIMO can be equipped with another order-of-magnitude number of antennas, maybe tens of thousands of antennas. The authors in \cite{[3],[15],[19],[37]} investigated UPA-based BS with tens of thousands of antennas. For instance, the BS in \cite{[34]} was equipped with $200\times 200=40000$ antennas, with $M_V=M_H=200$. Moreover, the majority of works considered UPA-based XL-MIMO with thousands of antennas \cite{[54],[14],[60],[68],[88]}. For instance, the authors in \cite{[68]} considered the BS with $140\times 56=7840$ antennas, where $M_H=140$ and $M_V=56$, respectively. Compared with ULA-based XL-MIMO, the hardware design of UPA-based XL-MIMO is much more complex since many antennas should be compactly deployed across a plane.

$\bullet$ \emph{Antenna spacing}: The antenna spacing for UPA-based XL-MIMO can be smaller than half-wavelength $\lambda$ \cite{[3],[15],[50],[54],[14],[60]}. For instance, the authors in \cite{[54]} considered various values of the antenna spacing with $\bigtriangleup _V=\bigtriangleup _H=\lambda/6, \lambda/12, \lambda/15$. Besides, the authors in \cite{[60]} investigated the UPA with $M_V=32$ microstrips, each of which embedded with $M_H=80$ antennas. The spacing between adjacent microstrips and the antenna spacing in each microstrip is $\bigtriangleup _V=\lambda/2$ and $\bigtriangleup _H=\lambda/5$, respectively. Another setting of the antenna spacing is to consider the plane with fixed-length sides (fixed $L_V$ and $L_H$), and the antennas are uniformly distributed across the plane \cite{[19],[65]}. Thus, the antenna spacing $\bigtriangleup _V=L_V/N_V$ and $\bigtriangleup _H=L_H/N_H$ can be adjusted in the simulation part by considering various numbers of antennas. The antenna spacing has a great effect on the system's performance. However, for UPA-based XL-MIMO with fixed physical size, decreasing the antenna spacing may not always benefit the system performance but significantly increase the design complexity \cite{ZheMag,[65]}. Thus, antenna spacing is an important factor in designing XL-MIMO. Moreover, relying on the 2D characteristics, both the vertical and horizontal antenna spacing can be correspondingly designed to meet different design requirements.

$\bullet$ \emph{Antenna characteristics}:
The physical size of each antenna element has also been considered for UPA-based XL-MIMO with patch antennas. For instance, the area of each antenna element in \cite{[34]} was $\lambda ^2/4\pi$. Besides, the authors in \cite{[3]} assumed that the element length was equal to the antenna spacing. Moreover, the authors in \cite{[54]} and \cite{[65]} considered the scenario that each user was also equipped with UPA-based XL-MIMO with patch antennas and point antennas, respectively. Multiple users with $144$ antennas each were investigated in \cite{[54]}. Moreover, the transmitter and receiver were equipped with a plane array of the same size. It was worth noting that the circular plane-based XL-MIMO was also a promising scheme, as discussed in \cite{[37]}.

$\bullet$ \emph{Aperture size}: When the BS is equipped with thousands or even tens of thousands of antennas, the aperture size becomes an important factor, especially in practical implementation. For the rectangular plane-based UPA, the aperture size of the plane is determined by $L_V=M_V\bigtriangleup _V$ and $L_H=M_H\bigtriangleup _H$. As discussed in Table~\ref{System_UPA}, the antenna spacing $\bigtriangleup _V$ and $\bigtriangleup _H$ in most works depend on the wavelength $\lambda$. Thus, the aperture size of the UPA is mainly determined by the number of antennas and the antenna spacing. In \cite{[34]}, $12.6\ \mathrm{m}\times 12.6\ \mathrm{m}$ UPA was considered with $M_V=M_H=200$, $\bigtriangleup _V=\bigtriangleup _H=\lambda/2$ and $f_c=2.4 \ \mathrm{GHz}$. Similar to ULA-based XL-MIMO, the high carrier frequency, and small antenna spacing must be taken into account in the implementation of UPA-based XL-MIMO with a moderate aperture size.

\subsection{Uniform Planar Array (UPA)--Based XL-MIMO}
\begin{table*}[t!]
  \centering
  \fontsize{8}{12}\selectfont
  \caption{System implementation features and antenna characteristics for UPA-based XL-MIMO. SB-MU and SB-SU denote the communication scenario with a single BS and multiple UEs, and the communication scenario with a single BS and a single UE, respectively. $\lambda$ denotes the wavelength.}
  \label{System_UPA}
   \begin{tabular}{ !{\vrule width1.2pt}  m{1.9 cm}<{\centering} !{\vrule width1.2pt}  m{0.5 cm}<{\centering} !{\vrule width1.2pt}  m{2.1 cm}<{\centering} !{\vrule width1.2pt}  m{0.8 cm}<{\centering} !{\vrule width1.2pt} m{1.75 cm}<{\centering} !{\vrule width1.2pt} m{1.55 cm}<{\centering} !{\vrule width1.2pt} m{2.1 cm}<{\centering} !{\vrule width1.2pt} m{2.2 cm}<{\centering} !{\vrule width1.2pt} m{1.2 cm}<{\centering} !{\vrule width1.2pt}}

    \Xhline{1.2pt}
        \rowcolor{gray!50} \bf Antenna element architecture  & \bf Ref.  & \bf Communication scenario  & \bf UL/DL  & \bf Number of BS antennas & \bf Antenna spacing & \bf Aperture size  & \bf Antenna characteristics & \bf Carrier frequency \cr
    \Xhline{1.2pt}
        \multirow{10}{*}{ Patch antenna} & \cite{[3]}  & SB-MU  & DL & $200\times 200$  &$0,\lambda/4$ $\lambda/2 ,3\lambda/4$ &$50\lambda\times 50\lambda$ & The element length is equal to antenna spacing. &--\\
        \cline{2-9} & \cite{[15]}  & SB-MU  & DL & $10^3\sim 10^6$  &$\lambda/4$ &$7.5\ \mathrm{m}\times 7.5\ \mathrm{m}$ & -- &$10\ \mathrm{GHz}$\\
        \cline{2-9} & \cite{[19]}  & SB-MU  & UL & $4096$ &Uniformly distributed in the plane &$1.2\ \mathrm{m}\times 1.2\ \mathrm{m}$ & Distributed architecture with square-shape sub-panels &$4\ \mathrm{GHz}$\\
        \cline{2-9} & \cite{[34]}  & SB-MU  & UL & $200\times 200$ &$\lambda/2$ &$12.6\ \mathrm{m}\times 12.6\ \mathrm{m}$ & The area of each element is $\lambda ^2/4\pi$. &$2.4\ \mathrm{GHz}$\\
        \cline{2-9} & \cite{[54]}  & SB-MU  & DL & $3600$ &$\lambda/6$, $\lambda/12$, $\lambda/15$ &$0.3\ \mathrm{m}\times 0.3\ \mathrm{m}$ & UPA based UEs with $144$ antennas&$10\ \mathrm{GHz}$\cr\Xhline{1.2pt}
        \multirow{12}{*}{Point antenna} & \cite{[14]}  & SB-MU  & UL & $40 \times 40$  &$\lambda/4$ &$0.1\ \mathrm{m}\times 0.1\ \mathrm{m}$ &Multi-antenna users &$30\ \mathrm{GHz}$\\
        \cline{2-9} & \cite{[37]}  & SB-MU  & DL & $124980$  &$\lambda/2$ &A circular planar array with the radius of $10 \ \mathrm{cm}$ & Circular planar array based BS &$300\ \mathrm{GHz}$\\
        \cline{2-9} & \cite{[60]}  & SB-MU  & UL & $80 \times 32$ &$\bigtriangleup _V={\lambda}/{2}$ $\bigtriangleup _H={\lambda}/{5}$ &$0.18\ \mathrm{m}\times 0.18\ \mathrm{m}$ &$\bigtriangleup _V\ne \bigtriangleup _H$ &$26\ \mathrm{GHz}$\\
        \cline{2-9} & \cite{[65]}  & SB-SU  & -- & $30\times30$ &Uniformly distributed in the plane &$10\lambda \times 10\lambda$ & Plane array based transmitter and receiver &--\\
        \cline{2-9} & \cite{[68]}  & SB-MU  & DL & $140\times 56$ &$\bigtriangleup _V={\lambda}/{2}$ $\bigtriangleup _H={\lambda}/{5}$ &$0.3\ \mathrm{m}\times 0.3\ \mathrm{m}$ &$\bigtriangleup _V\ne \bigtriangleup _H$&$28\ \mathrm{GHz}$\cr\hline
    \Xhline{1.2pt}
    \end{tabular}
  \vspace{0cm}
\end{table*}

$\bullet$ \emph{Carrier frequencies}:
Various carrier frequencies can be adopted in UPA-based XL-MIMO from $2.4 \ \mathrm{GHz}$ to $300 \ \mathrm{GHz}$. The authors in \cite{[15]} and \cite{[19]} considered UPA-based XL-MIMO systems on the Sub-6GHz, where $4 \ \mathrm{GHz}$ and $2.4 \ \mathrm{GHz}$ carrier frequencies were investigated in \cite{[15]} and \cite{[19]}, respectively. Besides, the mmWave band can also be implemented \cite{[54],[60],[68]}. For instance, the carrier frequency of $30 \ \mathrm{GHz}$ was considered in \cite{[54]}. The authors in \cite{[37]} investigated the XL-MIMO system operated in the THz band with $300 \ \mathrm{GHz}$ carrier frequency.
\begin{table*}[t!]
  \centering
  \fontsize{8}{12}\selectfont
  \caption{System implementation features and antenna characteristics for CAP-based XL-MIMO. SB-MU and SB-SU denote the scenario with single BS and multiple users and the scenario with single BS and single user, respectively.}
  \label{System_CAP}
   \begin{tabular}{ !{\vrule width1.2pt}  m{1.9 cm}<{\centering} !{\vrule width1.2pt}  m{0.5 cm}<{\centering} !{\vrule width1.2pt}  m{2.1 cm}<{\centering} !{\vrule width1.2pt}  m{0.8 cm}<{\centering} !{\vrule width1.2pt} m{2 cm}<{\centering} !{\vrule width1.2pt} m{3.9 cm}<{\centering} !{\vrule width1.2pt} m{2.5 cm}<{\centering}!{\vrule width1.2pt}}

    \Xhline{1.2pt}
        \rowcolor{gray!50} \bf Aperture architecture & \bf Ref.  & \bf Communication scenario  & \bf UL/DL  & \bf Aperture size  & \bf Antenna characteristics & \bf Carrier frequency \cr
    \Xhline{1.2pt}

        \multirowcell{6}{1D line \\segment}& \cite{[44]}  & SB-SU  & UL & $5\ \mathrm{m}$  &1D CAP line segment-based transmitter with the length of  $0.3\ \mathrm{m}$  &$30\ \mathrm{GHz}$\\
        \cline{2-7} & \cite{[13]}  & SB-SU  & UL DL & $5\ \mathrm{m}$  & CAP line segment based transmitter and receiver &$300\ \mathrm{GHz}$, $60\ \mathrm{GHz}$, $28\ \mathrm{GHz}$\\
        \cline{2-7} & \cite{[61]}  & SB-MU  & UL & $1\ \mathrm{m}$ &Multiple single-antenna users &$30\ \mathrm{GHz}$ \\
        \cline{2-7} & \cite{[9]}  & SB-SU  & UL & $3\ \mathrm{m}$ &1D CAP line segment-based transmitter with the length of $0.2\ \mathrm{m}$ &$30\ \mathrm{GHz}$ \cr\Xhline{1.2pt}
        \multirow{12}{*}{2D plane} & \cite{[24]}  & SB-MU  & UL & Circular plane with the radius of $10\ \mathrm{m}$  &CAP circular plane, centralized and distributed deployment &-- \\
        \cline{2-7} & \cite{[11]}  & SB-SU  & UL &The relative receiver area is considered. &The transmitter is equipped with a $5\times 5 \ \mathrm{cm}^2$ CAP plane. The rectangular receiver planes with various length-width ratios are considered. &$28\ \mathrm{GHz}$\\
        \cline{2-7} & \cite{[41]}  & SB-SU  & -- & $30\lambda \times 30\lambda$ & CAP plane based transmitter and receiver&--\\
        \cline{2-7} & \cite{[70]}  & SB-MU  & DL & $1\ \mathrm{m}^2$ &CAP plane based users &$2.4\ \mathrm{GHz}$ \\
        \cline{2-7} & \cite{[64]}  & MB-MU  & UL & Circular planes with the radius of $10\ \mathrm{m}$  &Centralized architecture and distributed architecture  &$2\ \mathrm{GHz}$\cr\hline

    \Xhline{1.2pt}
    \end{tabular}
  \vspace{0cm}
\end{table*}

\subsection{Continuous Aperture (CAP)-Based XL-MIMO}
For the CAP MIMO, with the aid of meta-materials, extremely dense antennas are embedded in a compact space so that the array aperture for the CAP MIMO is approximately continuous \cite{[23],[24],[11],[70],[41]}. Specifically, an ideal CAP MIMO comprises an infinite number of infinitesimal antennas and adopts the form of a spatially-continuous EM volume. This hardware design is also called ``\emph{holographic MIMO}" \cite{[41],[39],[38],[40],[9],[44]} or ``\emph{LISs}" \cite{[23],[24],[11],[25],[64],[13]}. Note that ULA-based XL-MIMO and UPA-based XL-MIMO take the form of discrete antennas over particular antenna spacing. However, CAP-based XL-MIMO adopts the form of the spatially-continuous aperture with an approximately infinite number of infinitesimal antennas. Furthermore, CAP-based XL-MIMO can generate any continuous current with any distribution on its spatially-continuous volume. We summarize the system implementation characteristics and antenna features for CAP-based XL-MIMO in Table~\ref{System_CAP}. Some classical aspects are discussed as follows:

$\bullet$ \emph{Aperture architecture}: Among the existing works for CAP-based XL-MIMO, two representative aperture architectures have been investigated: \emph{1D CAP line segment} and \emph{2D CAP plane}. For 1D CAP-based XL-MIMO, line segments with continuous aperture can be considered \cite{[13],[9],[44],[61]}. For 2D CAP-based XL-MIMO, 2D planes with arbitrary shapes are commonly adopted \cite{[24],[11],[70],[41],[64],[39],[25]}. Among these works, the rectangular or square CAP planes have been widely considered \cite{[23],[11],[70],[39],[41],[64],[25]}. For instance, the authors in \cite{[11]} studied a general scenario of rectangular receiver planes with various length-width ratios. Besides, the circular CAP plane can be adopted \cite{[24],[64]}. From a mathematical perspective, the 1D CAP line segment can be regarded as a special case of ULA-based XL-MIMO with infinitesimal antenna spacing. Similarly, the 2D CAP plane can also be regarded as a special case of UPA-based XL-MIMO with infinitesimal antenna spacing. Thus,  CAP-based XL-MIMO analysis can be performed based on the ULA/UPA-based XL-MIMO analysis by adjusting the antenna spacing to be infinitesimal.

$\bullet$ \emph{Aperture size}:
Note that the aperture size of the aforementioned ULA-based XL-MIMO and UPA-based XL-MIMO is determined by the number of antennas and the antenna spacing. However, different from the two aforementioned XL-MIMO schemes, the aperture size of CAP-based XL-MIMO is determined by the aperture length for 1D CAP-based XL-MIMO and the area or side length for 2D CAP-based XL-MIMO, due to its continuous aperture. For 1D CAP-based XL-MIMO, line segments with a few meters in length were considered \cite{[13],[9],[44],[61]}. For instance, the authors in \cite{[9]} and \cite{[13]} investigated 1D CAP-based XL-MIMO with the length of $5\ \mathrm{m}$. For 2D CAP-based XL-MIMO, the authors in \cite{[24]} and \cite{[64]} considered the circular CAP plane with the radius of $10\ \mathrm{m}$. Moreover, the authors in \cite{[11]} investigated the relative receiver area, where the ratio between the square of transmitting distance and the receiver area was considered. Thus, various values of the receiver area could be adjusted. Compared with ULA-based XL-MIMO and UPA-based XL-MIMO with discrete aperture, the aperture size for CAP-based XL-MIMO can be accurately determined by the size length or the area due to the continuous aperture.

$\bullet$ \emph{Antenna characteristics}:
Some works studied the scenario with both the transmitter and receiver equipped with CAP-based line segments \cite{[13],[9],[44]} or CAP-based planes \cite{[11],[70],[41]}. For instance, the authors in \cite{[44]} and \cite{[9]} considered that the transmitter was equipped with the length of $0.3\ \mathrm{m}$ and the length of $0.2\ \mathrm{m}$ CAP line segment, respectively. Besides, the transmitter in \cite{[11]} was equipped with a $5\times 5 \ \mathrm{cm}^2$ CAP plane. The authors in \cite{[24]} and \cite{[64]} investigated both the centralized and distributed architecture. More specifically, in \cite{[64]}, the centralized architecture with one large CAP plane and the distributed architecture with a few randomly deployed CAP planes were studied. Thus there were multiple BSs equipped with CAP planes in this distributed architecture.

$\bullet$ \emph{Carrier frequencies}:
Similarly, CAP-based XL-MIMO can operate on various carrier frequencies from $2 \ \mathrm{GHz}$ to $300 \ \mathrm{GHz}$. The Sub-6GHz carrier frequency was studied in \cite{[70]} and \cite{[64]} with the carrier frequency being $2.4 \ \mathrm{GHz}$ and $2 \ \mathrm{GHz}$, respectively. Besides, the authors in \cite{[11],[13],[44],[61],[9]} investigated CAP-based XL-MIMO system operated on the mmWave band. For instance, the carrier frequency of $30 \ \mathrm{GHz}$ was experimented in \cite{[13],[44],[61],[9]}. The authors in \cite{[13]} considered various carrier frequencies, such as the mmWave band ($28 \ \mathrm{GHz}$) and THz band ($300 \ \mathrm{GHz}$)~\cite{du2022performance,du2021millimeter}.

\subsection{Design Interconversion and Analysis Comparison}

So far, four XL-MIMO hardware designs have been introduced respectively. All these four XL-MIMO hardware designs are potential and can be mutually interconverted as shown in Fig. \ref{fig1:four_models}. More interestingly, UPA-based XL-MIMO with patch antennas can be regarded as a general design. In contrast, three other schemes can be viewed as special cases of UPA-based XL-MIMO with patch antennas from a mathematical perspective:

$\bullet$ \emph{UPA-based XL-MIMO (patch antennas)} $\rightarrow$ \emph{UPA-based XL-MIMO (point antennas)}: As discussed above, UPA-based XL-MIMO with point antennas is a special case of UPA-based XL-MIMO with patch antennas. From a mathematical perspective, the sizeless point antenna can be adjusted from the patch antenna with a particular size by letting $A\rightarrow 0$ or assuming that only a particular point across the patch antenna receives the impinging EM wave. This scheme is a reasonable assumption of UPA-based XL-MIMO with patch antennas, which is convenient for the system analysis.

$\bullet$ \emph{UPA-based XL-MIMO (patch/point antennas)} $\rightarrow$ \emph{ULA-based XL-MIMO}: ULA-based XL-MIMO can be viewed as a special case of UPA-based XL-MIMO with only one column or one row of antenna elements ($M_H=1$ or $M_V=1$). Most works considered ULA-based XL-MIMO with point antennas, but ULA-based XL-MIMO with patch antennas was considered in \cite{[32]}.

$\bullet$ \emph{UPA/ULA-based XL-MIMO (patch/point antennas)} $\rightarrow$ \emph{2D/1D CAP-based XL-MIMO}: While CAP-based XL-MIMO employs a spatially-continuous aperture, diverging in design and processing protocol from discrete array aperture XL-MIMO systems, it is beneficial to establish clear mathematical and physical connections between them. 1D CAP-based XL-MIMO can be regarded as a special case of ULA-based XL-MIMO by letting the antenna spacing be infinitesimal. Moreover, when the element size of the patch antenna is correspondingly equal to the antenna spacing ($A_H=\bigtriangleup _H$ and $A_V=\bigtriangleup _V$), UPA-based XL-MIMO with patch antennas can become  2D CAP-based XL-MIMO. Besides, 2D CAP-based XL-MIMO can also be viewed as a special case of UPA-based XL-MIMO with point antennas by considering an infinitesimal antenna spacing.

Comparing XL-MIMO hardware design performance is crucial for assessing their potential and advancing their practical application \cite{2023arXiv230406141X,[29]} from the perspective of effective DoF (EDoF) and system capacity. The authors in \cite{[29]} explored ULA-based XL-MIMO and 1D CAP line segment-based XL-MIMO. With both transmitter and receiver utilizing XL-MIMO, the study began with a ULA-based design, calculating EDoF from the channel matrix as indicated in \cite[eq. (3)]{[29]}. For the 1D CAP line segment-based XL-MIMO, the approach involved defining a source auto-correlation kernel for channel description and deriving EDoF through asymptotic analysis within the established framework.

Based on the EDoF analysis framework from \cite{[29]}, the authors in \cite{2023arXiv230406141X} first computed the EDoF for both the ULA and 1D CAP line segment analytically, then derived their capacity based on \cite[eq. (5)]{2023arXiv230406141X}.
Notably, each antenna element of the ULA had a particular length. Furthermore, by comparing the discrete ULA with the 1D CAP line segment under equivalent array aperture conditions, the authors found that the EDoF of the discrete ULA converged to that of the 1D CAP line segment as the number of antennas increased.

The analysis framework in \cite{[29]} and \cite{2023arXiv230406141X} provided guidelines for the performance comparison for different XL-MIMO hardware designs. However, \cite{[29]} and \cite{2023arXiv230406141X} are limited to 1D arrays. The comparison of UPA-based XL-MIMO and 2D CAP plane-based XL-MIMO performance remains an open area for research. The EDoF for UPA-based XL-MIMO is derivable from its discrete channel matrix, while the EDoF for 2D CAP plane-based XL-MIMO requires a new definition of the source auto-correlation kernel, building on the 1D scenarios presented in \cite{[29]} and \cite{2023arXiv230406141X}.

Appropriate comparison criteria are essential for fair performance comparisons between different XL-MIMO hardware designs. To maintain comparability, specific characteristics and settings across all designs must be uniform for fairness. For ULA-based XL-MIMO and 1D CAP line segment-based XL-MIMO, the performance comparison should be carried out with the same array aperture, i.e., the same array length. When comparing UPA-based XL-MIMO with 2D CAP plane-based XL-MIMO, due to the abundant geometric variables, several comparison principles could be considered:
\textit{1) Same side length}: in this case, both the UPA and 2D CAP plane have the same physical size (i.e., side length, plane area, and array aperture); \textit{2) Same plane area}: for the fixed same plane area, the side length and the array aperture for the UPA and 2D CAP plane can be different; and \textit{3) Same array aperture}: the array aperture plays an important role in near-field communications. As such, the array aperture can also be a comparison criterion. 
Under Criteria 2) and 3), the side length and plane array can be adjusted to gain insights for the practical implementation of XL-MIMO. Given that UPA-based XL-MIMO and ULA-based XL-MIMO have discrete array apertures and particular numbers of antennas, the performance comparison can be implemented by assuming the same number of antennas or array apertures. 

The aforementioned criteria for comparing various XL-MIMO hardware designs lay a foundation for systematic evaluation. However, the performance behaviors of these XL-MIMO designs and the effectiveness of the comparison metrics require further study in future research.

{\textit{{\textbf{Lessons Learned:}}}}
We summarize the lessons learned and significant insights and compare XL-MIMO with conventional mMIMO from the perspective of the system model design. We review the system implementation features and antenna characteristics for various XL-MIMO hardware designs in Table \ref{System_ULA}-Table \ref{System_CAP}. To offer more insights into the four XL-MIMO hardware designs, we compare their number of antennas, antenna spacing, aperture size, and analysis framework, aiming to guide future XL-MIMO research and aid in system analysis, optimization, and practical application:

$\bullet$ \emph{Number of antennas}: Compared with conventional mMIMO, the number of antennas of XL-MIMO increases by an order of magnitude. For ULA-based XL-MIMO, thousands of antennas are employed. For UPA-based XL-MIMO, the number of antennas can even approach tens of thousands as many antennas can be simultaneously deployed across the vertical and horizontal directions. However, as studied in \cite{[65], ZheMag}, with the fixed physical size, there exists the EDoF performance saturation: the EDoF would increase and then become flat with the number of antennas increasing. Thus, it will be interesting to study the optimal number of antennas for different applications. For CAP-based XL-MIMO, thanks to the meta-material, continuous or approximately continuous array aperture can be realized by deploying extremely dense antennas in a compact space. Thus, an infinite number of infinitesimal antennas can be considered in an ideal CAP-based XL-MIMO. However, due to the extremely large number of antennas, the design complexity and deployed technology for XL-MIMO impose much higher requirements than that of conventional mMIMO.

$\bullet$ \emph{Antenna spacing}: For ULA-based XL-MIMO, two choices for the antenna spacing are mainly considered: the well-studied half-wavelength antenna spacing and the adjustable antenna spacing by limiting the total array length and adjusting the number of antennas. For UPA-based XL-MIMO, many works consider that the antenna spacing is much smaller than the half-wavelength. More interestingly, UPA-based XL-MIMO can have different antenna spacing in the vertical direction and the horizontal direction instead of limiting them to equal, which provides various design flexibility for the practical implementation of XL-MIMO. Meanwhile, the infinitesimal or approximately infinitesimal antenna spacing can be realized for CAP-based XL-MIMO. Note that the much smaller antenna spacing, compared with that of conventional mMIMO, can provide the ability to deploy an extremely large number of antennas in a compact space with a fixed physical size. However, the small antenna spacing can introduce severe mutual coupling effects and increase the design precision, which should be noticed in XL-MIMO.

$\bullet$ \emph{Aperture size}: For ULA-based XL-MIMO and UPA-based XL-MIMO, the aperture size is usually determined by the array length and the array side length, respectively. Due to the discrete aperture, the aperture size of these two XL-MIMO schemes is determined by the number of antennas and the antenna spacing. Moreover, the antenna spacing is usually dependent on the carrier frequency. Thus, the high carrier frequency and small antenna spacing are advocated to derive the moderate aperture size for implementing ULA-based XL-MIMO and UPA-based XL-MIMO. For CAP-based XL-MIMO, the aperture size can be precisely determined by the aperture length for 1D CAP-based XL-MIMO and the area or side length for 2D CAP-based XL-MIMO, respectively. Compared with mMIMO, the aperture size for XL-MIMO is much larger, which can be more clearly observed from Table~\ref{System_ULA}$\sim$Table~\ref{System_CAP}.

$\bullet$ \emph{Analysis framework}: Based on the above observations, important characteristics for four XL-MIMO hardware designs have been discussed. When studying the signal processing and system optimization for these XL-MIMO designs, different analysis frameworks should be considered due to the different characteristics. For UPA-based XL-MIMO with patch antennas, from the perspective of the whole array, the analysis is performed in a discrete fashion due to the discrete features of each antenna element. However, from the perspective of each antenna element, an accurate analysis scheme is to capture the EM wave characteristics across each patch antenna element with a certain physical size. Thus, the analysis framework for each antenna element should be performed through exact integrals across each patch antenna element, such as the computation of channel gain for each antenna element \cite{[4],9723331}. Besides, one simplified analysis framework for UPA-based XL-MIMO with patch antennas can be carried out by assuming that only one point across the antenna element is impinged by EM waves \cite{ZheMag,[14]}. Under this assumption, UPA-based XL-MIMO with patch antennas reduces to UPA-based XL-MIMO with point antennas. More interestingly, for CAP-based XL-MIMO, due to the continuous array aperture, the analysis should be performed based on the integral across the whole array. However, when performing the analysis, these XL-MIMO hardware designs can be mutually linked to each other as illustrated in Fig. \ref{fig1:four_models}.

Based on these significant findings and observations, the researchers can derive insights into XL-MIMO hardware designs and consider proper XL-MIMO schemes for further research, such as performance analysis and optimization.

\section{Channel Modeling}\label{channel}
As discussed above, compared with conventional mMIMO, XL-MIMO embraces several new features in the hardware structure. In addition, XL-MIMO increases the number of antennas and introduces fundamental changes in EM characteristics. The significant increase in the array aperture would make the receiver be in the near-field region. Thus, several EM characteristics, which are not considered in conventional mMIMO, should be investigated, such as the spherical wave characteristic, spatial non-stationarity, polarization feature, and mutual coupling feature. Moreover, the channels in XL-MIMO show different characteristics than conventional mMIMO. Thus, it is vital to derive suitable channel models to capture the EM fundamentals of XL-MIMO, which can facilitate signal processing and performance analysis In this section, we first discuss several electromagnetic features in XL-MIMO. Then, we introduce five distance boundaries and three EM regions in XL-MIMO. Finally, we introduce the major channel modeling, including LoS channel modeling, NLoS channel modeling, and hybrid propagation channel modeling.

\subsection{Electromagnetic Characteristics for XL-MIMO}\label{EMCharacteristics}
\begin{figure}[t]
\setlength{\abovecaptionskip}{-0.1cm}
\centering
\includegraphics[scale=0.42]{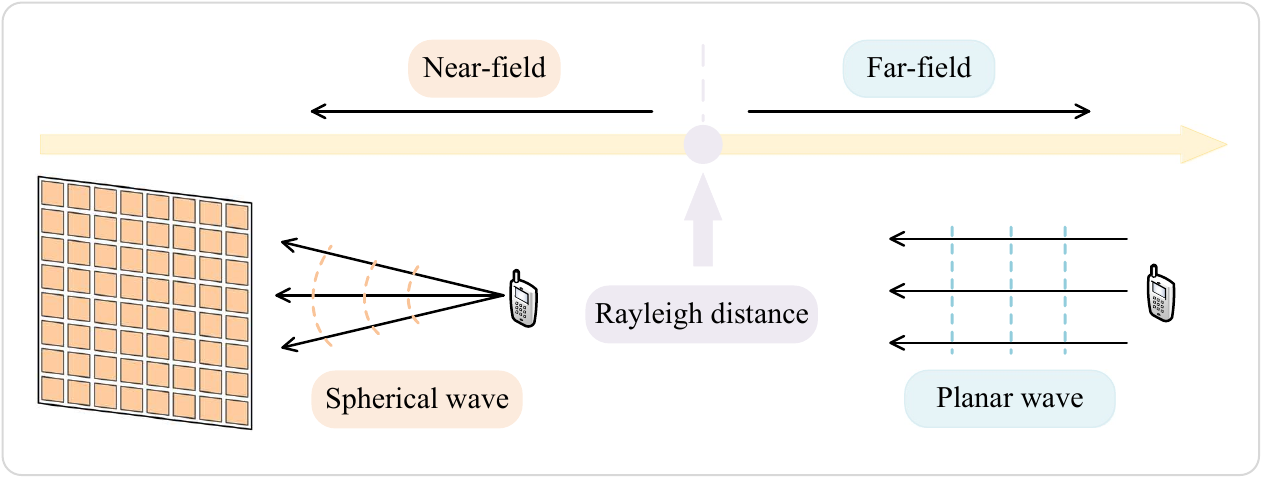}
\caption{The EM wave characteristics for the near-field and far-field regions, which are bounded by the Rayleigh distance. The spherical wave and planar wave characteristics should be considered in the near-field and far-field regions, respectively. 
\label{Rayleigh_Distance}}
\end{figure}

$\bullet$ \emph{Spherical Wave Characteristics}:
With the sharp increase in the number of antennas for XL-MIMO, the fundamental EM characteristics are different from those of conventional mMIMO. Note that the EM region can be generally divided into far-field and near-field regions, bounded by the Rayleigh distance \cite{Rayleigh}. The EM wave characteristics for the near-field and far-field are depicted in Fig. \ref{Rayleigh_Distance}. The array aperture and the carrier frequency determine the Rayleigh distance, which will be introduced later in Section~\ref{Distances}. In conventional mMIMO, the Rayleigh distance can be neglected, due to the moderate array aperture, so the receivers are usually assumed to be located in the far-field region. In the far-field region, the EM wave can be simply modeled based on the plane wave assumption  \cite{ZheMag,9903389,2022arXiv221201257G}. All antenna elements in the array for the plane wave embrace the same signal amplitude and angle of arrival/departure (AoA/AoD). Thus, in the far-field based on the plane-wave characteristics, only the angle characteristic can be relied on to perform the signal processing and system design.

However, for XL-MIMO, the significant increase in the number of antennas and the array aperture size would lead to the non-negligible Rayleigh distance and make the receiver be located in the near-field region. Thus, the plane wave assumption in conventional mMIMO is less applicable to the channel modeling of XL-MIMO. More specifically, the EM wave in XL-MIMO should be modeled with the vectorial spherical wave characteristics \cite{[32],9903389,[101]}. For the vectorial spherical wave, the distances and AoA/AoD between the transmitter antenna elements and the receiver antenna elements would vary over the antenna array \cite{[32],[4]}.
Thus, for XL-MIMO with a large aperture, it is necessary to consider spherical wave-based channel modeling to describe the actual EM characteristics, which embraces distance and angle properties. The authors in \cite{[4]} summarize the fundamental properties for the channel modeling in the near-field: 1) The propagation distances vary over the array elements due to the large array aperture; 2) The antenna areas, determined by the actual aperture and the EM wave impinged angles, should be considered; 3) Due to the various EM wave impinged angles, the signal strength losses caused from the polarization mismatch vary over the array.

\begin{figure}[t]
\setlength{\abovecaptionskip}{-0.1cm}
\centering
\includegraphics[scale=0.36]{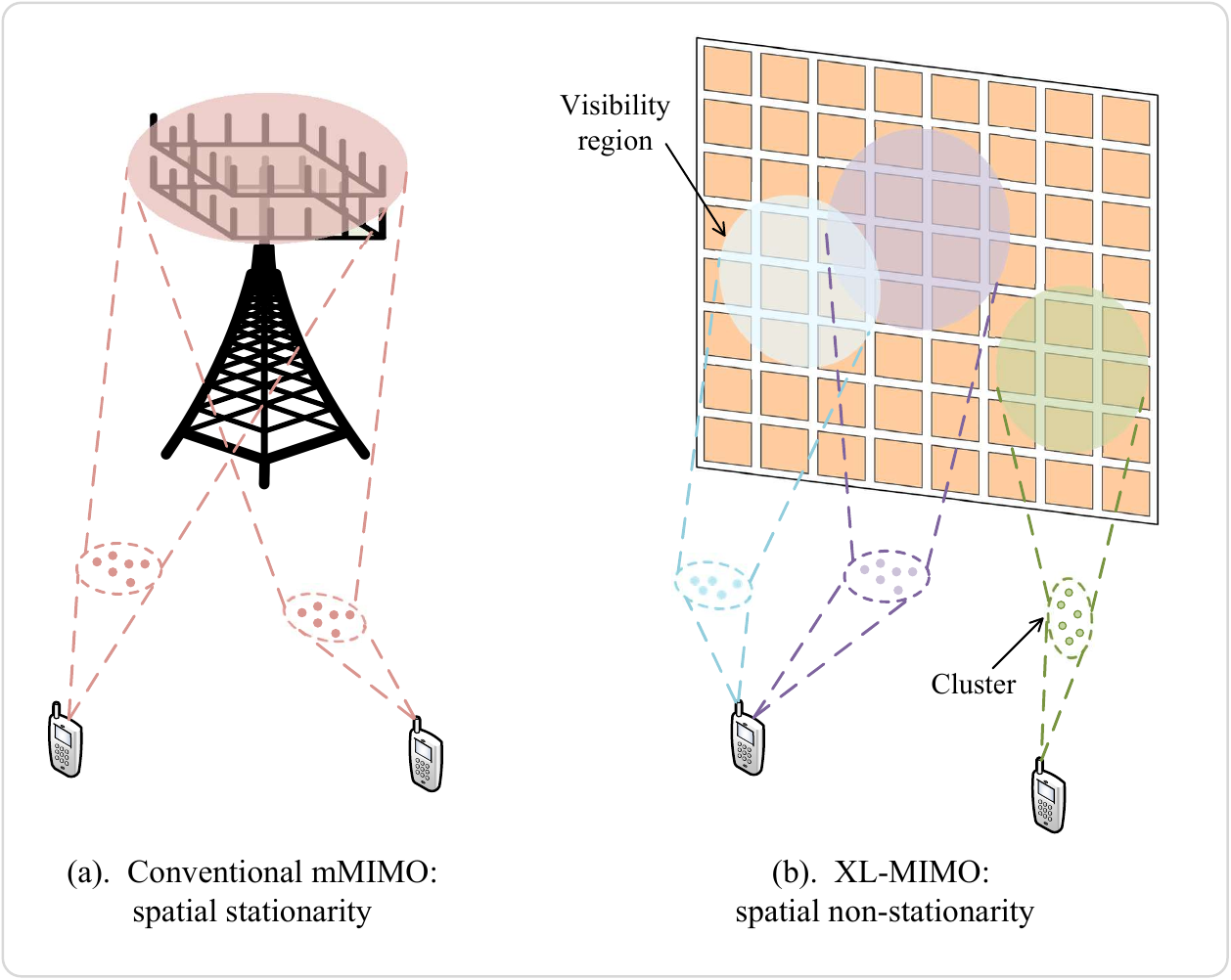}
\caption{The diagram for the spatial stationary mMIMO and spatial non-stationary XL-MIMO.
\label{Non_Stationary}}
\vspace{-0.7cm}
\end{figure}

$\bullet$ \emph{Spatial Non-Stationarity}: In conventional mMIMO, the channel is spatially stationary due to the moderate array dimension. However, in XL-MIMO, the extremely large array aperture would result in the spatial non-wide sense stationary properties \cite{[1],[2],[20],[21],[27],[31],9170651}. When the array aperture is extremely large, different regions of the array would observe the propagation environment in different views, which means that the regions can observe the signal transmitted from a certain propagation path but with different powers or the signal transmitted from different propagation paths. Different terminals can be viewed by different regions of the array, and each terminal's power is focused on a particular region of the array. This particular visible region of the array from a given terminal is defined as the visibility region (VR) of this given terminal \cite{9170651,6393523,[98],[1]}. The spatial stationary mMIMO and spatial non-stationary XL-MIMO diagram are shown in Fig. \ref{Non_Stationary}. For conventional mMIMO, as shown in Fig. \ref{Non_Stationary} (a), all array elements are visible to the terminals, and the terminal's power can spread over the whole array. Thus, the channel for conventional mMIMO shows the stationary spatial property. For XL-MIMO, as shown in Fig. \ref{Non_Stationary} (b), it can be found that the clusters are not visible over the whole array, and the power of each terminal is predominately focused on a limited region of the array, i.e., VRs. Thus, the channel for XL-MIMO embraces the spatial non-stationarity.

The spatial non-stationarity would lead to a departure from conventional channel models in mMIMO and affect the system's performance. Many works have endeavored to study the spatial non-stationarity for XL-MIMO \cite{[43],[62],[72],[81],[85],BokaiICC}. For example, the authors in \cite{[98]} proposed a realistic low-complexity spatial non-stationary channel modeling structure and verified the feasibility and validity based on the practical channel measurements and Ray-tracing simulations. It is worth noting that the channel for a particular terminal is assumed to be approximately stationary in its VRs and zero outside the VRs. To derive an analyzable structure, a simple choice is to decompose the array into a few sub-arrays, where each sub-array is visible to particular terminals. For this sub-array scheme, $\mathbf{1}$ and $\mathbf{0}$ vectors indicate whether a particular sub-array is the VR for a given terminal. This approach is adopted in \cite{[1],[2],[21],[62],BokaiICC} to make the spatial non-stationarity-aided XL-MIMO systems more tractable, which facilitates channel modeling and performance analysis for XL-MIMO. 

$\bullet$ \emph{EM Polarization Property}: As a well-known property, EM polarization transverse waves transmit with a particular oscillation orientation. Most works for conventional mMIMO omitted this EM polarization property for simplicity. However, it is necessary to consider this property since it can capture the actual EM characteristics in XL-MIMO \cite{8976425,9806042,5498959}. Several studies have investigated the EM polarization property. In \cite{5498959,li2014frequency,8263240,8016370}, dual-polarization was analyzed. More specifically, the authors in \cite{5498959} and \cite{li2014frequency} considered the dual-polarized UPA with square patch antennas. In \cite{8263240}, the dual-circularly polarized circular metasurface with various printed elements was studied. The authors in \cite{8016370} considered the dual linear polarization based on the metasurface, where slot-shaped sub-wavelength unit cells were located across the metasurface. Moreover, the triple polarization was studied in \cite{[11],[44],[65],[70],[97],[9]}, modeled by the dyadic Green's function. The authors in \cite{[11]} and \cite{[70]} considered the tri-polarized 2D CAP plane. The tri-polarized 1D CAP line segment was investigated in \cite{[44]} and \cite{[9]}. Moreover, the authors in \cite{[65]} and \cite{[97]} studied the UPA with point antennas over the triple polarization. As observed in \cite{[65]}, the consideration of multiple polarizations benefited the EDoF performance compared with that of the single polarization.

$\bullet$ \emph{Mutual Coupling Property}:
When the antenna elements are closely packed with a small antenna spacing as in XL-MIMO, the mutual coupling property will significantly affect the practical design and system performance \cite{940505,8350292}. The mutual coupling property denotes that the voltage on each antenna element would depend not only on the incident field but also on the voltages on other antenna elements \cite{940505}. This property becomes significant in a scenario with small antenna spacing. The mutual coupling property will deteriorate the channel and substantially degrade the signal-to-interference-noise ratio (SINR) and the convergence of processing algorithms. Most existing works on XL-MIMO and mMIMO omitted the mutual coupling property. However, studies considered the mutual coupling property \cite{[56],[58],[96],6522419}. The authors in \cite{[56]} and \cite{[96]} described the mutual coupling property with the aid of the mutual coupling matrix and investigated the effects of the mutual coupling property on the array gain and channel characteristics.
Moreover, the authors in \cite{6522419} investigated the effects of the mutual coupling property in the scenario with fixed physical spaces. The mutual coupling matrix in \cite{6522419} was determined by the antenna impedance, load impedance, and mutual impedance. Besides, various coupling properties, such as the coupling between antenna elements, the coupling between users and antenna elements, and the coupling between users, were comprehensively studied in \cite{[58]}. The modeling methods for the mutual coupling effect in these works provide fundamentals for the further analysis of XL-MIMO systems considering the mutual coupling property.

{\textit{{\textbf{Lessons Learned:}}}} Four representative electromagnetic characteristics for XL-MIMO have been introduced: spherical wave characteristics, spatial non-stationarity, EM polarization property, and mutual coupling property. These characteristics are usually omitted in conventional mMIMO for ease of analysis. However, for XL-MIMO, these characteristics have been proven to be vital for accurate modeling and analysis. Thus, four electromagnetic characteristics have been thoroughly reviewed, which can embrace vital insights for the modeling and design of XL-MIMO systems with these characteristics considered.

\subsection{Distance Boundaries and EM Regions}\label{Distances}
\begin{figure*}[t]
\setlength{\abovecaptionskip}{-0.1cm}
\centering
\includegraphics[scale=0.6]{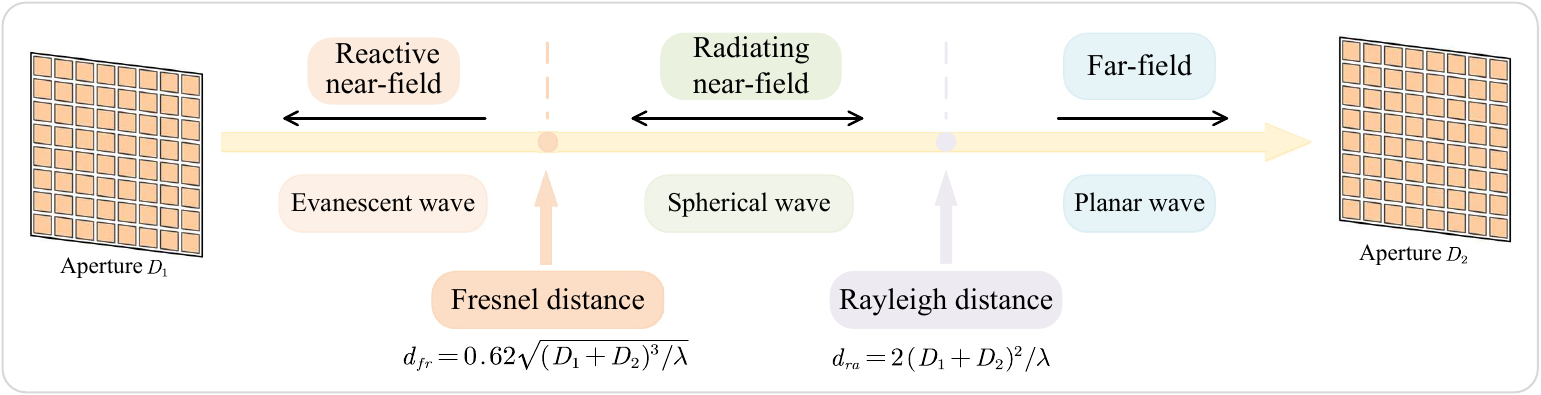}
\caption{Three EM regions and their respective distance boundaries. In the reactive near-field region, the evanescent wave is the strongest, where the power is concentrated in the surroundings of the source. In the radiating near-field region, the spherical wave propagation characteristics should be paid attention to. In the far-field region, the planar wave can be assumed due to the large propagation distance.
\label{Three_regions}}
\end{figure*}
As discussed above, the receivers in XL-MIMO are very likely located in the near-field region. The well-known boundary to divide the near-field region and the far-field region is the Rayleigh distance \cite{9903389,[32],[101]}. Nonetheless, other distance boundaries and EM regions have also been investigated based on different principles and perspectives to explore the EM characteristics in XL-MIMO. In this subsection, we introduce these distance boundaries and EM regions and provide insights into the EM characteristics in XL-MIMO.

$\bullet$ \emph{Rayleigh Distance}: As a well-known distance boundary, the Rayleigh distance (also called the Fraunhofer distance) is defined to divide the near-field region and the far-field region. The Rayleigh distance is motivated by considering the phase discrepancy caused by the wave's curvature \cite{9723331}. The Taylor expansion can accurately denote the EM wave's phase. The phase can be approximated in the far field by its first-order Taylor expansion. However, this approximation will lead to phase discrepancy. The maximum allowable phase discrepancy caused by the most important neglected term, i.e., the second-order Taylor expansion term, is ${\pi}/8$ \cite{balanis2015antenna}. When the largest phase discrepancy ${\pi}/8$ appears, the distance between the BS array center and the UE array center is defined as \emph{Rayleigh distance} $d_{ra}$. In other words, based on the first-order Taylor expansion, the largest phase discrepancy larger than ${\pi}/8$ will appear when the propagation distance is shorter than the Rayleigh distance. When the communication distance is larger than the Rayleigh distance, the UE is regarded as being located in the far-field region (also called Fraunhofer region), where the planar wave assumption can be implemented. On the contrary, when the communication distance is smaller than the Rayleigh distance $d<d_{ra}$, the UE is regarded as being located in the near-field region, where the spherical wave should be considered. Besides, the far-field region (also called Fraunhofer region) is defined as the region farther than the Rayleigh distance $d\geqslant d_{ra}$.

Regarding the computation of the Rayleigh distance, for the scenario with XL-MIMO based BS and single-antenna UE, the Rayleigh distance is $d_{ra}=2D^2/\lambda$, where $D$ is the maximum array length of the BS and $\lambda$ is the wavelength \cite{balanis2015antenna,9903389,7942128}. Note that $D$ is also called array aperture, which is the maximum array length for the array. For instance, for a UPA with side length being $L_H$ and $L_V$, the array aperture is taken to be its diagonal as $D=\sqrt{L_{H}^{2}+L_{V}^{2}}$. Moreover, for the scenario where both the BS and UE are equipped with XL-MIMO with $D_1$ and $D_2$ array aperture, respectively, the Rayleigh distance can also be computed as $d_{ra}=2(D_1+D_2)^2/\lambda$ based on the criterion that the maximum allowable phase discrepancy is ${\pi}/8$ \cite{[35],7942128}. The distance in this arrangement is called double-side Rayleigh distance (DS-RD) \cite{[35]} since both the transmitter and receiver are XL-MIMO arrays and thus, the ``double-side" is considered.

$\bullet$ \emph{Fresnel Distance}: When the communication distance is smaller than the Rayleigh distance, the maximum phase discrepancy will be larger than ${\pi}/8$ with the first-order Taylor expansion. This significant phase discrepancy is undesirable in many scenarios. To further derive a distance boundary smaller than the Rayleigh distance, the second-order Taylor expansion can be implemented to approximate the phase. When this approximation is considered, the most critical neglected term, i.e., the third-order Taylor expansion term, can tolerate only the maximum allowable phase discrepancy with ${\pi}/8$. Based on this criterion, for the scenario with XL-MIMO based BS and single-antenna UE, the \emph{Fresnel distance} can be computed as $d_{fr}=0.62\sqrt{{{D^3}/{\lambda}}}$. Based on the second-order Taylor expansion, the most significant phase discrepancy ${\pi}/8$ will appear when the propagation distance is shorter than the Fresnel distance. Besides, the Fresnel distance for the scenario where both the BS and UE are equipped with XL-MIMO with $D_1$ and $D_2$ array aperture, respectively, can be computed as $d_{fr}=0.62\sqrt{{{(D_1+D_2)^3}/{\lambda}}}$ by replacing $D$ by the sum of $D_1$ and $D_2$ \cite{balanis2015antenna}. 

Specifically, based on the Rayleigh distance and Fresnel distance, three EM regions are distinguished as shown in Fig. \ref{Three_regions}. The region, which is farther than the Rayleigh distance $d\geqslant d_{ra}$, is defined as the \emph{far-field region}, as discussed above. The near-field region, which is smaller than the Rayleigh distance $d<d_{ra}$, can be further divided into the reactive near-field region and the radiating near-field region, bounded by the Fresnel distance. More specifically, the region, which is greater than the Fresnel distance $d_{fr}$ and smaller than the Rayleigh distance $d_{ra}$, i.e., $d_{fr}\leqslant d\leqslant d_{ra}$, is called the \emph{radiating near-field region} or \emph{Fresnel region}. In this region, the spherical wave characteristics should be paid attention to. Furthermore, the region with a distance smaller than the Fresnel distance $d_{fr}$, i.e., $d\leqslant d_{fr}$, is called the \emph{reactive near-field region}. In this region, the evanescent waves are dominant \cite{[28],balanis2015antenna,2023arXiv230304003L}. Note that the channel power for evanescent waves is concentrated in the surroundings of the transmitter and thus the evanescent waves cannot be regarded as the conventional EM waves for propagating since they cannot be observed at a particular distance of many wavelengths. For instance, the Rayleigh distance and Fresnel distance are $d_{ra}=267 \ \mathrm{m}$ and $d_{fr}=10 \ \mathrm{m}$, respectively, with the array aperture $D_1=D_2=1 \ \mathrm{m}$, and the carrier frequency of $10 \ \mathrm{GHz}$. Thus, the receivers in XL-MIMO systems are very likely located in the near-field, especially in the radiating near-field. The majority of works on XL-MIMO focused on the radiating near-field region. Note that the Fresnel distance is a further distance boundary to divide the near-field region. It is also reasonable and tractable to use the classical Rayleigh distance to distinguish the near-field region with spherical wave characteristics and the far-field region with planar wave characteristics due to the small region of the reactive region and its uncommon evanescent wave characteristics.

\begin{table*}[t!]
  \centering
  \fontsize{8}{12}\selectfont
  \caption{Insightful distance boundaries and their characteristics.}
  \label{Distance_boundaries}
   \begin{tabular}{ !{\vrule width1.2pt}  m{2.6 cm}<{\centering} !{\vrule width1.2pt}  m{1.6 cm}<{\centering} !{\vrule width1.2pt}  m{2.4 cm}<{\centering} !{\vrule width1.2pt} m{2.4 cm}<{\centering} !{\vrule width1.2pt} m{3.2 cm}<{\centering} !{\vrule width1.2pt} m{3. cm}<{\centering} !{\vrule width1.2pt}}

    \Xhline{1.2pt}
        \rowcolor{gray!50} \bf Distance boundary & \bf Ref.  & \bf Hardware design architecture & \bf Definition criterion & \bf Distance formula & \bf Distance characteristics \cr
    \Xhline{1.2pt}
    Rayleigh distance & \cite{[35]}, \cite{balanis2015antenna}, \cite{7942128}  & Arbitrary array aperture for both the BS and UE  & The maximum allowable phase discrepancy ${\pi}/8$ based on the first-order Taylor expansion &$d_{ra}=\frac{2(D_1+D_2)^2}{\lambda}$ &Classical distance boundary to distinguish the (radiating) near-field and far-field regions\cr\hline
    Fresnel distance & \cite{balanis2015antenna}, \cite{7942128}  & Arbitrary array aperture for both the BS and UE & The maximum allowable phase discrepancy ${\pi}/8$ based on the second-order Taylor expansion & $d_{fr}=0.62\sqrt{\frac{(D_1+D_2)^3}{\lambda}}$ & The distance boundary to distinguish the reactive near-field and the radiating near-field regions\cr\hline
    Bj{\"o}rnson distance & \cite{9723331}, \cite{2022arXiv220903082R}  &A square UPA-based BS with patch antennas and a point transmitter &The normalized antenna array gain is close to $1$. &$d_B=2\sqrt{N}L$ &$d_B$ can achieve almost maximum antenna array gain with a much smaller value than $d_{ra}$.\cr\hline
    Effective Rayleigh distance & \cite{[7]}  & A ULA-based BS and a single antenna UE &The normalized coherence is greater than $95\%$. &$d_{eff}=\epsilon \cos ^2\theta \cdot \frac{2D^2}{\lambda}$ &$d_{eff}$ is defined motivated by the perspective of array gain and directly influences the transmission rate.\cr\hline
    Uniform power distance & \cite{[32]} & A UPA-based BS with patch antennas and a single antenna UE &The power ratio between the weakest and strongest antenna elements is greater than a particular threshold. &Numerically computed based on a particular threshold & The uniform power distance embraces both the amplitude discrepancy and the phase discrepancy. \cr\hline

    \Xhline{1.2pt}
    \end{tabular}
  \vspace{0cm}
\end{table*}

$\bullet$ \emph{Bj{\"o}rnson Distance}: A significant distance boundary called the Bj{\"o}rnson distance was proposed in \cite{9723331} and \cite{2022arXiv220903082R}. The square UPA-based XL-MIMO with patch antennas was considered, where $N$ identical square patch antenna elements with area $A$ were incorporated, serving a point transmitter. The authors in \cite{9723331} and \cite{2022arXiv220903082R} defined \emph{normalized antenna array gain}, which was the ratio of a receive antenna array gain to the largest antenna array gain. The largest antenna array gain was achieved in the far-field region over the perpendicular planar wave. Thus, the normalized antenna array gain was close to $1$ with the propagation distance increasing to the Rayleigh distance. Based on the setting in \cite{9723331} and \cite{2022arXiv220903082R}, the Rayleigh distance could be computed as $d_{ra}=2NL^{2}/\lambda$, where $L=\sqrt{2A}$ was the diagonal of each square patch antenna element.

It is worth noting that the normalized antenna array gain is close to $1$ with a sufficiently large propagation distance. The authors in \cite{9723331} and \cite{2022arXiv220903082R} proposed the \emph{Bj{\"o}rnson distance} as $d_B=2\sqrt{N}L$. The normalized antenna array gain is close to $1$  when the propagation distances are beyond $d_B$. Different from the Rayleigh distance, the Bj{\"o}rnson Distance grows with the square of the number of antennas $\sqrt{N}$ instead of $N$. For the scenario with $N=10^4$ and $A=(\lambda/4)^2$, the authors in \cite{2022arXiv220903082R} observed that about $96\%$ of the maximum antenna array gain could be achieved for $d=d_B$ and the normalized antenna array gain was almost $1$ when the propagation distance reaches the Rayleigh distance. The Rayleigh distance is much larger than the Bj{\"o}rnson distance as $d_{ra}/d_B\approx 35$. This interesting finding implies that the Bj{\"o}rnson distance can achieve almost maximum antenna array gain with a much smaller value compared with the Rayleigh distance.

$\bullet$ \emph{Effective Rayleigh Distance}: The classical Rayleigh distance is introduced from the perspective of the maximum allowable phase discrepancy, which cannot directly impact the transmission rate. Accordingly, the array gain can embrace the near-field effects and directly determine the transmission rate. Thus, it is necessary to define a straight distance bound to capture the near-field effects from the perspective of the array gain. A refined Rayleigh distance was proposed in \cite{[7]}, called the effective Rayleigh distance.

The authors in \cite{[7]} first proposed a piecewise-far-field channel model to approximate the near-field channel based on ULA-based XL-MIMO and a single-antenna UE. The whole array was divided into multiple sub-arrays with a small number of antennas each. Furthermore, the UE can be regarded as located in the far-filed region from the view of each sub-array, even if the UE is located in the near-field region of the whole array. Based on this assumption, the piecewise-far-field channel model was proposed to approximate the near-field channel. Then, a new metric called the normalized coherence was defined, determined by the near-filed channel and its corresponding far-field approximation, which could directly influence the transmission rate. Correspondingly, the authors defined the effective Rayleigh distance $d_{eff}$ based on the criterion that the normalized coherence was greater than $95\%$ when the propagation distance was larger than $d_{eff}$. Accordingly, the authors computed the effective Rayleigh distance as $d_{eff}=\epsilon \cos ^2(\theta) \cdot 2D^2/\lambda$, where $\epsilon=0.367$ and $\theta$ was the angle between the center of ULA-based BS and the UE. As observed, the effective Rayleigh distance is smaller than the classical Rayleigh distance and is related to the direction angle $\theta$. The effective Rayleigh distance is defined from the perspective of array gain and directly influences the transmission rate. The effective Rayleigh distance provides essential insights for rate analysis and practical communications in XL-MIMO.
\begin{figure*}[t]
\centering
\includegraphics[scale=1]{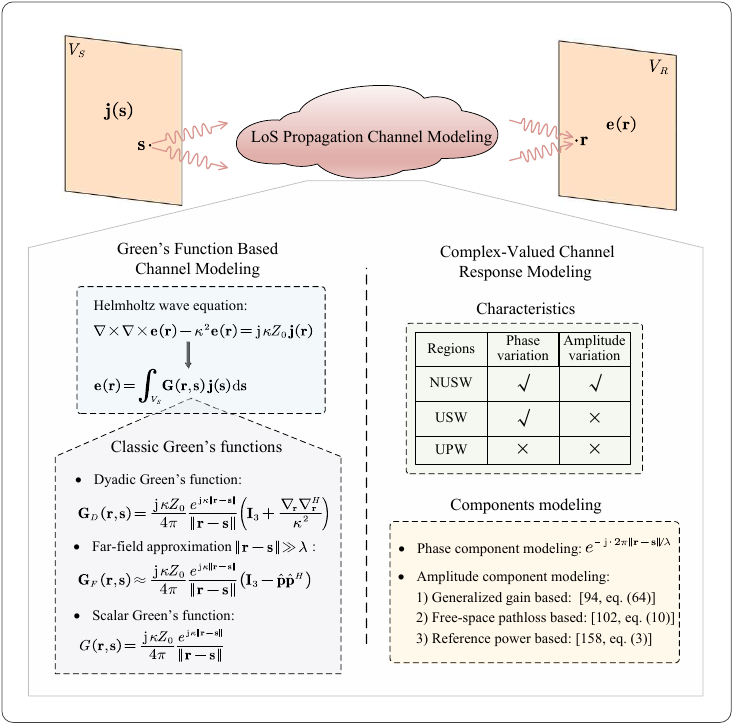}
\caption{The diagram for two methods for the LoS propagation channel modeling: Green's function based channel modeling and complex-valued channel response representation. $\mathbf{j}\left( \mathbf{s} \right) $ is the current distribution density at any arbitrary transmit point $\mathbf{s}$ inside the source volume $V_S$ and the electric field $\mathbf{e}\left( \mathbf{r} \right)$ at any arbitrary receive point $\mathbf{r}$ inside the receive volume $V_R$. $\mathbf{I}_3$ denotes a $3\times 3$ identity matrix. $\kappa =2\pi /\lambda $ is the wavenumber, $Z_0=376.73\Omega $ is the intrinsic impedance of spatial medium in free space, $\nabla _{\mathbf{r}}$ is the first-order partial derivative operator with respect to $\mathbf{r}$ and $\hat{\mathbf{p}}=\left( \mathbf{r}-\mathbf{s} \right) /\left\| \mathbf{r}-\mathbf{s} \right\| $.
\label{LoS}}
\end{figure*}

$\bullet$ \emph{Uniform-Power Distance}: As discussed above, the Rayleigh distance concerns the phase discrepancy caused by the wave's curvature but neglects the effect of the amplitude/power difference. However, the propagation distance would affect both the phase and amplitude. Thus, the conventional Rayleigh distance to distinguish the near-field and far-field regions is insufficient. The authors in \cite{[33]} proposed a refined distance boundary, called the \emph{uniform-power distance} (UPD) $d_{UPD}$, where UPA-based XL-MIMO with patch antennas served single UE. To construct this distance boundary, the authors defined the power ratio between the weakest and the strongest antenna elements. Then, UPD could be numerically derived as the minimized distance value which made the power ratio greater than a particular threshold (the authors considered $\alpha =90\%$ in \cite{[33]}). The UPD considered both the amplitude discrepancy and the phase discrepancy and was defined from the perspective of channel response characteristics and thus could be relied on to define three field regions with different channel characteristics \cite{[32]}, which will be carefully introduced in Sec. \ref{complex}.

We summarize the distance boundaries in Table~\ref{Distance_boundaries}. It is worth noting that the first two distance boundaries, i.e., Rayleigh distance and Fresnel distance, are shown to be applicable to arbitrary array aperture for both the BS and UE. However, the last three distance boundaries, i.e., Bj{\"o}rnson distance, effective Rayleigh distance, and critical distance were proposed based on particular scenarios, such as the scenario with a ULA-based BS and a single-antenna UE in \cite{[7]} and \cite{[33]}. Thus, the derivations for these three distance boundaries based on the generalized scenario with the arbitrary array aperture at the BS and the UE remain an open problem.

{\textit{{\textbf{Lessons Learned:}}}} To capture the EM characteristics for XL-MIMO, many distance boundaries can be defined based on different principles and perspectives as shown in Table~\ref{Distance_boundaries}. All these boundaries can give the researchers useful perspectives for the design and analysis of XL-MIMO and can be correspondingly chosen to model the EM characteristics for XL-MIMO. Besides, to promote further research on XL-MIMO, other distance boundaries can also be proposed to describe the EM characteristics for XL-MIMO systems better. The introduction of these distance boundaries can be motivated by the distance boundaries reviewed in this part.

\begin{table*}[t!]
  \centering
  \fontsize{8}{12}\selectfont
  \caption{Green's function based LoS channel modeling. SB-MU and SB-SU denote the scenario with single BS and multiple users and the scenario with single BS and single user, respectively.}
  \label{Green}
   \begin{tabular}{ !{\vrule width1.2pt}  m{2. cm}<{\centering} !{\vrule width1.2pt}  m{1.3 cm}<{\centering} !{\vrule width1.2pt}  m{3.8 cm}<{\centering} !{\vrule width1.2pt}  m{2.15 cm}<{\centering} !{\vrule width1.2pt}  m{6.1 cm} <{\centering} !{\vrule width1.2pt}}

    \Xhline{1.2pt}
        \rowcolor{gray!50} \bf Type & \bf Ref.  & \bf Hardware design architecture & \bf Communication scenario & \bf Characteristics \cr
    \Xhline{1.2pt}

        \multirowcell{15}{Dyadic Green's \\ function based} & \cite{[65]}  & UPA with point antennas & SB-SU & UPA-based transmitter and receiver \\
        \cline{2-5} & \cite{[97]}  & UPA with patch antennas & SB-MU  &UPA-based transmitter and receivers \\
        \cline{2-5} & \cite{[11]}, \cite{[70]}  & 2D CAP plane & SB-MU & 2D CAP plane based transmitter and receivers\\
        \cline{2-5} & \cite{[44]}  & 1D CAP line segment & SB-SU & 1D CAP line segment based transmitter and receiver\\
        \cline{2-5} & \cite{[9]}  & 1D CAP line segment & SB-SU & A practical case with unparallel 1D CAP line segments with arbitrary orientation and position is studied.\\
        \cline{2-5} & \cite{[29]}  & 1D CAP line segment and ULA-based XL-MIMO & SB-SU & The half space is considered and an approach to quickly compute the Green function based on the Sommerfeld identity is proposed.\\
        \cline{2-5} & \cite{2023arXiv230108411W}  & 1D CAP line segment and ULA-based XL-MIMO & SB-SU & Three combinations of the transmitter/receiver with 1D CAP line segment or ULA-based XL-MIMO is considered and a non-asymptotic analysis is implemented.
        \cr\Xhline{1.2pt}
        \multirowcell{5}{Scalar Green's \\ function based} & \cite{[38]}  & 2D CAP plane & Only one receiver without any transmitter & The 4D Fourier plane-wave representation for the LoS channel is derived based on the Weyl's identity.\\
        \cline{2-5} & \cite{9765526} & 2D CAP plane &SB-SU &The 2D Fourier plane-wave representation for the LoS channel is derived based on the Weyl's identity.\\
        \cline{2-5} & \cite{[13]} & 1D CAP line segment &SB-SU &Unparallel 1D CAP line segments with arbitrary orientation and position is considered.
        \cr\hline

    \Xhline{1.2pt}
    \end{tabular}
  \vspace{0cm}
\end{table*}

\subsection{LoS Propagation Channel Modeling}\label{LOSchannel}
Due to the extremely large array aperture for XL-MIMO, which shortens the transmission range, an LoS propagation link is common between the transmitter and the receiver. Furthermore, the LoS propagation channel is predominant in the EM transmission. Thus, it is necessary to derive a significant LoS propagation channel model based on different requirements. Besides, the receivers for XL-MIMO are likely located in the near-field region. Thus, it is necessary to consider the LoS propagation channel model capturing the near-field EM characteristics introduced above.

Many studies have modeled the LoS propagation channel. Among many existing works for XL-MIMO, two major LoS propagation channel modeling schemes are studied including Green's function based channel modeling and the complex-valued channel response representation channel modeling. The primary motivation of the Green's function based channel modeling scheme is to numerically solve Maxwell's equations, which can depict the electric field between each transmitting and receiving point. As for the complex-valued channel response representation scheme, the complex-valued channel response between each transmitting and receiving point is constructed, composed of the amplitude and phase components. These two LoS propagation channel modeling schemes can provide significant channel models for the analysis of XL-MIMO.

\subsubsection{Green's Function Based Channel Modeling}
As illustrated in Fig. \ref{LoS}, to characterize the radiated information-carrying EM waves in space, the homogeneous Helmholtz wave equation is applied to capture the relationship between the current distribution density $\mathbf{j}\left( \mathbf{s} \right) $ at any arbitrary transmit point $\mathbf{s}$ inside the source volume $V_S$ and the electric field $\mathbf{e}\left( \mathbf{r} \right)$ at any arbitrary receive point $\mathbf{r}$ inside the receive volume $V_R$. Then, the dyadic Green's function $\mathbf{G}_{D}\left( \mathbf{r},\mathbf{s} \right) \in \mathbb{C} ^{3\times 3}$ is introduced to explicitly express the relationship between $\mathbf{j}\left( \mathbf{s} \right) \in \mathbb{C} ^{3}$ and $\mathbf{e}\left( \mathbf{r} \right)\in \mathbb{C} ^{3}$ \cite{[65],[70],[44],WavesandFields,4685903}. Interestingly, the dyadic Green's function can be regarded as the system impulse response. Moreover, when the distance between the receive point $\mathbf{r}$ and the transmit point $\mathbf{s}$ is much greater than the wavelength, i.e.,  $\left\| \mathbf{r}-\mathbf{s} \right\| \gg \lambda$, the dyadic Green's function $\mathbf{G}_{D}$  reduces to its far-field approximated form $\mathbf{G}_{F}\left( \mathbf{r},\mathbf{s} \right) \in \mathbb{C} ^{3\times 3}$, which is also well-studied \cite{[65],[11]}. Note that the dyadic Green's function includes triple polarization. Another well-studied Green's function is the scalar Green's function  $G\left( \mathbf{r},\mathbf{s} \right)$, which is a scalar without the consideration of polarization effects. As shown in Fig. \ref{LoS}, the dyadic Green's function $\mathbf{G}_{D}\left( \mathbf{r},\mathbf{s} \right)$ can be derived by multiplying $G\left( \mathbf{r},\mathbf{s} \right)$ by the operator 
$( \mathbf{I}_3+{\nabla _{\mathbf{r}}\nabla _{\mathbf{r}}^{H}}/\kappa ^2 ) $ \cite{[65],[70]}. 

Meanwhile, as observed in \cite[Fig. 3]{[65]}, the consideration of three polarizations in the dyadic Green's function based channel can achieve about one-fold EDoF improvement compared with that of the scalar Green's function based channel since the dyadic Green's function follows a rigorous EM wave-physics solution. Meanwhile, the dyadic Green's function can capture the near-field evanescent wave characteristics with three polarizations and thus is applicable in  near-field based XL-MIMO systems. However, the scalar Green's function is more tractable and easier to analyze compared with the dyadic Green's function so that the Green's function in future research can be selected based on different research objectives.

The dyadic Green's function based channel modeling and the scalar Green's function based channel modeling have been widely studied, as summarized in Table~\ref{Green}. For the dyadic Green's function based channel modeling, the authors in \cite{[65]} and \cite{[97]} investigated UPA-based XL-MIMO with point antennas and patch antennas, respectively. The dyadic Green's function based channel for 2D CAP plane-based XL-MIMO was studied in \cite{[11]} and \cite{[70]}, where both the transmitter and the receiver were equipped with a 2D CAP plane. Moreover, the authors in \cite{[44]} and \cite{[9]} considered 1D CAP line segment-based XL-MIMO. A practical case with unparallel 1D CAP line segments with arbitrary orientation and position was studied. Besides, the authors in \cite{[29]} and \cite{2023arXiv230108411W} considered both ULA-based XL-MIMO and 1D CAP line segment-based XL-MIMO. The authors in \cite{[29]} modeled and analyzed the system from the perspective of the asymptotic analysis, but a non-asymptotic analysis was implemented in \cite{2023arXiv230108411W}. Moreover, the authors in \cite{[29]} considered the half-space and proposed an approach to quickly compute the Green function in the half-space based on the Sommerfeld identity. For the scalar Green's function based channel modeling, the authors in \cite{[39]} and \cite{9765526}  2D CAP plane-based XL-MIMO and applied the Weyl's identity to derive the Fourier plane-wave representation for the scalar Green's function based LoS channel. Moreover, 1D CAP line segment-based XL-MIMO was investigated, where unparallel 1D CAP line segments with arbitrary orientation and position were studied in \cite{[13]}. This Green's function based channel modeling scheme is motivated by the actual propagation of EM waves and can capture the actual EM characteristics, which is advocated by many studies for the channel modeling of XL-MIMO. 

\subsubsection{Complex-Valued Channel Response Representation} \label{complex}
To derive the LoS channel model, which is tractable and easy to analyze, the complex-valued channel response representation scheme is a notable method. The complex-valued channel response comprises the amplitude component and the phase component, which would show different characteristics in different EM regions discussed above. Moreover, many methods for the modeling of the amplitude component can be implemented based on different requirements and precisions, whereas the phase component is usually modeled in a unified form. With the aid of the complex-valued channel response modeling scheme, the analysis of XL-MIMO can be performed in a tractable fashion, and some insightful results can be obtained.

The complex-valued channel response can model the EM features between a particular transmitting and receiving point. Here, the complex-valued channel response would show different features in different field regions bounded by UPD introduced in Sec. \ref{Distances}. Based on the UPD in \cite{[32]}, non-uniform spherical wave (NUSW), uniform spherical wave (USW), and uniform planar wave (UPW) regions are defined.

\begin{itemize}
\item \emph{NUSW region}: NUSW region denotes the region which is smaller than the UPD $d<d_{UPD}$. The channel response indicates this region's non-uniform spherical wave feature. Both the amplitude variation and phase variation across the receiving aperture are noticeable. Thus, the exact propagation distance between the particular transmitting and receiving points should be considered in the modeling of the channel in this region.
\item \emph{USW region}: In this region, which is greater than the UPD and smaller than the Rayleigh distance $d_{UPD}<d<d_{ra}$, the channel response indicates the uniform spherical wave feature, where the phase variation over the receiving aperture is noticeable, whereas the amplitude variation over the receiving aperture can be neglected. In other words, the amplitude is equal across the receiving aperture. The phase should be precisely modeled based on the exact propagation distance between the transmitting and receiving points.
\item \emph{UPW region}: The channel response in this region, $d>d_{ra}$, displays a uniform planar wave feature, which is widely applied in conventional mMIMO. Since the propagation distance is much larger than the array aperture, both amplitude and phase variations across the receiving aperture can be negligible. Furthermore, the channel response can be modeled based on the planar wave assumption, where all antenna elements experience the same signal amplitude and AoA/AoD. The amplitude component depends only on the propagation distance between the center of the receiver and the center of the transmitter, and the incident angle determines the phase variation.
\end{itemize}

Interestingly, among the existing works, three major modeling methods for the amplitude component have been implemented based on different requirements and principles. In contrast, the phase component is modeled in a unified style $e^{-\mathrm{j}\frac{2\pi}{\lambda}\left\| \mathbf{r}-\mathbf{s} \right\|}$ as illustrated in the right part of Fig. \ref{LoS}. Three modeling methods for the amplitude component are comprehensively discussed as follows.

$\bullet$ \emph{Generalized Gain Based Modeling}: For the near-field channel modeling, three fundamental properties are vital: 1) The propagation distances vary across the array aperture; 2) The effective variable areas should be considered due to the different incident angles; 3) The losses caused by the polarization mismatch vary across the array aperture due to the received signal from various incident angles.

\begin{figure}[t]
\centering
\includegraphics[scale=0.8]{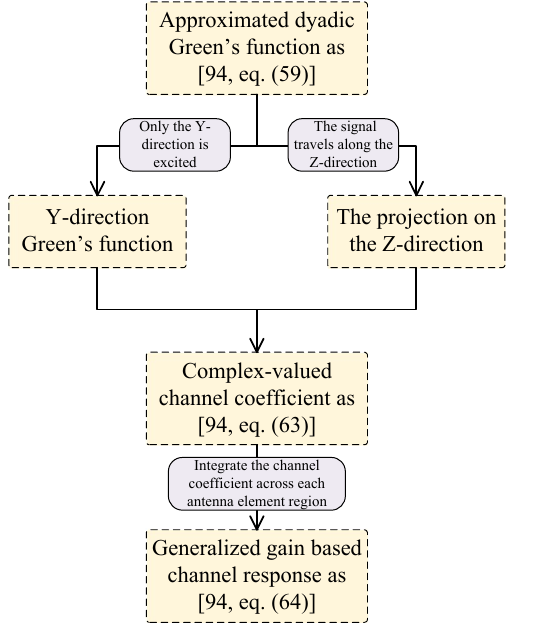}
\caption{The diagram for the generalized gain based modeling method \cite{[4]}.
\label{Response_Green_Function}}
\end{figure}

To capture all these fundamental properties, a generalized channel gain representation can be performed to form the amplitude component as shown in \cite{[4],2022arXiv220903082R,9723331}. The process of this modeling scheme is summarized in Fig. \ref{Response_Green_Function}. The authors in \cite{[4],2022arXiv220903082R,9723331} considered one receiver equipped with UPA-based XL-MIMO with patch antennas at the $XY$-plane and one lossless isotropic transmit point antenna. Firstly, the dyadic Green's function was introduced to express the relationship between the current distribution density $\mathbf{j}\left( \mathbf{s} \right)$ and the electric field $\mathbf{e}\left( \mathbf{r} \right)$ as \cite[eq. (59)]{[4]}. Moreover, a well-approximated dyadic Green's function was considered as \cite[eq. (60)]{[4]}, which was tight when the propagation distance was beyond the Fresnel distance. Secondly, the authors assumed that the signal traveled along the $Z$-direction and only the $Y$-direction of the $\mathbf{j}\left( \mathbf{s} \right)$ was excited as \cite[eq. (61)]{[4]}. Then, the $Y$-direction Green's function and the projection on the $Z$-direction were considered to model the complex-valued channel coefficient as \cite[eq. (63)]{[4]}. Finally, the complex-valued channel response for each antenna element was derived by integrating the channel coefficient across each antenna element region as \cite[eq. (64)]{[4]}. The above three fundamental properties for the near-field channel modeling were clearly presented explicitly in the expression of the channel coefficient as discussed in \cite[eq. (69)]{[4]}.

$\bullet$ \emph{Free-Space Pathloss Based Modeling}: As discussed above, the generalized channel gain representation can embrace three fundamental properties of the near-field channel modeling. Another choice for the amplitude component is to consider only the well-known \emph{free-space pathloss}, which is determined by the propagation distance and the wavelength. The authors in \cite{[7]} and \cite{[8]} studied the amplitude component for the scenario with one BS equipped with ULA-based XL-MIMO and one single-antenna UE based on the free-space path loss. Furthermore, a piecewise-far-field model with piecewise-linear phase characteristics was proposed in \cite{[7]} to approximate the near-field channel with high accuracy as shown in \cite[Fig. 3]{[7]}. This approximation was considered as a piecewise-linearization of the classical near-field channel model. The authors in \cite{[15]} and \cite{[14]} considered the scenario with one BS equipped with UPA-based XL-MIMO with patch antennas and multiple UEs equipped with multiple and single antennas, respectively. Due to the far smaller size of each patch antenna element than the propagation distance between each element and the UE, the propagation distance was assumed to be constant across each patch antenna element. Thus, the area of each patch antenna element was introduced when depicting the amplitude component as \cite[eq. (10)]{[14]} and \cite[eq. (12)]{[15]}.

$\bullet$ \emph{Reference Power Based Modeling}: As for this useful modeling method for the amplitude component, the reference gain was defined as the channel power $\beta _0$ at the reference distance $d_0=1 \ \mathrm{m}$ \cite{[33],[34],[92]}. Then, the amplitude component was denoted by the ratio of $\beta _0$ to the propagation distance. The authors in \cite{[33]} and \cite{[92]} investigated the BS-equipped ULA-based XL-MIMO and multiple or single UEs, respectively. Furthermore, UPA-based XL-MIMO with point antennas was considered in \cite{[34]} with multiple UEs.

{\textit{{\textbf{Lessons Learned:}}}} For the LoS propagation, the Green function based modeling \cite{[65],[97],[39],9765526} and complex-valued channel response representation schemes \cite{2022arXiv220903082R,[7],[33]} are introduced. As for the Green function based modeling scheme, which is available for both the discrete and continuous aperture, the diagram and characteristics are illustrated in Fig. \ref{LoS} and Table~\ref{Green}. For the complex-valued channel response representation scheme, which is available for the discrete aperture, three general modeling schemes are reviewed: generalized gain based modeling, free-space pathloss based modeling, and reference power based modeling, which can be considered based on different criteria and requirements. Note that the Green function based modeling scheme can capture the essential EM characteristics, and the complex-valued channel response representation scheme is more tractable and easier to analyze. The relationship between these two modeling schemes is inspired in Fig. \ref{Response_Green_Function}. Note that the Green's function based modeling scheme can accurately capture the EM wave characteristics.

\subsection{NLoS Propagation Channel Modeling}\label{NLOSchannel}
Due to the complex EM propagation environments, the NLoS propagation channel is also significant for XL-MIMO. The stochastic NLoS propagation channel model can represent a class of scattered propagation environments. When modeling the NLoS propagation channel, the complex scattered propagation environments and the near-field characteristics should be considered. Besides, as discussed above, the spatial non-stationarity should also be investigated since different regions of the array can experience different propagation parameters and environments when the array dimension is extremely large.

Several works modeled the NLoS propagation channel. Among these works, two NLoS propagation channel modeling schemes can be summarized: Fourier plane-wave representation based modeling and array response vector representation based modeling. More specifically, the 4D Fourier plane-wave representation for the NLoS propagation channel response over arbitrary scattering mechanisms can be derived based on the theory that every spherical wave can be decomposed exactly into an infinite number of plane waves \cite{chew1999waves,weyl1919ausbreitung}. On the other hand, the NLoS channel response can be represented by the superposition of a particular number of array response vectors or the array response matrices determined by the array response vectors. The array response vectors can incorporate the spherical wave characteristics in the near-field region.

\begin{figure*}[t]
\centering
\includegraphics[scale=0.6]{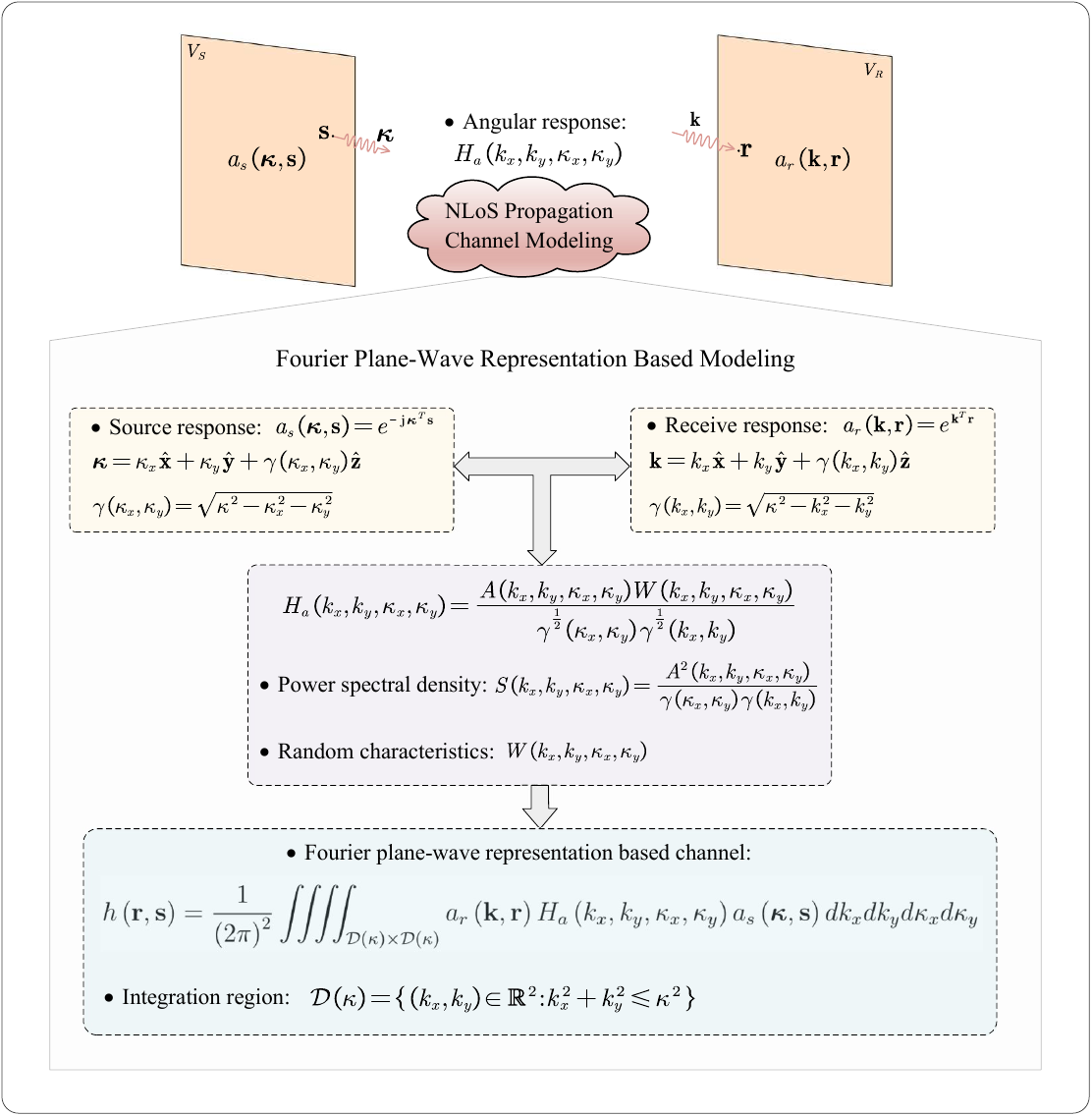}
\caption{The diagram for the Fourier plane-wave representation based NLoS channel modeling method. The Fourier plane-wave representation is composed of three parts: source response, received response, and angular response. $\boldsymbol{\kappa }$ and $\mathbf{k}$ are the corresponding wave vectors. $S\left( k_x,k_y,\kappa _x,\kappa _y \right) $ denotes the power spectral density and $W\left( k_x,k_y,\kappa _x,\kappa _y \right) $ represents the random characteristics.
\label{Fourier}}
\end{figure*}

\begin{figure*}[t]
\centering
\includegraphics[scale=0.9]{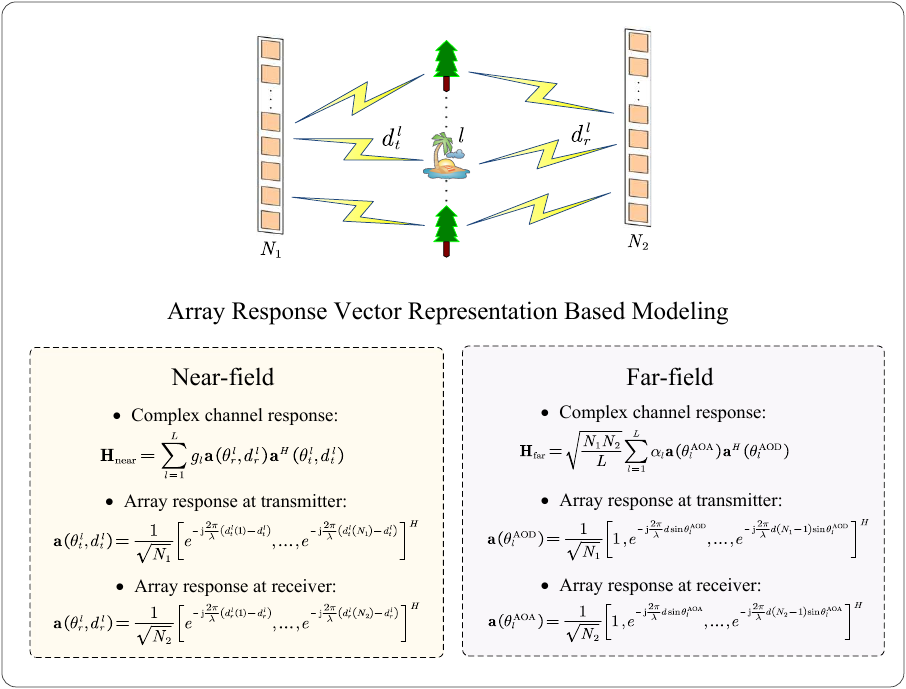}
\caption{Illustration of the array response vector representation based NLoS channel modeling. $N_1$, $N_2$, and $L$ denote the numbers of antennas for the transmitter array and the receiver array, and the number of paths, respectively. For the near-field, $\theta _{t}^{l}$ and $\theta _{r}^{l}$ denote the angle of the $l$-th path of the transmitter and receiver, respectively. $d_{t}^{l}$ and $d_{r}^{l}$ represent the distances between the $l$-th scatter and the center of the antenna array of the transmitter and receiver, respectively. $d_{t}^{l}\left( n_1 \right)$ and $d_{r}^{l}\left( n_2 \right)$ are the distances between the $l$-th scatter and the $n_1$-th element of the transmitter array and $n_2$-th element of the receiver array, respectively. For the far-field, $\theta _{l}^{\mathrm{AOD}}$ and $\theta _{l}^{\mathrm{AOA}}$ are the AoD and AoA for the $l$-th path, respectively. $d$ is the antenna spacing.
\label{NLOS_Vec}}
\end{figure*}

\subsubsection{Fourier Plane-Wave Representation Based Modeling}
Based on the theory that every spherical wave can be decomposed exactly into an infinite number of plane waves, one representative NLoS propagation channel modeling method, i.e., \emph{Fourier Plane-Wave Representation} can be adopted \cite{[41],9765526,[39]} as illustrated in Fig. \ref{Fourier}. Firstly, the 2D Fourier plane-wave representation can represent the incident field based on the Weyl's identity \cite{weyl1919ausbreitung}. Then, the received field does not need any external stimulus and is thus viewed as locally source-free. As a result, the received field obeys the homogeneous Helmholtz equation and can also be denoted by the 2D Fourier plane-wave representation. Moreover, these two 2D Fourier plane-wave representations can be coupled through a scattering kernel integral operator, which reflects the scattering mechanism and links all incident plane waves and every received plane wave. Thus, a 4D Fourier plane-wave representation for the NLoS propagation channel response over arbitrary scattering mechanisms can be given through three parts:
\begin{itemize}
\item \emph{Source response}: depicts the excitation current at the particular transmitted point to the certain transmit propagation direction, which is determined by the wavenumber of the transmitted field and the coordinate of the particular transmitted point;
\item \emph{Received response}: projects the certain received propagation direction to the induced current at an arbitrary received point, which is determined by the wavenumber of the received field and the coordinate of the particular received point;
\item \emph{Angular response}: maps each source propagation direction onto each received propagation direction. It can also be viewed to depict the channel coupling between every pair of the transmitted propagation direction and the received propagation direction, which is composed of the 4D power spectral density, capturing the scattering propagation environment and the random characteristics for the channel.
\end{itemize}
Furthermore, the discrete Fourier plane-wave series expansion can be applied by discretizing the 4D Fourier plane-wave representation based channel model to obtain the tractable and easily analyzed channel. Relying on the fact that the angular response is non-zero only within the particular wavenumber support, the discretized plane waves are defined within the lattice ellipses, and the discrete Fourier plane-wave series expansion can be derived by sampling on the finite integration area. Note that the approximation error reduces as the appropriate array size becomes large \cite{[41],9765526,[39]}.

The authors in \cite{[41]} and \cite{9765526} provided the above fundamental theories for the Fourier plane-wave representation based channel modeling with one CAP-based transmitter and one CAP-based receiver. The authors in \cite{[54]} extended the scenario in \cite{[41]} and \cite{9765526} for the single UE to the scenario, where the BS and multiple UEs were all equipped with UPA-based XL-MIMO with patch antennas. It was found that the number of antennas per plane should be larger than the number of sampling lattice ellipses to fully guarantee to capture the EM characteristics. Moreover, based on \cite{[41]} and \cite{[54]}, the Fourier plane-wave representation based channel modeling for the scenario with multiple BSs and UEs was investigated in \cite{HaoLei} by assuming that the random characteristics for each BS-UE path were independent. Besides, the authors in \cite{CompliantChannel} proposed an EM-compliant channel model based on the Fourier plane-wave series expansion based model in \cite{[41]}, where angular power spectrum, the distortion of antenna patterns, and the decrease of antenna efficiency were jointly investigated to derive a more practical channel model for XL-MIMO. All the above works focus on small-scale fading but neglect the effect of large-scale fading gain~\cite{du2019distribution}. Thus, it would be interesting to consider the Fourier plane-wave representation based channel modeling with large-scale fading in the future.

\subsubsection{Array Response Vector Representation Based Modeling}
Another promising NLoS channel modeling method is the array response vector representation based modeling. In this modeling method, the NLoS channel response is denoted by the superposition of a particular number of array response vectors or the array response matrices determined by the array response vectors \cite{[5],[6],[35],[61],[73],[87],[88]}. For convenience, we discuss the scenario where the transmitter and the receiver are equipped with ULA-based XL-MIMO. In this scenario, the NLoS channel comprises several array response matrices. Each array response matrix is determined by the complex NLoS path gain, the array response vector at the transmitter, and the array response vector at the receiver, as illustrated in Fig. \ref{NLOS_Vec}. In the near-field, the array response vector is determined by both the AoA/AoD and the distance, which can model the spherical wave characteristics. However, in the far-field, the array response vector is only determined by the AoA/AoD rather than the distance.

The authors in \cite{[6]} modeled the NLoS channel for the scenario with one BS equipped with ULA-based XL-MIMO and one single-antenna UE based on the near-field array response vector representation. Then, the near-field polar-domain representation for the near-field array response channel was proposed, which provided fundamental preliminaries for the channel estimation for XL-MIMO. Furthermore, the authors in \cite{[35]} investigated the NLoS channel for the scenario where both the BS and UE were equipped with ULA-based XL-MIMO. The double-side near-field channel model was accordingly defined. Moreover, the authors in \cite{[87]} and \cite{[88]} modeled the NLoS channel based on the far-field array response vector representation. 

\begin{figure*}[t]
\setlength{\abovecaptionskip}{-0.1cm}
\centering
\includegraphics[width = 0.9\textwidth]{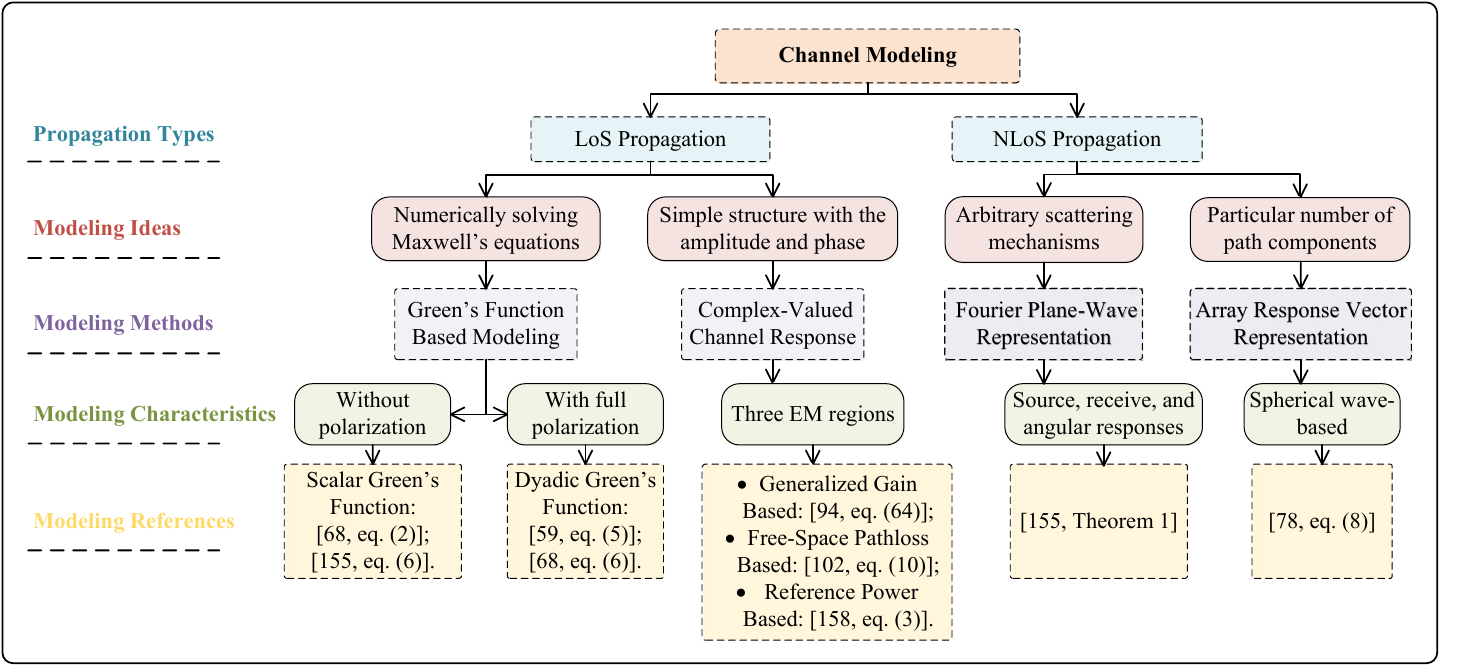}
\caption{Fundamentals of the channel modeling aspects for XL-MIMO.}
\label{Channel}
\end{figure*}

{\textit{{\textbf{Lessons Learned:}}}} To capture the fundamentals of the channel modeling for the LoS or NLoS propagation, a tutorial is given in Fig. \ref{Channel}, which can provide insights for the channel modeling for XL-MIMO \cite{ZheMag}. Certain channel modeling schemes can be chosen based on this tutorial. Firstly, the channel propagation type, including the LoS propagation and the NLoS propagation, should be chosen.

For the LoS propagation channel modeling, two major modeling methods based on different modeling ideas can be studied: Green's function based modeling method and complex-valued channel response representation method. The Green's function based modeling method is derived by numerically solving Maxwell's equations, which can capture the actual EM characteristics and is available for both the discrete aperture and the continuous aperture. Moreover, the dyadic Green's function and the scalar Green's function, distinguished by studying the polarization characteristics or not, can be applied by referring to \cite[eq. (5)]{[70]}, \cite[eq. (6)]{[65]} and \cite[eq. (2)]{[65]}, \cite[eq. (6)]{9765526}, respectively. The complex-valued channel response representation method is adopted by simply considering the amplitude component and the phase component, which is only available for the discrete aperture but is tractable and easy to analyze. Relying on this method, the characteristics for three classical EM regions can be indicated. Note that the phase component is usually modeled in a unified way and three major methods for the modeling of the amplitude component can be applied: generalized gain based method \cite[eq. (64)]{[4]}, free-space pathloss based method \cite[eq. (10)]{[14]}, and reference power based method \cite[eq. (3)]{[92]}.

For the NLoS propagation channel modeling, two major modeling methods can be adopted by the characteristics for scattering mechanisms. A novel Fourier plane-wave representation based modeling method can be applied to arbitrary scattering mechanisms. This method can be applied to both the discrete aperture and the continuous aperture, which is composed of the source, receive, and angular responses, referring to \cite[Theorem 1]{9765526}. For the array response vector representation based method, the array response contributed by a particular number of path components can be derived, which holds only for the discrete aperture and can capture the spherical wave characteristics \cite[eq. (8)]{[35]}.

\subsection{Hybrid Propagation Channel Modeling}\label{HybridChannel}
Channel modeling for the LoS propagation and the NLoS propagation has been well reviewed above. These channel modeling schemes are exact/approximate solutions to Maxwell's equations in specific scenarios. In practice, it is also possible to have a hybrid propagation channel that embraces both the LoS and NLoS propagation links. Thus, it is necessary to investigate the accurate and tractable channel modeling schemes for the hybrid propagation path channel. Moreover, channel modeling schemes assume that all scatters are either located in the far-field or near-field regions. In practice, a more generalized hybrid field should be developed, where some scatters are located in the near-field region, and others are located in the far-field region. Thus, it is necessary to investigate these two hybrid propagation channel modeling schemes to capture practical channel characteristics in XL-MIMO.

\begin{algorithm}[t]
\caption{A tutorial to generate the hybrid propagation path channel based on the complex-valued channel response representation method.}
\label{Hybrid_Complex}
\begin{algorithmic}[1]
\Require
Geometric characteristics, position coordinates, and EM parameters for XL-MIMO schemes.
\Ensure
Hybrid propagation path channel based on the complex-valued channel response representation.
\For{The complex-valued channel response representation method}
\State \textbf{LoS propagation component}: Generate the LoS propagation component in the complex-valued response form as \cite[eq. (25)]{[35]}.
\State \textbf{NLoS propagation component}: Generate the NLoS propagation component based on the complex-valued array response representation as \cite[eq. (21)]{[35]}.
\State \textbf{Hybrid propagation path channel}: Generate the hybrid propagation path channel by the combination of above two components as \cite[eq. (27)]{[35]}.
\EndFor
\end{algorithmic}
\end{algorithm}

\subsubsection{Hybrid Propagation Path Channel Modeling}
Many works have studied hybrid propagation channel modeling. Two major modeling schemes include the complex-valued channel response representation based modeling scheme and the Fourier plane-wave representation based modeling scheme.

$\bullet$ \emph{Complex-Valued Channel Response Based Modeling}: One promising modeling scheme is the complex-valued channel response based modeling scheme. For this modeling scheme, the LoS propagation component is modeled in the complex-valued response form, and the array response vector representation based modeling scheme depicts the NLoS propagation component with a particular number of scattering paths. Thus, the hybrid propagation path channel based on the complex-valued channel response modeling scheme can be derived by combining the above two components \cite{[35]}. For the scenario where both the BS and the UE were equipped with ULA-based XL-MIMO in \cite{[35]}, the authors firstly modeled the LoS propagation component as the complex-valued response form as \cite[eq. (25)]{[35]}. Furthermore, the NLoS propagation component could also be modeled in the array response vector representation as \cite[eq. (21)]{[35]}. Thus, the hybrid propagation path channel was derived as \cite[eq. (27)]{[35]}. To explain the hybrid propagation path channel modeling based on the complex-valued channel response representation, we provide a tutorial as Algorithm~\ref{Hybrid_Complex}.

$\bullet$ \emph{Fourier Plane-Wave Representation Based Modeling}:
Another promising modeling scheme can be implemented with Fourier plane-wave representation. The Fourier plane-wave representation method relies on that every spherical wave can be exactly into an infinite number of plane waves. Accordingly, both the LoS and NLoS propagation channels can be incorporated based on this modeling scheme \cite{9765526}. As studied in \cite{9765526}, the LoS propagation channel could be firstly derived based on the scalar Green's function through \cite[eq. (5)]{9765526}. Then, the scalar Green's function based channel could be exactly given by the 2D Fourier plane-wave representation based on the Weyl's identity as \cite[Lemma 1]{9765526}. Moreover, the NLoS propagation could also be derived based on the 4D Fourier plane-wave representation as \cite[Theorem 1]{9765526}, which had also been discussed above in detail. Thus, the hybrid propagation channel can be derived by combining these two components in a unified manner. To better demonstrate this modeling scheme, a tutorial, describing the hybrid propagation path channel modeling based on the Fourier plane-wave representation method, is given in Algorithm~\ref{Hybrid_Fourier}.

\begin{algorithm}[t]
\caption{A tutorial to generate the hybrid propagation path channel based on the Fourier plane-wave representation method.}
\label{Hybrid_Fourier}
\begin{algorithmic}[1]
\Require
Geometric characteristics, position coordinates, and EM parameters for XL-MIMO schemes.
\Ensure
Hybrid propagation path channel based on the Fourier plane-wave representation method.
\For{The Fourier plane-wave representation method}
\State \textbf{LoS propagation component}:  Generate the LoS propagation component by the 2D Fourier plane-wave representation method as \cite[Lemma 1]{9765526}.
\State \textbf{NLoS propagation component}: Generate the NLoS propagation component based on the 4D Fourier plane-wave representation method as \cite[Theorem 1]{9765526}.
\State \textbf{Hybrid propagation path channel}: Generate the hybrid propagation path channel by the combination of the above two components.
\EndFor
\end{algorithmic}
\end{algorithm}

\subsubsection{Hybrid-Field Channel Modeling}
In practical communications, some of the scatters are located in the near-field region and others are located in the far-field region. Thus, to capture this practical channel characteristic, the concept of hybrid-field channel has been introduced. The authors in \cite{[55],9940281} investigated the hybrid-field channel model, where the BS was equipped with ULA-based XL-MIMO, and the UE was equipped with a single antenna. Here, the NLoS channel was modeled based on the array response vector representation based modeling in \cite{[6]}, where both the scatters located in the near-field and the scatters located in the far-field were considered. More specifically, the channel components contributed by the scatters located in the near-field and those contributed by the other scatters were modeled separately based on the array response vector representation for the near and far fields, respectively. Then, the hybrid-field channel could be represented by the combination of near-field and far-field channel components, where an adjustable parameter was also introduced to control the proportion of two types of path components. This hybrid-field channel model provides fundamentals for hybrid-field based communications.

{\textit{{\textbf{Lessons Learned:}}}}
We discuss the hybrid propagation channel modeling from the perspective of the hybrid propagation path channel modeling and hybrid-field channel modeling. The hybrid propagation path channel modeling embraces both the LoS and NLoS propagation links. Two major methods can be implemented to model the hybrid propagation path channel: complex-valued channel response based modeling and Fourier plane-wave representation based modeling. Moreover, we provide two useful tutorials to inspire the hybrid channel modeling as shown in Algorithm~\ref{Hybrid_Complex} and Algorithm~\ref{Hybrid_Fourier}.

Based on these observations and insights, the EM characteristics in XL-MIMO, which differ from that of conventional mMIMO, are discussed. More importantly, the fundamentals for channel modeling are thoroughly reviewed, which are useful for performance analysis and signal processing design in XL-MIMO and provide vital guidance for future research on XL-MIMO.

\section{Signal Processing}\label{SignalProcessing}
Next, we focus on signal processing issues for XL-MIMO, in which we should capture the channel characteristics for XL-MIMO as presented in the previous sections to design efficient signal processing schemes. More specifically, channel estimation, beamforming scheme design, and deep learning-empowered processing are reviewed in this section. Again, due to the extremely large array aperture, signal processing schemes for XL-MIMO systems would involve very high computational complexity. Thus, to promote practical implementation and meet the green communication demands for future communications, low-complexity signal processing schemes for XL-MIMO should be developed. Thus, in this section, we summarize and motivate low-complexity signal processing schemes for XL-MIMO. Many low-complexity signal processing schemes with different channel characteristics, design ideas, and design algorithms are reviewed.

\subsection{Channel Estimation}\label{ChannelEstimation}
\begin{table*}[t!]
  \centering
  \fontsize{8}{12}\selectfont
  \caption{Characteristics for the channel estimation schemes for XL-MIMO. ULA, UPA (patch antennas), and UPA (point antennas) denote ULA-based XL-MIMO, UPA-based XL-MIMO with patch antennas, and UPA-based XL-MIMO with point antennas, respectively. }
  \label{CE}
   \begin{tabular}{                                                     !{\vrule width1.2pt}  m{1.9 cm}<{\centering}                         !{\vrule width1.2pt}  m{0.5 cm}<{\centering}                         !{\vrule width1.2pt}  m{2.5 cm}<{\centering}                         !{\vrule width1.2pt}  m{1.9 cm}<{\centering}                         !{\vrule width1.2pt} m{1.6 cm}<{\centering}                          !{\vrule width1.2pt} m{2.3 cm}<{\centering}                          !{\vrule width1.2pt} m{4.1 cm}<{\centering}   !{\vrule width1.2pt} }

    \Xhline{1.2pt}
        \rowcolor{gray!30} \bf Classification  &  \bf Ref. &  \bf Channel model &  \bf Channel characteristics  &  \bf Hardware design  &  \bf Channel estimation algorithm & \bf Algorithm characteristics \cr
    \Xhline{1.2pt}
        \multirow{10}{*}{ \shortstack{Polar domain \\ based schemes}} & \cite{[6]}  & Array response vector representation  & LoS + NLoS & ULA  &P-SOMP, P-SIGW  & Based on the sparsity in the polar domain and CS algorithms\\
        \cline{2-7} & \cite{[35]}  & Array response vector representation  & LoS + NLoS, DS-NF & ULA   & Hierarchical parameter and OMP-based algorithm & The LoS and NLoS path components are estimated separately.\\
        \cline{2-7} & \cite{[55]}  & Array response vector representation  & NLoS, hybrid-field & ULA  & HF-OMP & Based on the sparsity in the polar domain and CS algorithms\\
        \cline{2-7} & \cite{[98]}  & Array response vector representation  & LoS + NLoS & ULA    &DL-OMP  & Joint dictionary learning and sparse recovery based channel estimation methods\cr\Xhline{1.2pt}

        \multirow{7}{*}{ \shortstack{Parameter based \\ schemes }} & \cite{[18]}  &  Pathloss based channel model  & LoS & UPA (patch antennas)    & Iterative Parametric  algorithm & Exploiting the specific structure of the radiated beams generated by the continuous aperture\\
        \cline{2-7} & \cite{[21]}  & Array response vector representation  & NLoS, spatial non-stationarity & ULA  & Subarray-Wise algorithm, Scatterer-Wise algorithm & Based on a refined OMP algorithm and the structure of sub-array and scatterer\\
        \cline{2-7} & \cite{[8]}  & Free-space pathloss based model  & LoS & ULA  & Near-field rainbow based beam training & Based on the controllable beam split effect and the beam training \cr\Xhline{1.2pt}

        \multirow{3}{*}{ \shortstack{JACE \\schemes}} &\cite{[85]}   & Bernoulli-Gaussian random variable  & NLoS, spatial non-stationarity & UPA (patch antennas)  &Bilinear Message Passing for JACE & Based on the bilinear Bayesian inference framework\\
        \cline{2-7} & \cite{[27]}  & Bernoulli-Gaussian random variable  & NLoS, spatial non-stationarity & UPA (patch antennas)    &JACE Algorithm & Based on Gaussian approximation and bilinear inference\cr\Xhline{1.2pt}

        \multirow{3.5}{*}{ \shortstack{Low-complexity\\ schemes} } &\cite{[50]}   & Array response vector representation  & NLoS & UPA (patch antennas)    & RS-LS & Exploiting the array geometry and utilizing the compact eigenvalue decomposition\\
        \cline{2-7} & \cite{9547795}  & Frequency domain channel response  & NLoS, spatial non-stationarity & ULA  &Turbo-OAMP & Based on the LBP and OAMP algorithm\cr\Xhline{1.2pt}
         
        \multirow{7}{*}{ \shortstack{Machine \\learning based\\  schemes} }  & \cite{LeiCE}  & Array response vector representation & LoS + NLoS & ULA  &P-MRDN, P-MSRDN & Based on the sparsity in the polar domain, the MRDN, and the ASPP architecture\\
        \cline{2-7} & \cite{[94]}  & Complex-valued channel response   & LoS + NLoS & UPA (patch antennas)   &A two-phase HSPM algorithm & Based on the DCNN network and relations of parameters between the reference and remaining sub-arrays\\
        \cline{2-7} & \cite{2022arXiv221115939Y}  & Complex-valued channel response & LoS + NLoS, hybrid-field & UPA (point antennas)   &FPN based schemes&Based on existing iterative channel estimators with each iteration implemented based on an FPN\cr\Xhline{1.2pt}
        
    \end{tabular}
  \vspace{0cm}
\end{table*}

By deploying an extremely large number of antennas, XL-MIMO can achieve high DoF and spectral efficiency performance.
However, the benefits of XL-MIMO performance improvement rely on the channel state information (CSI) quality. Therefore, it is important to investigate efficient and applicable channel estimation schemes.
Again, in XL-MIMO, several near-field features should be considered in channel estimation, e.g., spherical wave characteristics, spatial non-stationarity, EM polarization property, and mutual coupling property.
In this subsection, we introduce the channel estimation schemes for XL-MIMO from the perspective of polar domain based, parameter based, joint activity and channel estimation, low-complexity based, and machine learning based channel estimation schemes as summarized in Table~\ref{CE}.
However, although some channel estimation schemes consider the characteristics for XL-MIMO channel, the computational complexity of these schemes is also too high.
Therefore, it is very meaningful to develop and study channel estimation schemes with high accuracy and acceptable complexity.

\subsubsection{Polar Domain Based Channel Estimation Schemes}
In conventional mMIMO systems, compressive sensing (CS) algorithms can achieve high normalized mean square error (NMSE) performance by utilizing the sparsity of the angular-domain channel.
Specifically, the orthogonal matching pursuit (OMP) algorithm transforms the signal into the angular-domain representation to estimate the channel based on the angular-domain sparsity.
However, different from conventional mMIMO systems, the channels for XL-MIMO systems do not exhibit sparsity in the angular domain, due to different electromagnetic characteristics \cite{[6]}.
Then, the existing angular-domain based CS algorithms cannot be directly applied to XL-MIMO.
Moreover, there are several CS algorithms that embrace spherical wave characteristics based on the polar domain in the near-field.

The authors in \cite{[6]} proposed a polar-domain representation of the XL-MIMO channel based on the array response vector, which fully captured the near-field spherical wave characteristics.
Considering both the angular and distance information, this channel representation exhibited sparsity in the polar domain.
Then, the polar-domain simultaneous orthogonal matching pursuit (P-SOMP) algorithm was investigated to obtain both the LoS and NLoS channel estimates.
Furthermore, the polar-domain simultaneous iterative gridless weighted (P-SIGW) algorithm, which directly estimated the near-field channel parameters, was proposed to improve the estimation accuracy.
Notably, the P-SOMP and P-SIGW algorithms outperformed existing angular-domain based SOMP and SIGW algorithms, which relied on the angular-domain channel sparsity.

The authors in \cite{[35]} extended the single-side near-field (SS-NF) in \cite{[6]} to the double-side near-field (DS-NF) where both the BS and the UE were equipped with ULA-based XL-MIMO and were located in each other's near-field region.
The authors proposed a DS-NF channel estimation scheme based on the DS-NF channel model where the LoS path component was modeled under the geometric free space assumption and the NLoS path components are modeled by the near-field array response vector.
In the DS-NF channel estimation scheme, a hierarchical parameter estimation algorithm was proposed to estimate the LoS path component, and an OMP based algorithm was utilized to estimate NLoS path components.
It was numerically demonstrated that the DS-NF scheme achieved the best performance out of both the SS-NF OMP scheme in \cite{[6]} and the conventional far-field OMP scheme.

An array response vector based hybrid-field channel model, which considered both the far-field and near-field path components, was investigated in \cite{[55]}.
By exploiting the hybrid-field channel feature and the channel sparsity in the polar domain, the hybrid-field OMP (HF-OMP) channel estimation scheme was utilized for the far-field and near-field path component estimation.
Simulation results showed that the proposed hybrid-field OMP scheme could achieve better NMSE performance than the far-field and near-field OMP algorithms.

A dictionary, called the polar-domain transform matrix, was constructed to fully capture the near-field channel sparsity in the polar domain by considering both the angular and distance information \cite{[6],[35],[55]}.
However, this transformation resulted in a high storage requirement and a high computational complexity \cite{[98]}.
Furthermore, the authors in \cite{[98]} proposed a distance-parameterized angular-domain sparse near-field channel representation model over array response vector based channels.
Based on the fact that the angle and distance were coupled, the size of the dictionary in \cite{[98]} only depended on the angular resolution instead of the resolutions of both the angle and the distance.
Furthermore, a dictionary learning orthogonal matching pursuit (DL-OMP) algorithm was investigated, which estimated the array response vector and updated the dictionary iteratively.
The DL-OMP algorithm outperformed the uniform and nonuniform P-OMP algorithms from \cite{[6]} in terms of not only the NMSE performances but also both the angle and distance estimation accuracy.

\subsubsection{Parameter Based Channel Estimation Schemes}

Different from the polar domain based channel estimation schemes, which directly recover the XL-MIMO channel, the parameter based channel estimation schemes aim to estimate the key parameters.
More specifically, to fully capture the spherical wave characteristics, the parameter based channel estimation schemes first estimate the angle and the distance parameters. Then, these two parameters are utilized to reconstruct the XL-MIMO channel.

A pathloss based channel model dominated by an LoS path was proposed in \cite{[18]} over millimeter and terahertz (THz) wave bands.
By simplifying the channel for the far-field, there were two parameters needed to be estimated.
The near-field channel could be represented as the superposition of the far-field channels by partitioning the continuous aperture into tiles which holds the far-field assumption.
Moreover, an iterative channel estimation algorithm was proposed by exploiting the specific structure of the radiated beams generated by the continuous aperture.
It was verified that the training overhead and computational complexity of the proposed estimation scheme did not scale with the number of antennas. Moreover, the proposed scheme outperformed benchmark schemes, especially in the poor scattering environment.

To explore the near-field spatial non-stationarity, an array response vector representation of XL-MIMO non-stationary channel was described through a mapping between sub-arrays and scatterers in \cite{[21]}.
Based on the refined OMP algorithm, a subarray-wise scheme was proposed for channel estimation by iteratively estimating and refining the position information.
Moreover, a scatterer-wise channel estimation scheme was investigated by positioning each scatterer and detecting its visibility regions to further improve accuracy.
It was numerically demonstrated that the subarray-wise method could recover the channel with low complexity, and the scatterer-wise method could accurately identify almost all the mappings between sub-arrays and scatterers. However, the scatterer-wise method utilized the multiple sub-array gain to estimate the positions and VRs of the scatters, thereby achieving much more accurate positioning and mapping results. Therefore, the subarray-wise method was suitable for the low-complexity subarray-based transceiver design. The scatterer-wise method entailed a substantially efficient and comprehensive globalized transceiver design.

The authors in \cite{[8]} proposed a fast wideband beam training scheme by utilizing the near-field beam split effect in the wideband XL-MIMO.
By controlling the near-field beam split effect, the proposed scheme could enable beams at different frequencies to be focused on different desired locations.
Furthermore, the distance and angle information could be obtained by the proposed beam training scheme to estimate the complex channel response of the XL-MIMO system.
It was demonstrated that the proposed scheme could realize fast near-field CSI acquisition with a very low training overhead.

\subsubsection{Joint Activity and Channel Estimation Schemes}

In the joint activity and channel estimation (JACE) schemes, not only spatial non-stationarity but also user activity patterns are considered.
Due to the spatial non-stationarity, XL-MIMO forms VRs, which leads to a subarray-wise sparse structure of the channel.
These schemes suppose that the user activities are operated in grant-free access mode, in which only a fraction of the potential users is active during a given time slot.
Therefore, the channel possesses a doubly-sparse and user-specific structure which can be modeled by a nested Bernoulli-Gaussian distribution.

A generalized gain based model called the nested Bernoulli-Gaussian model was proposed to capture the spatial non-stationarity, user activity patterns and channel fading in the grant-free XL-MIMO system in \cite{[85]}.
Based on the nested Bernoulli-Gaussian model, the authors proposed a novel bilinear inference-based JACE algorithm by decomposing the nested Bernoulli-Gaussian random variable into a bilinear inference problem of two independent random quantities.
The proposed algorithm was effective for grant-free access and could achieve the genie-aided ideal estimation performance.
Furthermore, different from the most closely-related bilinear generalized approximate message passing algorithm, the authors in \cite{[27]} employed the belief combining strategy tailored for this specific bilinear problem and proposed an iterative JACE algorithm.
To further enhance the practicality of the proposed algorithm, an EM-based auto-parameterization method was proposed to estimate instantaneous sub-array activity factors in \cite{[27]}.
The effectiveness of the proposed algorithm was revealed by the performance assessments via Monte-Carlo simulations in the NMSE and activity error rate (AER) performance.

\subsubsection{Low-Complexity Channel Estimation Schemes}

Channel estimation schemes for XL-MIMO suffer from high computational complexity due to the sharp increase in the number of antennas. Therefore, the trade-off between performance and complexity is an important challenge. The conventional minimum mean-squared error (MMSE) channel estimation is rarely adopted in XL-MIMO due to its high complexity.
Then, the conventional least-squares (LS) scheme, which requires no prior statistical information, was used to estimate the channel of XL-MIMO with low computational complexity in \cite{[81]}, \cite{[80]}.
However, the LS estimation scheme achieved a poor performance at the low signal-to-noise-ratio (SNR) scenarios.
Thus, it is necessary to investigate low-complexity channel estimation schemes with high accuracy and acceptable complexity.
Next, we introduce two low-complexity schemes in detail.

The authors in \cite{[50]} modeled the channel by providing an exact integral expression for the spatial correlation matrix with non-isotropic scattering and directive antennas where the BS was equipped with UPA-based XL-MIMO.
A novel and low-complexity channel estimation scheme, called reduced-subspace LS (RS-LS), was proposed based on the MMSE and the LS estimation schemes.
The proposed RS-LS scheme utilized the compact eigenvalue decomposition of the spatial correlation matrix and exploited the array geometry. To circumvent the knowledge of the correlation matrix of a particular UE, the authors utilized the compact eigenvalue decomposition of the correlation matrix in an isotropic scattering environment to construct the conservative RS-LS estimator. It was shown that the RS-LS scheme outperformed the conventional statistics-unaware LS scheme and yielded similar NMSE performance as the MMSE estimation scheme with increasing SNR at a much lower complexity. Moreover, the conservative RS-LS estimator achieved a $6\ \mathrm{dB}$ NMSE gain lower than LS but still had about $5\ \mathrm{dB}$ NMSE performance gap higher than RS-LS due to the higher rank of its applied isotropic correlation matrix than the actual correlation matrix.

The uplink channel of a BS using ULA-based XL-MIMO was modeled as a Bayesian network based on the generalized gain in XL-MIMO systems in \cite{9547795}.
The proposed model captured the structured sparsity in the antenna-delay domain with the non-stationary property.
Then, a low-complexity channel estimation scheme called the turbo orthogonal approximate message passing (turbo-OAMP) algorithm was proposed to efficiently perform the Bayesian inference.
The turbo-OAMP consisted of a linear minimum-mean-square-error (LMMSE) estimator and a non-linear MMSE estimator designed based on the loopy belief propagation (LBP) and OAMP. Note that the turbo-OAMP employed the LBP to achieve the approximate Bayesian inference instead of the MMSE module in conventional OAMP. In contrast to the existing state-of-the-art baselines, such as the turbo-OAMP with i.i.d. Bernoulli Gaussian prior \cite{7817805}, sub-array OMP \cite{[21]} and variational Bayesian inference (VBI) with the Dirichlet process \cite{8742766}, the proposed turbo-OAMP algorithm achieved better NMSE performance with substantially reduced pilot overheads, especially in low SNR scenarios.

\subsubsection{Machine Learning Based Channel Estimation Schemes}\label{MLCE}
In XL-MIMO systems, the aforementioned channel estimation schemes consider the near-field features and achieve better NMSE performance in their scenarios compared with the conventional schemes.
However, some estimation schemes cannot achieve satisfactory estimation accuracy. The high accuracy of some schemes comes at a cost of higher complexity. Recently, machine learning based channel estimation schemes have been explored to excavate the characteristics for the channel with both improved channel estimation performance and low complexity. Machine learning-based channel estimation methods can be divided into two categories: model-driven \cite{[new1],[new3]} and data-driven \cite{[new2],[new4]}. The existing model-driven methods are more applicable and tractable than the data-driven ones with the preferable interpretability, simpler modeling, and lower complexity for training \cite{[new1]}. Three practical residual neural networks were proposed to obtain accurate channel state information in the RIS-aided mmWave communication system in \cite{JinYu}.
The proposed neural network models leveraged the low-rank structure of RIS cascaded channels based on the similarity between the image noise reduction and channel estimation.
Moreover, the proposed models could achieve better NMSE performance than conventional CS algorithms and existing conventional channel estimation methods (e.g., the alternating direction method of multipliers and the denoising convolution neural network) with lower complexity.

The authors in \cite{LeiCE} investigated an XL-MIMO system with a ULA-based BS and a single-antenna UE. The polar-domain representation of the XL-MIMO channel, which was based on the array response vector in \cite{[6]}, was considered. Based on the multiple residual dense network (MRDN) architecture \cite{JinYu}, a polar-domain multiple residual dense network (P-MRDN)-based channel estimation scheme was proposed to utilize the channel sparsity in the polar-domain in XL-MIMO systems. Furthermore, a polar-domain multi-scale residual dense network (P-MSRDN)-based channel estimation scheme was proposed by exploiting the multi-scale feature integration capabilities of atrous spatial pyramid pooling (ASPP) in \cite{chen2018encoder}. Notably, the proposed P-MRDN and P-MSRDN outperformed the LS method, the orthogonal matching pursuit (OMP) algorithm \cite{4385788}, the polar-OMP (P-OMP) method \cite{[55]}, and the MRDN \cite{JinYu}. Although the proposed P-MRDN and P-MSRDN had higher computational complexity than the P-OMP, all these schemes involved a similar order of magnitude of running time.

The authors in \cite{2022arXiv221115939Y} and \cite{10001564} investigated the THz XL-MIMO system with a UPA-based BS and a single-antenna UE. Moreover, the THz hybrid-field channel was considered, where the complex-valued channel response modeling scheme was implemented. A unified and theoretical DL-based channel estimation framework called fixed point networks (FPNs) was proposed. The proposed framework leveraged existing iterative channel estimators, where each iteration was implemented based on an FPN. It was shown that the proposed schemes perfectly matched the XL-MIMO scenario due to the low complexity and the scalability of other heavily out-of-distribution scenarios.

In the THz XL-MIMO system, a complex-valued channel response based model, called hybrid spherical- and planar-wave channel model (HSPM), was investigated by accounting for the spherical-wave propagation and the parameter-shift effect among sub-arrays in \cite{[94]}.
Then, a two-phase channel estimation scheme was proposed to capture the features of the HSPM.
In the first phase, a deep convolutional neural network (DCNN) was developed to estimate the channel paraments of the HSPM with reduced complexity and high estimation accuracy.
In the second phase, the proposed algorithm explored the relations of parameters between the reference and remaining sub-arrays.
Then, the channel matrix was reconstructed for the channel estimation.
Compared to other methods, such as the OMP \cite{6847111}, AMP \cite{6566160}, CNN \cite{8752012}, and RNN \cite{8639163} based methods, the proposed channel estimation scheme converged fast and achieved high accuracy with 5.2 dB NMSE improvement and had substantially low complexity.

In XL-MIMO, channel estimation schemes should aim at not only the NMSE performance but also the complexity due to the extremely large number of antennas.
Several machine learning based channel estimation schemes have been reviewed above, and all these schemes achieve high NMSE performance with low complexity. Therefore, the machine learning based algorithms are absolutely suitable for XL-MIMO scenarios to realize the low-complexity channel estimation. Note that \cite{LeiCE,[94]} are data-driven and \cite{2022arXiv221115939Y} is model-driven. As for an intelligent and efficient method, machine learning still needs to be further examined in order to be applied in channel estimation in the future.

{\textit{{\textbf{Lessons Learned:}}}} 
In this part, we summarize five channel estimation approaches for XL-MIMO, detailed in Table~\ref{CE}. Compared with conventional mMIMO, channel estimation schemes for XL-MIMO focus on near-field characteristics, such as the spherical wave. Moreover, the complexity of channel estimation schemes in XL-MIMO is much higher than that of conventional mMIMO. 
Given that channel estimation complexity in XL-MIMO exceeds that in conventional mMIMO, we highlight low-complexity and machine learning-based methods to advance efficient channel estimation design.

Low-complexity channel estimation schemes in XL-MIMO rely on channel characteristics and efficient algorithms, such as the compact eigenvalue decomposition of the spatial correlation matrix \cite{[50]}, and the use of Loopy Belief Propagation for approximate Bayesian inference \cite{9547795} to design less complex schemes. 
Machine learning-based methods leverage analogies between image denoising and channel estimation, as well as neural networks' strong predictive capabilities, often outperforming other benchmarks in accuracy \cite{LeiCE,[94],2022arXiv221115939Y}.
These machine learning-based methods also aim to reduce the computational load by substituting high-complexity modules in traditional schemes with neural networks, enhancing simplicity and accuracy. 
For example, \cite{2022arXiv221115939Y} introduced a fixed-point network to replace the non-linear estimator in the orthogonal approximate message passing (OAMP) algorithm, yielding a lower-complexity, higher-accuracy alternative. The promising results of such machine learning-based channel estimation methods motivate researchers for further exploration.

The remaining three channel estimation strategies, i.e., polar domain-based, parameter-based, and joint activity and channel estimation schemes, employ different channel characteristics and design ideas, resulting in relatively higher complexities. Specifically, the polar domain-based schemes encounter increased complexity since the sampling dimension in the polar domain is several times the number of antennas.
Parameter-based channel estimation schemes focus on estimating key channel parameters (i.e., the distance and angle information) based on near-field characteristics and then reconstructing the channel utilizing these parameters. 
The parameter-based schemes often rely on the specific structure of the antenna array or channel, such as the radiated beams \cite{[18],[8]}, the subarray architecture, and the properties of the scatterer \cite{[21]}. Compared to the channel estimation schemes that estimate an entire channel, the complexity of parameter-based channel estimation schemes can be slightly reduced. 

All these channel estimation schemes are designed based on different channel characteristics, such as polar domain sparsity, channel parameters, and spatial non-stationarity, and different algorithms, including OMP, OAMP, and MRDN. 
Further research leveraging these important characteristics and methods to design applicable channel estimation schemes for XL-MIMO is interesting.

\subsection{Beamforming Schemes Design}\label{Processing}

The beamforming of XL-MIMO systems is changed from far-field beam steering to near-field beam focusing due to different EM characteristics from the planar wave to the spherical wave. Linear signal beamforming schemes, such as maximal ratio (MR), zero-forcing (ZF), and minimum mean-square error (MMSE), are practical candidates for mMIMO systems. As the number of antennas increases, the algorithm's complexity needs to be reduced further. In doing so, a few low-complexity beamforming schemes have been proposed. In this part, we review the beamforming schemes design for XL-MIMO from the perspective of conventional linear beamforming schemes, low-complexity beamforming schemes, and optimization design for beamforming schemes.

\subsubsection{Conventional Linear Beamforming Schemes}
With the increase in the number of BS antennas in XL-MIMO systems, conventional linear beamforming schemes are able to achieve near-optimal performance, and thus are more attractive than nonlinear beamforming schemes. Due to the discrepancy between planar and spherical waves, the schemes addressing the far-field beam split no longer work well in the near-field, posing challenges to XL-MIMO communications.

$\bullet$ \emph{ZF Based Beamforming Schemes}: It is common to use linear ZF precoding, which offers near-optimal performance and low complexity. The authors in \cite{[3]} considered the scenario with UPA-based XL-MIMO with patch antennas serving two single-antenna users. They compared different precoding schemes and found that ZF precoding outperformed MR. More specifically, a reconfigurable holographic surface-assisted low-earth-orbit (LEO) satellite communication system was designed by the authors in \cite{[14]}. Moreover, they developed a holographic beamforming algorithm for sum rate maximization and derived a closed-form optimal holographic beamformer. To maximize ZF precoded XL-MIMO downlink SE, the authors in \cite{[48]} proposed efficient resource allocation (RA) procedures. With the aid of the proposed RA procedures, the ZF based quasi-distributed algorithm could outperform the centralized one with approximate operation settings.

 $\bullet$ \emph{MR Based Beamforming Schemes}: Besides ZF precoding schemes, the MR precoding was considered in \cite{[4], [49]} studying the uplink SE achieved by single-antenna UEs communicating with LIS. Other recent studies \cite{[32],[92],[93]} considered the scenario that the BS was equipped with ULA-based XL-MIMO or UPA-based XL-MIMO communicating with multiple single-antenna users. Besides, the authors in those papers derived a closed-form expression of the resulting SNR with MR beamforming.

$\bullet$ \emph{MMSE Based Beamforming Schemes}: The authors in \cite{[20]} proposed an efficient channel modeling scheme for XL-MIMO systems, where the BS was equipped with ULA-based XL-MIMO or UPA-based XL-MIMO. They found that, unlike ZF and MR combining schemes, MMSE combining scheme presented an optimal performance in any studied scenarios, resulting in unlimited capacity regardless of any number of users, even under crowded configurations. The authors in \cite{[34]} studied the near-field modeling and performance analysis for XL-MIMO communication. They took into account the performance of MR, ZF, and MMSE beamforming schemes. The authors in \cite{[61]} investigated the multi-user communication performance with both the near-field and far-field UE and scatterers based on the MR and MMSE combining schemes.

\subsubsection{Low-Complexity Beamforming Schemes}
Utilizing the extremely large number of antennas mitigates many problems thanks to improved diversity gain. However, it also creates new implementation issues, one of which is the extremely high computational complexity. As the dimension of channels grows in XL-MIMO systems, linear precoding schemes such as ZF precoding require matrix inversions of considerable size, resulting in high complexity. Thus, many low-complexity beamforming schemes have been proposed to address this issue as summarized in Table~\ref{low-complexity-beamforming}.

\begin{table*}[t!]
  \centering
  \fontsize{8}{12}\selectfont
  \caption{Characteristics for low-complexity beamforming schemes.}
  \label{low-complexity-beamforming}
   \begin{tabular}{!{\vrule width1.2pt}  m{2.1 cm}<{\centering}  !{\vrule width1.2pt}  m{0.45 cm}<{\centering}      !{\vrule width1.2pt}  m{0.75 cm}<{\centering}   !{\vrule width1.2pt} m{1.8 cm}<{\centering}    !{\vrule width1.2pt} m{4.3 cm}<{\centering}   !{\vrule width1.2pt} m{5.4 cm}<{\centering}   !{\vrule width1.2pt} }

    \Xhline{1.2pt}
        \rowcolor{gray!30} \bf Classification  &  \bf Ref.  &  \bf UL/DL &  \bf Channel characteristics  &  \bf Design ideas & \bf Design characteristics \cr
    \Xhline{1.2pt}
        \multirow{8.5}{*}{ \shortstack{RK algorithms \\ based schems}} & \cite{[72]}    & UL & NLoS, spatial non-stationarity  &
        Applying three heuristic acceleration methods, SwoR, GRK, and RSK, in RK based RZF receivers  & The SwoR acceleration method achieves the best benefit-cost ratio, and the GRK method is robust in the severe IUI scenario.\\
        \cline{2-6} &  \cite{[43]} & UL & NLoS, spatial non-stationarity   & Applying RK algorithms to each sub-array to design distributed receivers & Distributed low-complexity receivers with RK algorithms applied at each sub-array and DLDF method at the CPU\\
        \cline{2-6} & \cite{BokaiICC}   & DL & NLoS, spatial non-stationarity  & Applying SwoR-RK algorithms to design the low-complexity precoding schemes &SwoR-RK algorithm achieves about $51.3\%$ complexity reduction compared with the traditional RZF algorithm with the SE performance approaching that of the RZF algorithm\cr\Xhline{1.2pt}

        \multirow{7}{*}{ \shortstack{Message passing \\based schemes}} & \cite{[2]}    & UL    & NLoS, spatial non-stationarity & Moving some processing with high computational complexity to LPUs in parallel & Various initialization options for VMP, data fusion strategies, and final detected schemes in the CPU can be applied. \\
        \cline{2-6} & \cite{[79]}   & UL & NLoS  & Combining the BP and VMP to construct a decentralized receiver & All LPUs decode the symbols in parallel and can exchange information with respective adjacent LPUs. \\
        \cline{2-6} & \cite{2022arXiv220510620H} & UL & NLoS  & Introducing the GNN module in AMP algorithm to cancel the multi-user interference
         & The proposed AMP-GNN scheme embraces both the low complexity for AMP detectors and the efficiency of GNN modules. \cr\Xhline{1.2pt}

        \multirow{14}{*}{ \shortstack{Matrix inversion \\ algorithms based \\schemes}} &\cite{[52]}   & UL & NLoS, spatial non-stationarity  &
        Computing the matrix inversion in EP detector for the sub-array based architecture recursively & The sub-array based EP detector is implemented in parallel and the matrix inversion in EP detector can be computed recursively as \cite[Algorithm 2]{[52]}.\\
        \cline{2-6} & \cite{[75]}   & UL & NLoS, spatial non-stationarity    &Applying PE to approximate the matrix inversion in each EP iteration & 
        The proposed PE-EP detectors have much lower computational complexity than that of the original EP detectors and can achieve the performance approaching that of the original EP detectors.\\
        \cline{2-6} & \cite{[54]}  & DL & NLoS, Fourier plane wave representation based   &Leveraging NS expansion to replace the matrix inversion in the ZF precoding scheme. & Applying 4 NS iterations to approximate the matrix inversion is efficient in achieving a good trade-off for the performance and complexity.\\
        \cline{2-6} & \cite{2023arXiv230513925X}    & DL & NLoS, spatial non-stationarity    &Applying Jac-PCG iterative inversion method to the matrix inversion in the RZF precoding scheme & The Jac-PCG algorithm based scheme achieves about 54\% reduction compared with the original scheme with the approaching close SE performance. \cr\Xhline{1.2pt}
        
    \end{tabular}
  \vspace{0cm}
\end{table*}

$\bullet$ \emph{Randomized Kaczmarz Algorithms Based Beamforming Schemes}: 
The randomized Kaczmarz (RK) algorithm, known for addressing systems of linear equations (SLEs), has been adapted to simplify linear signal processing in mMIMO. This method, as seen in \cite{[72]} and \cite{[43]}, has been utilized to design uplink receivers for XL-MIMO systems to minimize computational complexity while balancing performance. Specifically, in \cite{[43]}, RK-based distributed receivers were studied with the sub-array architecture and spatial non-stationarity. The RK algorithm was applied at each sub-array to find the solution of the SLE. Then, the distributed linear data fusion (DLDF) receiver at the CPU was implemented as in \cite{[1]}.

The authors in \cite{[72]} proposed three accelerated RK based RZF receivers, where three heuristic acceleration methods, i.e., the sampling without replacement (SwoR) technique, the greedy RK (GRK) algorithm, and the randomized sampling Kaczmarz (RSK) algorithm, were applied. 
The GRK was a greedy scheme that applied the complete residual information of the SLE to accelerate the convergence further. Different from GRK, the RSK applied partial residual information to solve the SLE with a much smaller amount of information compared with that of the GRK. Numerical results demonstrated that the RK-RZF based on the SwoR acceleration method could reduce the computational complexity by nearly $20\%$ and $70\%$ in the mMIMO scenario without spatial non-stationarity and in the XL-MIMO scenario with spatial non-stationarity, respectively. 
In addition, the GRK-RZF scheme was seen to be robust in the scenario with severe inter-user interference (IUI). 

The authors in \cite{BokaiICC} studied low-complexity precoding design and proposed an SwoR-RK algorithm-based precoding scheme. The SE and bit error rate (BER) performance were analyzed with the sub-array architecture and the spatial non-stationarity. Numerical results demonstrated that the computational complexity given by the RK algorithm and the SwoR-RK algorithm achieved about $51.3\%$ reduction compared to that of the traditional RZF algorithm. Additionally, the proposed algorithm was seen to achieve performance close to that of RZF precoding while maintaining a balance between SE performance and computational complexity.

$\bullet$ \emph{Message Passing Based Beamforming Schemes}: Some algorithms based on message passing schemes have been proposed to reduce the processing complexity of XL-MIMO. 

Considering the channel spatial non-stationarity and the existence of VRs, the authors in \cite{[2]} proposed uplink distributed receivers based on variational message passing (VMP). As shown in \cite[Fig. 3]{[2]}, VMP was first implemented at each local processing unit (LPU). Then the decoded symbols from the LPUs were fused at the CPU to derive the final detected symbols. Various initialization options for VMP, data fusion strategies, and final detected schemes in the CPU were considered with different processing complexities. Based on these various strategy choices, a variety of potential VMP receivers could be applied in different practical applications. Numerical results demonstrated that the proposed scheme outperformed the fully centralized one and approached the genie-aided match filter \cite{burchill1995matched}. More interestingly, the LPU aided schemes alleviated the processing complexity in a parallel manner.

To overcome three major challenges in XL-MIMO systems: computational complexity, scalability, and non-stationarities, the authors in \cite{[79]} proposed an uplink decentralized receiver based on a combination of belief propagation (BP) and VMP. In this scheme, all LPUs decoded the symbols in parallel and could exchange information with respective adjacent LPUs. This decentralized receiver, which was scalable, was seen to outperform other decentralized receivers and could achieve almost the same performance as a centralized benchmark scheme.

Moreover, the authors in \cite{2022arXiv220510620H} proposed a low-complexity graph neural network (GNN) enhanced approximate message passing (AMP) scheme, called AMP-GNN, which was a model-driven DL-based detector. The neural network structure was obtained by unfolding the AMP detector and incorporating the GNN module. The proposed processing scheme could efficiently cancel for the multi-user interference and achieve both the low complexity for the AMP detector and the efficiency of the GNN module. And the proposed AMP-GNN could significantly improve the performance of the AMP detector.

$\bullet$ \emph{Matrix Inversion Algorithms Based Beamforming Schemes}: Nevertheless, implementing the above conventional beamforming schemes can be challenging due to the channel matrix inversion operation, which can add significant computational complexity to the systems with large antenna arrays. The authors in \cite{[75],[52]} proposed low-complexity expectation propagation (EP) detectors for XL-MIMO systems. The authors in \cite{[52]} investigated the uplink sub-array based architecture, where the sub-array based EP detector was implemented at each sub-array in parallel. When considering the sub-array based architecture, the matrix inversion at the EP detector could be computed recursively as \cite[Algorithm 2]{[52]}. Moreover, the authors in \cite{[75]} studied uplink PE-EP detectors, where the matrix inversion in each EP iteration was approximated by polynomial expansion (PE). Based on the low-complexity PE approximation, the PE-EP detectors had much lower computational complexity, i.e.,  $\mathcal{O} \left( K^2 \right) $, than that of the original EP detectors, i.e., $\mathcal{O} \left( K^3 \right) $, where $K$ was the number of single-antenna UEs. Numerical results validated that the proposed PE-EP detectors could achieve performance approaching that of the original EP detectors, which showed a good balance between the performance and complexity.

At the same time, the authors in \cite{[54]} focused on the downlink multi-user communications and modeled the electromagnetic channel in the wavenumber domain using the Fourier plane wave representation. Besides, they proposed the leveraging Neumann series (NS) expansion to replace the matrix inversion in the ZF precoding scheme. More specifically, in the NS based method, the matrix inversion was decomposed into the main diagonal matrix and off-diagonal matrix, and then the matrix inversion could be represented based on these decomposed matrices. Numerical results demonstrated that the NS based ZF precoding could achieve similar performance to that of the original ZF precoding while avoiding high complexity matrix inversion. Moreover, applying 4 NS iterations to approximate the matrix inversion was efficient in achieving a good trade-off for the performance and complexity.

To reduce the computational complexity for the matrix inversion, the authors in \cite{2023arXiv230513925X} explored several iterative methods for computing the matrix inversion in the precoding scheme design for XL-MIMO systems. Several iterative methods to calculate the matrix inversion were applied: Gauss-Seidel (GS), Jacobi Over Relaxation (JOR), Conjugate Gradient (CG), and Jacobi-Preconditioning Conjugate Gradient (Jac-PCG) methods. The authors compared these methods from the perspective of SE performance, computational complexity, and convergence speed as summarized in \cite[Table II]{2023arXiv230513925X}. Compared with the JOR method, the GS method exhibited higher SE performance and faster convergence. However, the computations in the GS method could not be implemented in parallel and had high complexity. The CG method had low computational complexity compared with that of the GS method but achieved bad SE performance. Among these methods, the Jac-PCG method could achieve a good trade-off between the computational complexity and the performance of XL-MIMO systems. More specifically, the Jac-PCG method had lower computational complexity and higher SE performance compared with that of the GS method and CG method, respectively. And the Jac-PCG method could achieve about $54\%$ reduction compared with that of the traditional RZF precoding scheme with the approaching close SE performance.

\subsubsection{Optimization Design for Beamforming Schemes}

\begin{table*}[tp]
  \centering
  \fontsize{8}{12}\selectfont
  \caption{Characteristics for optimization design of beamforming schemes}
  \label{Optimization}
      \begin{tabular}{ !{\vrule width1.2pt}  m{0.6 cm}<{\centering} !{\vrule width1.2pt}  m{2.1 cm}<{\centering} !{\vrule width1.2pt}  m{0.7 cm}<{\centering}  !{\vrule width1.2pt}  m{2.8 cm}<{\centering} !{\vrule width1.2pt}  m{2.3 cm}<{\centering} !{\vrule width1.2pt}  m{3.1 cm}<{\centering} !{\vrule width1pt} m{2.8 cm}<{\centering} !{\vrule width1.2pt}}

    \Xhline{1.2pt}
        \rowcolor{gray!50} \bf Ref.  &  \bf Hardware design &  \bf UL/DL &  \bf Channel characteristics &  \bf Optimization objective  &  \bf Optimization variables  &  \bf Optimization methods \cr\Xhline{1.2pt}
        \cite{xu2023resource} & \makecell[c]{ULA} & \makecell[c]{DL}& NLoS & Sum spectrum efficiency maximization &Digital precoding and analog precoding matrices &Riemannian manifold optimization, alternating optimization, reinforcement learning, and AI-generated optimization \cr\hline        
        \cite{[14]} & \makecell[c]{UPA\\ (point antennas)} & \makecell[c]{UL}& LoS, free-space pathloss based & Sum rate maximization &Holographic beamforming scheme &Lagrange multiplier \cr\hline
        \cite{[59]} & \makecell[c]{ULA} & \makecell[c]{--}& LoS, free-space pathloss based & Spectrum efficiency maximization (only one pair of the transmitter and receiver) &The digital precoder, analog precoder, adjustable selection matrix, power allocation matrix, and the number of data streams &Alternating optimization\cr\hline
        \cite{[60]} & \makecell[c]{UPA \\(point antennas)} & \makecell[c]{UL}& \makecell[c]{LoS + NLoS} & Sum-mean-square-error-minimization &Beam combining schemes &Alternating optimization\cr\hline
        \cite{[68]} & \makecell[c]{UPA\\ (point antennas)} & \makecell[c]{DL}& LoS, free-space pathloss based & Sum rate maximization &DMA configuration and digital precoding scheme  &Riemannian manifold optimization and Alternating optimization\cr\hline  
        \cite{[70]} & \makecell[c]{CAP} & \makecell[c]{DL}& LoS, dyadic Green's function based & Sum rate maximization &Pattern functions and receiving combiners &Alternating optimization\cr\hline           
        \cite{[63]} & \makecell[c]{UPA \\(point antennas)} & \makecell[c]{UL}& \makecell[c]{NLoS}& Energy efficiency maximization &Transmit precoding design and DMA tuning strategy &Alternating optimization \cr\hline\Xhline{1.2pt}
    \end{tabular}
  \vspace{0cm}
\end{table*}

As discussed above, conventional linear and low-complexity processing schemes can be implemented based on different requirements. However, to further capture the superiority of XL-MIMO, the beamforming schemes can be optimized to further improve the corresponding system performance \cite{xu2023resource,9575181}. Several works consider the optimization of signal beamforming schemes as summarized in Table~\ref{Optimization}.

Compared to the far-field, near-field communications exhibit many distinct characteristics, which make near-field resource allocation different and are often modeled as a multi-objective joint optimization problem. The authors in \cite{xu2023resource} compared optimization tools, e.g., numerical techniques, reinforcement learning methods, and AI-generated optimization, for addressing near-field resource allocation, emphasizing their strengths and limitations. More specifically, the beamforming precoding matrix and power control strategy were studied as use cases. For the 
fully-digital and hybrid fully-connected precoding architecture design, the authors in \cite{xu2023resource} compared an AI-generated optimization method \cite{2023arXiv230805384D} with optimal beamforming \cite{6832894}, Riemannian manifold optimization \cite{[68]}, alternating optimization \cite{[68]}, and reinforcement learning \cite{ZihengTWC}. It was observed that the AI-generated optimization method was suitable in the complex and ever-changing scenario.

The authors in \cite{[14]} considered the downlink transmission, where the holographic beamforming and the digital beamformer were implemented. The free-space LoS propagation channel was modeled, where the BS was equipped with UPA-based XL-MIMO with point antennas, serving multiple LEO satellites equipped with multiple antennas. The ZF digital beamformer was implemented, and the holographic beamforming scheme was optimized to maximize the sum rate. The Lagrangian multiplier-associated method was implemented to solve the optimization problems. The proposed scheme had guaranteed robustness since the tracking errors of the satellites’ positions could have few effects on the sum rate.

The distance-aware precoding (DAP) scheme was proposed in \cite{[59]} to improve the capacity by applying the near-field effect. The authors considered the scenario with the transmitter and the receiver equipped with ULA-based XL-MIMO, where the digital precoder, the analog precoder, and the adjustable selection matrix were considered when transmitting the signals. Furthermore, the optimization problem of maximizing the SE was formulated. Then, the DAP algorithm was proposed as \cite[Algorithm 1]{[59]} to optimize the digital precoder, analog precoder, adjustable selection matrix, power allocation matrix, and number of data streams to maximize the SE performance based on the alternating optimization. The authors demonstrated that the proposed DAP precoding scheme could achieve a two-times increase in the SE performance compared to classical hybrid precoding schemes.

The authors in \cite{[60]} investigated the uplink transmission for the scenario where one BS equipped with UPA-based XL-MIMO with point antennas served multiple single-antenna UEs. Besides, the baseband combiner and the weight matrix, a configurable weight matrix denoting the array's response, were optimized to maximize the achievable sum rate. The equivalent matrix-weighted MMSE was applied to transform the sum-rate maximization objective to a simpler equivalent objective, where the equivalent matrix-weighted MMSE problem was convex to the optimized variables \cite{5756489}. Then, the alternating optimization method was adopted to derive a feasible solution. The proposed framework could effectively alleviate the sum-rate loss introduced by near-field and dual-wideband effects.

The authors in \cite{[68]} considered the downlink transmission for the scenario where one BS was equipped with UPA-based XL-MIMO with point antennas serving multiple single-antenna UEs. Three typical antenna architectures: fully digital, (phase shifters based-) hybrid, and dynamic metasurface antenna (DMA) architectures, were investigated. The beam focusing problem for the sum-rate maximization was formulated, and the alternating design algorithm was proposed to optimize the DMA configuration and the digital precoding scheme jointly. The authors observed that one could reliably simultaneously communicate with multiple UEs located in the same angular direction with different distances with the aid of the proposed beam-focusing precoding schemes.

It is worth noting that XL-MIMO can also generate any electric current density distribution in a desired manner. Thus, it is important to design the pattern, i.e., the electric current density distribution, to achieve improved performance. A pattern-division multiplexing technique was proposed in \cite{[70]} for the scenario where one BS equipped with a 2D CAP plane serving multiple UEs also equipped with 2D CAP planes. The sum capacity maximization problem was formulated to optimize the pattern. Furthermore, the weighted MMSE approach was adopted to derive the equivalent transformation of the sum capacity maximization problem. Then, the pattern functions and receiving combiners were designed based on iterative optimization. The pattern-division multiplexing technique showed the potential for XL-MIMO to control the electric current density to approach outstanding performance.

To promote the practical applications for XL-MIMO, the power consumption problem is also an important factor due to the extremely large aperture of XL-MIMO. It is vital to consider the energy efficiency (EE) maximization problem for sustainability~\cite{10032267}. The authors in \cite{[63]} investigated the EE maximization problem for the uplink transmission. An algorithmic framework to design the transmit precoding of each multi-antenna UE and the tuning strategy was considered, considering the instantaneous or statistical channel state information (CSI). Then, the algorithmic framework was implemented using Dinkelbach's transform, alternating optimization, and deterministic equivalent methods.

{\textit{{\textbf{Lessons Learned:}}}} Although conventional linear beamforming schemes can be applied in XL-MIMO, they may cause extremely higher computational complexity than that of mMIMO. Thus, many low-complexity processing schemes for XL-MIMO are discussed. All these schemes inspire the low-complexity design and promote the practical implementation of XL-MIMO. To design the low-complexity beamforming schemes, novel array architectures, such as the sub-array architecture with distributed learning, and efficient algorithms, such as the RK algorithms, VMP algorithms, and low-complexity matrix inversion algorithms, are applied. To promote further research on low-complexity beamforming schemes for XL-MIMO, intelligent array architecture and network architecture, such as the distributed architecture \cite{9113273} and modular architecture \cite{[93]} are advocated to provide flexible and efficient processing. Further, other low-complexity algorithms and machine-learning networks can also be applied to design lightweight beamforming schemes for XL-MIMO. Additionally, optimization design for beamforming schemes has also been reviewed. These studies can motivate further research on optimization design for XL-MIMO systems with different optimization scenarios, objectives, and methods to exploit the potential of XL-MIMO fully.

\subsection{Machine Learning Empowered Signal Processing}\label{LearningEmpoweredProcessing}

With the spherical wave propagation model, the traditional signal processing methods are no longer effective in the analysis of XL-MIMO systems with high accuracy due to near-field communication. However, the machine learning based methods can capture the inherent characteristics of the channel. Hence, the related algorithms, e.g., the machine learning based channel estimation, machine learning based beamforming, distributed learning, and so on, are proposed for different application scenarios~\cite{zhang2023generative,ni2022integrating}. The machine learning based channel estimation has been reviewed and motivated in Sec. \ref{MLCE} so that we focus on the machine learning based beamforming and distributed learning in this part.

\subsubsection{Machine Learning Based Beamforming}

Apart from the commonly used linear receive beamforming schemes, i.e., MRC beamforming, ZF beamforming, and MMSE beamforming, the codebook-based beamforming with a predefined set of beamforming vectors is of more interest in traditional mMIMO systems \cite{[new5]}. This scheme with high computational complexity makes it challenging to be applicable to different scenarios, especially the near-field effect incurred by the extremely large-scale antenna arrays of XL-MIMO systems. In contrast, the hierarchical methods, e.g., the deep neural network (DNN) based scheme \cite{[new5]} and long short-term memory (LSTM) network \cite{[new6]}, could show an advantage in solving the non-convex problem with the aid of deep learning. In the near-field domain, the distance and angle restrict the codebook design. In \cite{[73]}, the above factors regarding the optimal codeword could be estimated with the convolutional neural network (CNN), which extracts features from the wide far-field beams. In \cite{[new5]}, the authors considered a DNN-aided codebook-based beamforming architecture for imperfect CSI under multipath, which could be applied to high-speed scenarios.

\subsubsection{Distributed Learning}
Recently, distributed learning, a method leading the deployment of computing towards the network edge, has been investigated in MIMO systems \cite{[new9],[new10], yang2023detfed}. In XL-MIMO systems, the data processing burden and complexity depend on the number of antennas since the high-dimension and complicated matrix manipulation were introduced. In XL-MIMO systems, the extremely large number of antennas leads to a significant data processing and computational burden. Distributed learning, which can relieve the processing burden at the CPU, is regarded as a solution to tackle these problems \cite{Zhilongmag,Ziheng,ZihengTWC}. 

To mitigate the significant processing complexity and energy use in XL-MIMO systems, the authors in \cite{Zhilongmag} explored a cell-free paradigm using multi-agent reinforcement learning (MARL) to develop distributed strategies. The authors proposed a multi-agent cell-free XL-MIMO system, where the BSs, UEs, and even the antennas could be viewed as agents to learn to allocate resources and transmit signals efficiently. Two use cases of the MARL-aided multi-agent cell-free XL-MIMO system were studied: antenna selection and power control. For instance, with the help of MARL, the multi-agent deep deterministic policy gradient (MADDPG) based antenna selection scheme could achieve a uniform SE performance and $26 \% $ EE improvement compared with the scenario without antenna selection.

The authors in \cite{Ziheng} leveraged fuzzy logic to reduce the computational complexity and proposed two signal processing architectures, namely, centralized training and centralized execution (CTCE) with fuzzy logic and centralized training and decentralized execution (CTDE) with fuzzy logic. Both of these schemes combined MARL and fuzzy logic to optimize the power control strategy for maximizing SE performance. Numerical results showed that leveraging fuzzy logic could effectively reduce the computational complexity by $7.59 \% $ and $18.44 \% $ for CTCE and CTDE, respectively, while achieving better realizability in practical application scenarios than conventional MARL-based algorithms.

Unlike approaches that lessen the training requirements for MARL networks \cite{Ziheng}, the authors in \cite{ZihengTWC} combined a decoupled architecture with a mechanism for prioritizing experience selection. This strategy enhances the convergence rate by segregating global rewards and focusing on experiences with greater losses during training. Numerical results demonstrated that the proposed method achieved a faster convergence rate and comparable performance with the conventional methods. Specifically, compared with conventional MADDPG algorithms, the convergence rate of the proposed algorithm demonstrated a $65.41 \% $and $61.88 \% $ improvement in static and dynamic scenarios, respectively.

{\textit{{\textbf{Lessons Learned:}}}} We motivate the emerging deep learning empowered signal processing, which can enhance XL-MIMO systems and is regarded as the promising solution for the existing difficulties in XL-MIMO, such as the near-field aided processing and extremely high computational complexity. In summary, based on the observations and insights in this part, we highlight different approaches for designing and optimizing signal processing schemes for XL-MIMO.

\section{Application Scenarios}\label{application}
In this section, we discuss several application scenarios in which the XL-MIMO technology can be utilized effectively to enhance the performance of wireless communication systems. These application scenarios demonstrate the adaptability and potential of XL-MIMO in addressing the challenges and requirements of various systems.

\subsection{Physical Layer Security Enhancement}
Improving the security at the physical layer is a crucial aspect of wireless communication systems~
\cite{anaya2022leakage,du2022reconfigurable}. The XL-MIMO technology offers substantial potential for enhancing physical layer security (PLS). In particular, XL-MIMO techniques capitalize on the unique properties of spherical-wavefront propagation, which becomes crucial when the distances between mobile users and the base station are small~\cite{[8]}.

For example, the authors in~\cite{anaya2022leakage} demonstrated that incorporating spherical-wavefront propagation in XL-MIMO configurations yields substantial PLS benefits. One of the advantages is the protection against eavesdroppers in similar angular orientations, which can otherwise threaten communication security under planar-wavefront propagation. In addition, the authors developed a leakage subspace precoding technique for secure precoding and user scheduling~\cite{[101]}. This novel method increases the spectral secrecy efficiency by more than 40 percent compared to traditional zero-forcing approaches under various eavesdropper collusion strategies~\cite{anaya2022spatial}. This finding highlights the capability of XL-MIMO technology to enhance the PLS of wireless communication systems.

The authors in~\cite{anaya2022spatial} analyzed the distinctions between plane-wave (PW) and spherical-wavefront (SW) models regarding PLS for XL-MIMO systems with multiple eavesdroppers. They considered both colluding and non-colluding strategies of eavesdroppers, providing a comprehensive understanding of the PLS implications for various scenarios. The investigation determined that the propagation characteristics associated with the SW model resulted in higher secrecy rates than that of the PW model~\cite{[41]}. The SW model reduces the regions where an eavesdropper can render the attack, emphasizing the need to consider the appropriate propagation model when designing and evaluating PLS in XL-MIMO systems.

In addition, the authors in~\cite{anaya2022spatial} investigated the effects of realistic propagation conditions on the achievable secrecy performance of MIMO systems in the presence of an eavesdropper (Eve), with a particular emphasis on the $\kappa -\mu$ shadowed fading model~\cite{lopez2017kappa}. They derived closed-form expressions for secrecy performance metrics such as secrecy outage probability, probability of strictly positive secrecy capacity, and average secrecy capacity under two conditions: 1) the transmitter knew the CSI of the legitimate receiver (Bob) but not Eve's CSI, and 2) the transmitter knew the CSI of both the Bob's and Eve's channels~\cite{anaya2022leakage}. The study also highlighted the effects of various propagation conditions and the number of antennas on the performance of secrecy.

{\textit{{\textbf{Lessons Learned:}}}}
Based on the analysis of multiple studies, it is evident that XL-MIMO technology significantly enhances the PLS of next-generation wireless communication systems. Compared to planar-wavefront propagation, the distinctive properties of spherical-wavefront propagation contribute to enhanced secrecy performance and protection against eavesdropping threats~\cite{[89]}. The significance of considering the appropriate propagation model when designing and evaluating PLS in XL-MIMO systems and the potential of novel precoding and scheduling strategies to enhance secrecy spectral efficiency further are among the most important lessons learned. Understanding the influence of genuine propagation conditions, such as the $\kappa - \mu$ shadowed fading model~\cite{sanchez2021information,du2022rethinking}, on secrecy performance metrics enables the development of more secure and resilient wireless networks. 

\subsection{UAV Communications}
Unmanned aerial vehicles (UAVs) are increasingly being utilized for communication purposes in modern wireless networks due to their flexibility, mobility, and ability to provide aerial coverage and LoS channels~\cite{du2022performanceuav}. Integrating UAVs with XL-MIMO systems offers opportunities to enhance communication performance, energy efficiency, and coverage in various application scenarios. We review recent research on UAV communications, focusing on developing advanced channel models and system performance analysis for XL-MIMO-aided UAV communications.

The authors in~\cite{ma2021non} introduced a geometric 3D non-stationary channel model for mmWave band-wideband UAV MIMO communications. This model is based on a multilayer cylinder reference model and considers stationary and moving clusters encompassing the transmitter (Tx) and receiver (Rx). The proposed model accommodates local and remote clusters in propagation environments and uses a continuous-time Markov model to represent the clusters' dynamic properties. The authors in~\cite{ma2021non} also examined the space-time-frequency correlation function, quasi-stationary interval, and Doppler power spectrum and derived closed-form expressions for the survival probabilities of clusters. Alternatively, the authors in~\cite{[20]} proposed a double-scattering XL-MIMO channel model for accurately representing spatial non-stationarity in dynamic environments. The model includes two types of dispersion clusters, one near the BS and the other near the UE. The authors also evaluated the model for uniform linear and planar array (ULA and UPA) antenna configurations. Through Monte-Carlo simulations, performance metrics such as SINR, condition number (CN), and SE were analyzed for linear combiners MRC, ZF, and MMSE. Considering XL-MIMO-aided UAV communications, both~\cite{ma2021non} and~\cite{[20]} provide valuable insights for developing advanced channel models and system performance analysis. Specifically, the mmWave UAV MIMO channel model proposed in \cite{ma2021non} can be combined with the double-scattering XL-MIMO channel model in~\cite{[20]} to create a comprehensive channel model that addresses both UAV communication scenarios and XL-MIMO system characteristics. The proposed channel models and their analyses can serve as guidelines for designing effective UAV communication systems for future space-air-ground integrated networks (SAGINs) and 6G wireless networks.

In addition, the authors in~\cite{deshpande2022energy} investigated using RIS in UAVs to provide energy-efficient communication to ground users in densely populated urban areas. The objective is to reduce the network's overall energy consumption while maintaining a specific QoS for the users. In this context, the optimization problem entails joint UAV trajectory and RIS phase decisions in order to minimize the downlink transmission power of the UAV and base station. The authors proposed a successive convex approximation (SCA) method to resolve this problem. Simulation results demonstrate that the proposed algorithm can provide a guaranteed minimum rate while minimizing the transmission power of the UAV and BS. The proposed optimization framework in~\cite{deshpande2022energy} can be extended to XL-MIMO systems when immense antenna arrays and their potential to improve communication efficiency are taken into account~\cite{[91]}. By incorporating XL-MIMO features into the optimization problem, the energy consumption of UAVs and BSs can be reduced further.

{\textit{{\textbf{Lessons Learned:}}}}
Several key lessons can be learned from the aforementioned studies on UAV communications in XL-MIMO systems. Advanced channel models, such as those incorporating mmWave and double-scattering characteristics, are crucial for accurately representing UAV communication scenarios in XL-MIMO systems. Additionally, integrating RIS with UAVs in XL-MIMO systems can significantly improve energy efficiency and communication performance in dense urban areas~\cite{[36],[63]}. This highlights the importance of developing collaborative optimization frameworks considering UAV trajectory and network resources to efficiently design XL-MIMO UAV communication systems~\cite{8811733,[27]}.

\subsection{Integrated Sensing and Communications (ISAC)}
ISAC plays an increasingly crucial role in modern mobile networks as it offers comprehensive support for a wide range of applications, e.g., Metaverse~\cite{wang2022wireless} and AI-generated content~\cite{wang2023guiding}, by merging communication and sensing functionalities.
XL-MIMO technology has shown great potential for enhancing the performance of ISAC networks. Researchers have investigated various aspects in this context, such as radar sensing, near-field radio sensing, and accurate channel modeling. By combining the insights from these studies, we can better understand how XL-MIMO systems can be optimized for better performance in mobile networks.

The authors in~\cite{[86]} investigated radar sensing using XL antenna arrays in mobile communication networks, introducing a flexible model that accounts for both spherical wavefront and amplitude variations among array elements. This research surpasses conventional uniform plane wave (UPW) models. It offers valuable insights for designing and implementing XL-MIMO radar and XL-phased-array radar modes in mobile networks~\cite{[41]}. In a related study, the authors in~\cite{wang2023cram} investigated near-field radio sensing utilizing XL antenna arrays. Using a uniform spheric wave (USW) sensing model, they developed closed-form expressions for the Cramér-Rao Bounds (CRBs) for angle and range estimations in near-field XL-MIMO radar mode and XL-phased array radar mode~\cite{wang2023cram}. This study highlights the practical limitations of antenna array size in near-field sensing and provides insight into the tradeoffs between array size and estimation precision.

 {\textit{{\textbf{Lesson Learned:}}}}
XL-MIMO's inherent properties distinctly enable it to integrate radar sensing with communication functions in mobile networks. The very large spatial extent of XL-MIMO arrays not only enhances communication performance but also provides finer spatial resolution. This dual function is an important advantage of XL-MIMO over conventional systems. Spherical wavefronts and spatial non-stationarities, intrinsic to the mobile environment, must be modeled for maximum efficacy.

\subsection{Internet-of-Things}
IoT networks are integral to modern smart city applications, home automation, healthcare monitoring, and more, necessitating the capability to support massive numbers of interconnected devices and ensure consistent, reliable communication. As these networks continue to expand in both scope and density, traditional communication systems can become strained under the sheer number of connections. XL-MIMO technology has shown promising results in addressing these challenges by providing enhanced capacity and performance. The following studies provide insights into how XL-MIMO can be tailored for IoT networks, considering QoS, capacity improvement, and near-field properties~\cite{zhang2021optimizing,yang2023bu}.

The authors in \cite{[78]} addressed the issue of sustaining minimum QoS in congested IoT networks with limited resources. In high-density XL-MIMO systems, they introduced a QoS-aware joint user scheduling and power allocation method for downlink channels. This method consists of two sequential steps: a clique-based scheduling algorithm for user scheduling and optimal power allocation subject to transmit power budget and minimum achievable rate per user constraints. The paper presented a generalized non-stationary multi-state channel model based on spherical-wave propagation to evaluate the proposed technique. The multipath fading model and the path loss rule account for users in different channel conditions encountering different propagation characteristics \cite{[new5]}. By integrating these factors, the authors in \cite{[78]} contributed significantly to the design and optimization of IoT networks in densely populated environments.

In a related study, the authors in~\cite{[59]} explored the viability of XL-MIMO communication as an attractive technology for capacity enhancement in forthcoming 6G networks. They propose a DAP framework that exploits the near-field effect as a new capacity enhancement strategy. The DAP framework optimizes the number of activated RF chains and precoding matrices to align with the increased DoFs~\cite{[48]}, thereby promoting the scalability of IoT networks and enabling the integration of a large number of devices across a variety of applications while maintaining reliable and efficient connectivity.

{\textit{{\textbf{Lesson Learned:}}}}
Emerging IoT applications demand a communication framework capable of efficiently handling an immense number of devices. Navigating through high-density connections involves more than mere numerical management as it demands adept communication orchestration, particularly using QoS-aware joint user scheduling and distance-aware precoding architectures. XL-MIMO’s capacity extends beyond throughput enhancement, ensuring each IoT device, from low-power sensors to high-demand surveillance cameras, attains its communication needs efficiently and with minimal latency. This involves a strategic adaptation of communication parameters via techniques like adaptive modulation and coding (AMC) and is crucial in maintaining reliable, low-latency communication across a densely populated network. Utilizing sophisticated algorithms safeguards the communication integrity amongst an array of IoT devices, each with varied communication and performance requirements, highlighting the pivotal role of XL-MIMO in managing spectral and spatial resources effectively.

\subsection{Edge Computing}
Edge computing is a promising solution to support new multimedia services that demand ultra-low latency and extensive computational capabilities in resource-constrained end-user devices. In this context, XL-MIMO can provide reliable access links and guarantee uniform QoS. The following studies explore the potential of edge computing and XL-MIMO.

The authors in~\cite{2022arXiv220109057D} introduced a MIMO-enabled mobile edge network designed to meet the stringent requirements of advanced services. They created a joint communication and computing resource allocation (JCCRA) problem to minimize energy consumption while meeting the delay constraints~\cite{[36]}. They proposed a distributed cooperative solution strategy based on the multi-agent, deep deterministic policy gradient (MADDPG) algorithm. Compared to heuristic baselines, simulation results demonstrate the efficacy of the proposed strategy, emphasizing a reduction in total energy consumption of up to five times. This study serves as a guide that an integration of XL-MIMO and edge computing has a potential to facilitate energy-efficient, low-latency communication and processing in future wireless networks.

In a related study, the authors in \cite{[new9]} examined the issue of proactively caching popular content in IoT-enabled multi-cell MIMO systems to achieve low latency and decreased backhaul congestion. They modeled the fluctuations in content popularity (CP) over time as a Markov process and used reinforcement Q-learning to determine the optimal content placement strategy. However, traditional Q-learning encounters problems with many Q-updates. Thus, the authors proposed two (linear and non-linear) function approximations-based Q-learning approaches that require only a constant number of variable updates. Simulations confirmed the convergence of these approximation-based methods and their successful learning of optimal content placement.

 {\textit{{\textbf{Lesson Learned:}}}}
Analyzing recent studies reveals the crucial integration of edge computing and XL-MIMO in meeting future wireless communication needs. Specifically, XL-MIMO enhances edge computing by providing distinct spatial diversity via its large-scale antenna arrays, which is critical for ultra-reliable, low-latency communication in applications like real-time analytics and augmented reality. In a technical light, XL-MIMO’s notable spatial resolution, coupled with edge computing, supports optimal content caching strategies through machine learning approaches, such as approximated Q-learning. This combination effectively manages local data, reduces backhaul traffic, and augments user experience by decreasing latency and bolstering data retrieval reliability, while also paving the way for predictive analytics at the edge, enhancing system efficacy and responsiveness.

\subsection{Massive Connectivity}
The integration of XL-MIMO systems and edge computing is redefining wireless connectivity by enhancing spatial multiplexing and distributing computational resources optimally, thereby bolstering spectral efficiency and minimizing latency. A representative example of this shift is integrating satellite communication systems with terrestrial networks. This convergence heralds an anticipation of a 6G-centric future. Central to this evolution is the increasing role of Massive Machine-Type Communication (MTC) designed to serve the burgeoning number of devices and systems. The challenge of ensuring massive connectivity becomes paramount, emphasizing the need for innovative access mechanisms and advanced transmission technologies.

XL-MIMO stands out as a key enabler for massive connectivity, characterized by extensive antennas and large aperture arrays, which introduce challenges due to near-field access channels and associated high costs~\cite{mei2023massive}. A proposed solution involves an uplink grant-free massive access scheme that employs a mixed-analog-to-digital converter (ADC) architecture, effectively balancing access performance with power conservation. A significant insight from~\cite{mei2023massive} is the development of a two-stage orthogonal approximate message-passing algorithm, skillfully managing joint activity detection and channel estimation through spatial and angular domain structured sparsity.

XL-MIMO systems introduce the concept of the visible region (VR), wherein users perceive different array parts due to the spatial non-stationarity, leading to variable VRs and necessitating carefully designed connectivity mechanisms~\cite{liu2023location}. The VR model therein highlights the linkage between a user's physical location and its VR, while introducing methods like the VR-Net, which employs neural networks. Simulation results have indicated that while all proposed schemes achieve a high VR recognition accuracy with a substantial number of beacon users, VR-Net, despite its simplicity, demonstrates notable resilience and efficacy while outperforming the traditional Voronoi cell partitioning technique, especially when the number of training samples is limited.

Achieving true massive connectivity involves more than addressing access and spatial challenges as mMTC introduces additional complexities related to spatial non-stationarities and visibility regions across expansive XL arrays as detailed in~\cite{alves2023noma}. The proposed NOVR-XL solution merges non-orthogonal multiple access (NOMA) with XL-MIMO, specifically for random access in saturated mMTC XL-MIMO contexts. Notably, NOVR-XL allows multiple overlapping users to engage with the same XL subarray, improving the average sum rate from 160 to 520 bits/channel use when the average number of UEs contending by the pilot is 2, outperforming the strongest user collision resolution scheme.

\section{Future Directions}\label{future}
Several prospective future research directions can be identified to further develop and implement the emerging XL-MIMO technology.

\subsection{AI-Aided Resource Allocation Scheme}
The performance of XL-MIMO systems can be substantially enhanced by employing machine learning and artificial intelligence techniques~\cite{wang2023cram,du2023generative}. This includes the development of intelligent channel estimation methods, beamforming strategies, user scheduling, and resource allocation schemes that adapt to the ever-changing wireless communication environment. For example, XL-MIMO systems with their extremely large-scale antenna arrays can benefit from AI-based beamforming strategies since AI solutions can exploit the near-field environment effects dynamically to improve the SINR and SE \cite{[59]}. Additionally, AI-enabled user scheduling algorithms, such as the one proposed in \cite{[78]}, can optimize resource allocation for IoT networks by considering the high-density user distribution typical of XL-MIMO scenarios. Future research could delve deeper into the potential of generative AI, e.g., generative diffusion models~\cite{du2023generative,lin2023unified}, in generating the optimal resource allocation schemes according to the user behavior and network conditions in XL-MIMO systems.

\subsection{Energy Efficiency and Green Communication}
As XL-MIMO systems increase in antenna numbers and complexity, energy consumption becomes a significant issue. Ensuring sustainability and widespread adoption of XL-MIMO technology requires the development of energy-efficient processing algorithms, power allocation schemes, and antenna selection strategies~\cite{[59],2022arXiv220109057D}.  Specifically, distributed signal processing algorithms can offload complex computations to more energy-efficient Baseband Units (BBUs) in the cloud, alleviating the energy demands at the base station. Distributed computing techniques, when integrated into XL-MIMO architectures, can utilize the computing power of edge devices, further optimizing energy utilization~\cite{[new9]}. For instance, the authors in \cite{[new9]} demonstrated an application of reinforcement Q-learning for optimal content caching in IoT-enabled multi-cell mMIMO systems, achieving considerable energy efficiency gains. 
Furthermore, energy-aware activation of RF chains in XL-MIMO systems was proposed in ~\cite{[59]}. By leveraging energy-aware algorithms, the system can dynamically adjust the number of active RF chains based on real-time communication requirements. This strategy allows for a flexible response to the increased DoFs in the near-field region, ensuring an optimal balance between energy consumption and system performance.

\subsection{Semantic Communications}
Advancements in XL-MIMO system performance can be realized by adopting innovative communication paradigms such as semantic communications (SemCom)~\cite{yang2022semantic}. SemCom can improve network efficiency and reduce the burden of wireless data transmission by extracting and transmitting task-related semantic information from source data~\cite{luo2022semantic,lin2022blockchain}. In particular, the semantic information encoder can extract semantic features from the source data. These extracted features, being more compact than the original data, reduce the necessity for unnecessary wireless transmissions, thereby promoting the deployment of URLLC~\cite{xia2023wiservr}.
Additionally, the implementation of the XL-MIMO technique can enhance the efficacy of the SemCom system. More specifically, an attention feature module incorporated into the semantic encoder can utilize these extracted features to produce a series of scaling parameters~\cite{du2022attention,lin2023blockchain}. These parameters, symbolizing the significance of the semantic features, can guide resource allocation and beamforming design~\cite{wang2023semantic}. For instance, under conditions of constrained network resources, intricate beamforming schemes can prioritize the transmission of more critical semantic information, consequently ensuring error-free communication~\cite{getu2023making}.

\subsection{XL-MIMO-Aided Wireless Network Security}
The deployment of XL-MIMO technology introduces new security and privacy challenges due to its unique near-field characteristics and large-scale antenna arrays~\cite{[20],[86],du2022reconfigurable}. Addressing these issues necessitates the development of secure signal processing schemes, authentication mechanisms, and encryption techniques to counter diverse threats, including jamming and eavesdropping~\cite{zou2016survey}. Specifically, enhanced physical layer security can be achieved by harnessing the large-scale antenna arrays inherent in XL-MIMO systems to generate secure communication channels, thereby complicating the interception of transmissions by eavesdroppers and offering a degree of defense against jamming attacks~\cite{[20]}. Concurrently, the near-field effects unique to XL-MIMO systems provide an opportunity to devise innovative location-based authentication schemes, thereby ensuring network access is restricted to legitimate users within a specific spatial region and mitigating risks related to covert communications. Future research could potentially further exploit machine learning-based anomaly detection and intrusion prevention mechanisms tailored to the unique characteristics of XL-MIMO systems, thereby advancing the state of security in XL-MIMO-enabled wireless networks.

\subsection{Testbeds and Experimental Evaluation}
To validate the theoretical results and performance claims of XL-MIMO systems, it is necessary to conduct extensive experimental evaluations through the development of testbeds and prototypes~\cite{[78],[86],wang2023cram}. This can provide invaluable feedback for refining XL-MIMO technology's design, implementation, and optimization. For example, the work in~\cite{2022arXiv220109057D} demonstrates the effectiveness of a distributed approach for minimizing energy consumption in CF mMIMO-enabled mobile edge networks, which could serve as a basis for further experimental evaluation. Similarly, the DAP architecture proposed in~\cite{[59]} could benefit from experimental validation to confirm its effectiveness in exploiting the near-field effects for capacity improvement in 6G XL-MIMO networks. The development of XL-MIMO testbeds would also facilitate the exploration of various aspects, such as channel modeling, antenna array configurations, and user distribution, under realistic network conditions. In addition, the establishment of standardized benchmarks for the evaluation and comparison of different XL-MIMO algorithms and architectures would significantly contribute to the advancement of this technology. Overall, experimental evaluations and testbed development are crucial for bridging the gap between theoretical research and practical implementation of XL-MIMO systems.

\section{Conclusions}
The 6G of wireless systems is expected to push further the limits to support extreme connectivity, ultra-low latency, massive capacity, enhanced coverage, and green communications for connecting the physical and human worlds. By offering the unique characteristics of communicating in the radiative near-field, XL-MIMO is considered one of the promising technologies to support the ambitious goals of 6G. In this context, this paper has provided a comprehensive review of XL-MIMO, an integral part of 6G wireless communications.
Four promising XL-MIMO hardware designs and their characteristics were first introduced. Then, we presented various channel modeling schemes for XL-MIMO, which provided essential fundamentals and insights for the research on XL-MIMO. 
Furthermore, a comprehensive review and motivation of signal processing schemes, particularly those with low complexity, were conducted to facilitate the practical implementation of XL-MIMO. Last, many XL-MIMO-empowered application scenarios and future directions for XL-MIMO were highlighted. Our survey serves as a guideline for primary XL-MIMO research works in future 6G communications from the perspective of hardware design schemes, channel modeling, low-complexity signal processing design, XL-MIMO-empowered application scenarios, and promising future directions.

\section{Acknowledgment}
We would like to thank Hao Lei, Bokai Xu, Zhilong Liu, and Ziheng Liu for their discussions, comments, and feedback during the writing.

\bibliographystyle{IEEEtran}
\bibliography{IEEEabrv,Ref}

\end{document}